\documentclass[prb,epsf,aps,url]{revtex4}

\usepackage[toc,page]{appendix}
\usepackage{hyperref}
\usepackage{graphicx,pdflscape}
\usepackage{float}
\usepackage{color,subfigure}
\usepackage{bm}
\usepackage{alltt,dsfont}
\usepackage{appendix}
\usepackage{amsmath,amssymb}
\usepackage{placeins,float}
\usepackage{atveryend}
\makeatletter
\let\origcitation\citation
\AtEndDocument{\def\mycites{}%
  \def\citation#1{\g@addto@macro\mycites{#1^^J}\origcitation{#1}}}
\AtVeryEndDocument{\newwrite\citeout\immediate\openout\citeout=\jobname.cit
  \immediate\write\citeout{\mycites}\immediate\closeout\citeout}
\makeatother

\newcommand {\apgt} {\ {\raise-.5ex\hbox{$\buildrel>\over\sim$}}\ }
\newcommand {\aplt} {\ {\raise-.5ex\hbox{$\buildrel<\over\sim$}}\ }

\newcommand{\gssim}{\apgt}

\newcommand{\G}{{\cal{G}}}
\newcommand{\GH}{{\bf g}}

\newcommand{\tJ}{$t$-$J$  }
\newcommand{\beq}{\begin{eqnarray}}
\newcommand{\eeq}{\end{eqnarray}}
\newcommand{\barray}{\begin{eqnarray}}
\newcommand{\earray}{\end{eqnarray}}
\newcommand{\nn}{\nonumber}

\renewcommand{\AA}{{{\mathcal A}}}

\newcommand{\disp}[1]{Eq.~(\ref{#1})}

\newcommand{\refdisp}[1]{Ref.~(\onlinecite{#1})}
\newcommand{\figdisp}[1]{Fig.~(\ref{#1})}

\newcommand{\chem}{{\bm \mu}}

\usepackage{color}
\usepackage{bm}
\usepackage{alltt,dsfont}
\usepackage{appendix}
\usepackage{amsmath,amssymb}
\usepackage{hyperref}
\hypersetup{
    colorlinks,
    citecolor=red,
    filecolor=cyan,
    linkcolor=blue,
    urlcolor=magenta
}

\usepackage{atveryend}
\makeatletter
\let\origcitation\citation
\AtEndDocument{\def\mycites{}%
  \def\citation#1{\g@addto@macro\mycites{#1^^J}\origcitation{#1}}}
\AtVeryEndDocument{\newwrite\citeout\immediate\openout\citeout=\jobname.cit
  \immediate\write\citeout{\mycites}\immediate\closeout\citeout}
\makeatother

\begin{document}

\title{Extremely correlated fermi liquid of $t$-$J$ model in two dimensions}
\author{ Peizhi Mai and B. Sriram Shastry}
\affiliation{Physics Department, University of California,  Santa Cruz, CA 95064 }
\date{\today}
\begin{abstract}
We study the two-dimensional $t$-$J$ model with  second neighbor hopping parameter $t'$ and in a broad range of doping $\delta$ using a closed set of equations from the {\em Extremely Correlated Fermi Liquid} (ECFL) theory. We obtain asymmetric energy distribution curves and symmetric momentum distribution curves of the spectral function, consistent with  experimental data. We further explore the Fermi surface and local density of states for different parameter sets. Using  the spectral function, we calculate the resistivity, Hall number and spin susceptibility. The curvature change in the resistivity curves with varying $\delta$ is presented and connected to intensity loss in  {\em Angle Resolved Photoemission Spectroscopy} (ARPES) experiments. We also discuss the role of the super-exchange $J$ in the spectral function and the resistivity in the optimal to overdoped density regimes.

\end{abstract}
\maketitle


\section{Introduction}

The $t$-$J$ model where extreme correlations are manifest, plays a fundamentally important role in understanding the physics of correlated matter, including high Tc superconductors\cite{PWA,tJ-review1}. Despite the large progress\cite{DMFT2,DMFT3,DMFT4,dqmc,qca,ctqmc,Mills1,Mills2} made in numerically solving $t$-$J$ model and the related Hubbard model, very few analytical techniques are reliable to obtain the low temperature physics in this model for a broad range of dopings due to its inherent difficulties including non-canonical algebra for Gutzwiller projected fermions and the lack of an obvious small parameter for perturbation expansion. 

To tackle this challenge, we have recently developed the {\em extremely correlated Fermi liquid} (ECFL) theory\cite{ECFL,Pathintegrals}. It is a non-perturbative analytical theory employing Schwinger's functional differential equations of motion to deal with lattice fermions under extreme correlation $U\to \infty$. The ECFL theory uses a systematic expansion of a bounded parameter $\lambda \in [0,1]$, analogous to the expansion parameter $\frac{1}{2S}$ in the Dyson-Maleev representation of spins \cite{dyson} via canonical Bosons, and therefore provides a controlled calculation for $t$-$J$ model. With recent advances in the theory\cite{Edward-Sriram}, it is possible to represent the ECFL equations to any order in $\lambda$ in terms of diagrams which are generalizations of the Feynman graphs, without  having to consider previous orders. 

The second order ${\cal O}(\lambda^2)$ ECFL theory gives a closed set of equations for the  Green's function and has been described in detail in \refdisp{SP}. It has been benchmarked successfully \cite{Sriram-Edward,WXD} against the exact results from the single impurity Anderson model and the  dynamical mean field theory (DMFT) \cite{badmetal,HFL,DMFT1,DMFT2}, in the case  of the infinite dimensional large-U Hubbard model. The benchmarking has also been carried out in one dimensional $t$-$J$ model, where $k$-dependent behavior is inevitable, against the density matrix renormalization group (DMRG) technique. ECFL and DMRG compare well\cite{PSS} in describing the spin-charge separation in Tomonaga-Luttinger liquid and the relevant strongly $k$-dependent self-energy. 

Recently in \refdisp{SP}, we have applied the second order ECFL theory into studying the 2-d $t$-$J$ model with a second neighbor hopping parameter $t'$. We calculated the spectral function peak, quasi-particle weight, resistivity from hole-doping ($t'\leq 0$) to electron-doping ($t'>0$). The high thermal sensitivity in spectral function and small quasiparticle weight indicate a suppression of an effective Fermi-liquid temperature scale. The curvature of resistivity vs $T$ changes between concave and convex upon a sign change in $t'$, implying a change of the effective Fermi-liquid temperature\cite{WXD}.   We also compute the optical conductivity and the non-resonant Raman susceptibilities in \refdisp{raman}.

In the present work, we perform a more detailed study in 2-d $t$-$J$ model. Apart from the spectral function peak height, we compute the energy distribution curves (EDC) and momentum distribution curves (MDC) which are measured in the Angle-resolved photoemission spectroscopy (ARPES) \cite{Gweon}. For the first time from a microscopic theory, we obtain an asymmetric EDC line shape and a rather symmetric MDC line shape, which are consistent with experimental observation\cite{Gweon}. The self-energy is also calculated. It is independent of $k$ in the infinite-d limit\cite{Sriram-Edward} and has strong $k$-dependence in 1-d\cite{PSS}. In 2-d our calculation gives a weakly $k$-dependent self-energy in the normal (metallic) state.  
For this reason, we expect the vertex correction to be modest. Then we compute the resistivity within the bubble scheme neglecting the  vertex corrections. Unlike \refdisp{SP}, here we focus on the doping dependence of resistivity vs $T$ curves at different $t'$, corresponding to experimental observation \cite{Ando}. Spin susceptibility and the NMR spin-lattice relaxation rate are also calculated with the ECFL Green's function and related to experiment\cite{Walstedt,Walstedt-Book}. At the end, we discuss the effect of the super-exchange interaction and justify our choice of $J$.

This work is organized as follows: First we summarize the ECFL formalism to calculate electron Green's function and introduce parameter region in Section \ref{mp}. In Section \ref{results}, we discuss the ECFL spectral properties, resistivity, Hall response and spin susceptiblity at a fixed typical superexchange $J$, as well as the effect of changing $J$. Section \ref{conclusion} includes a conclusion and some remarks. 

\section{Method and Parameters}\label{mp}

\subsection{Summary of second-order ECFL theory}
In the ECFL theory \cite{ECFL} the one-electron Greens function in momentum space is expressed as the product of an auxiliary Greens function $\GH$ and a caparison function $\widetilde{\mu}$:   
\beq \G(k)=
  \GH(k) \times \widetilde{\mu}(k),  \label{eq1} \eeq 
where  $k\equiv(\vec{k}, i \omega_n)$ and $\omega_n=  (2n+1) \pi k_B T$ is the Matsubara frequencies. Here $\GH(k)$ is a canonical Fermion propagator vanishing as $1/\omega$ as  $\omega\to \infty$,  and $\widetilde{\mu}(k)$ plays a role of adaptive spectral weight due to the non-canonical nature of the problem. In the minimal version of second order theory \cite{Sriram-Edward} including superexchange $J$, they can be written explicitly  as
\beq
\widetilde{\mu}(k)= 1-\lambda \frac{n}{2} +\lambda \Psi(k) \label{mu2}
\eeq
 \beq 
 \GH^{-1}(k)=i\omega_n+{ \chem}- \frac{u_0}{2}+\frac{\lambda}{4}nJ_0 - \widetilde{\mu}(k)  \varepsilon'_k- \lambda \chi(k), \label{eq2}
 \eeq
where $\chem$ is  the  chemical potential, and $\varepsilon'_k= \varepsilon_k -\frac{u_0}{2} $. Here  $u_0$ is a Lagrange multiplier\cite{u0} guaranteeing the shift invariance of \tJ model at every order of $\lambda$. To elaborate,  $u_0$ absorbs any arbitrary uniform shift of the band $\varepsilon_k \to \varepsilon_k +c$,  a constant shift which should not change the results. The band dispersion includes next nearest neighbor hopping is $\varepsilon_k= - 2t (\cos(k_x a_0)+ \cos(k_ya_0)) - 4 t' \cos(k_x a_0) \cos(k_y a_0)$, and $\Psi$ and $\chi$ are two self energy parts. These are given by \cite{Sriram-Edward}
\beq
 \Psi(k)  = -   \sum_{pq}(\varepsilon'_p+\varepsilon'_q+J_{k-p})\GH(p)\GH(q)\GH(p+q-k),  \label{eq3}
\eeq
$\chi=\chi_0+\lambda\chi_1$ with $\chi_0= - \sum_p \GH(p) (\varepsilon'_p+ \frac{1}{2} J_{k-p})$, and 
\beq
 &&\chi_1(k)=
 -  \sum_{pq} (\varepsilon'_p+\varepsilon'_q +\frac{1}{2}(J_{k-p}+J_{k-q}))\nn \\
 &&\times(\varepsilon'_{p+q-k}+  J_{k-q}) \GH(p)\GH(q)\GH(p+q-k). \label{eq4}
\eeq
 where $\sum_k \equiv \frac{k_B T}{N_s} \sum_{k_x,k_y, \omega_n}$, $N_s$ is the number of sites and 
 $J_{k}=2J (\cos k_x a_0+\cos k_y a_0)$ is the nearest neighbor exchange.

Denoting the particle and hole density per-site by $n$ and $\delta=1-n$ respectively,  the two chemical potentials $\chem$ and $ u_0$ are determined through the number sum rules \beq
\sum_{k} \GH(k)\, e^{i \omega_n 0^+} = \frac{n}{2}= \sum_{k} \G(k)\, e^{i \omega_n 0^+}\,  . \label{sumrule}
 \eeq
  After analytically continuing $i\omega_n\to \omega+ i 0^+$ we determine the   interacting electron spectral function $\rho_{\G}(\vec{k},\omega)= -\frac{1}{\pi} \Im m \G(\vec{k},\omega)$. The set of Equations~(1-6)  was solved iteratively on $L\times L$ lattices with $L=61, 131, 181$ and a frequency grid with $N_\omega=2^{14}$ points. $L>61$ is usually for $t'\geq0$ at low temperatures where the spectral function peak is higher and sharper than the negative $t'$ cases; therefore it requires better $k$ resolution.

\subsection{Parameters in the programs}
In this calculation, we set $t=1$ as the energy scale and $t'$ is varied between $-0.4$ and $0.4$. We fix the superexchange to $J=0.17$ unless otherwise specified because $J$, usually is estimated to be  in the region from $0$ to $0.4$, and has a small effect on the $k$-dependent behavior and barely influences the averaged physical quantities like resistivity, since the calculation includes a summation in $k$ space. This argument will be further justified in the last part of Section \ref{results}. Besides, we also explore a large region of doping $\delta$ from $0.11$ to $0.3$, where the second order ECFL theory is reliable\cite{Sriram-Edward}, and present the $\delta$-dependent behavior at different $t'$. {\em  If not specified, $\omega$ is in  units of} $t$. According to \refdisp{tJ-review1}, we assume $t=0.45$ eV when using the absolute temperature scale.

\subsection{The sign of $t'$}
The significance of the sign of $t'$ should be kept in mind, and the case $t'>0$ is believed to correspond to electron-doped cuprate superconductors whereas $t'<0$ is the hole-doped cuprates. The hole-doped case appears highly non-Fermi liquid like as compared to the electron-doped case in experiments, and our earlier calculations as well as the present ones give a microscopic understanding of this important basic fact. We   emphasize   that, despite this, the $t'>0$ case is also strongly correlated, when we view the T-dependence of the spectral features, where  the effective Fermi scale is much reduced from the bare (band structure) value.

\begin{figure}[!]
\subfigure[\;\; $t'=0.4$]{\includegraphics[width=.44\columnwidth]{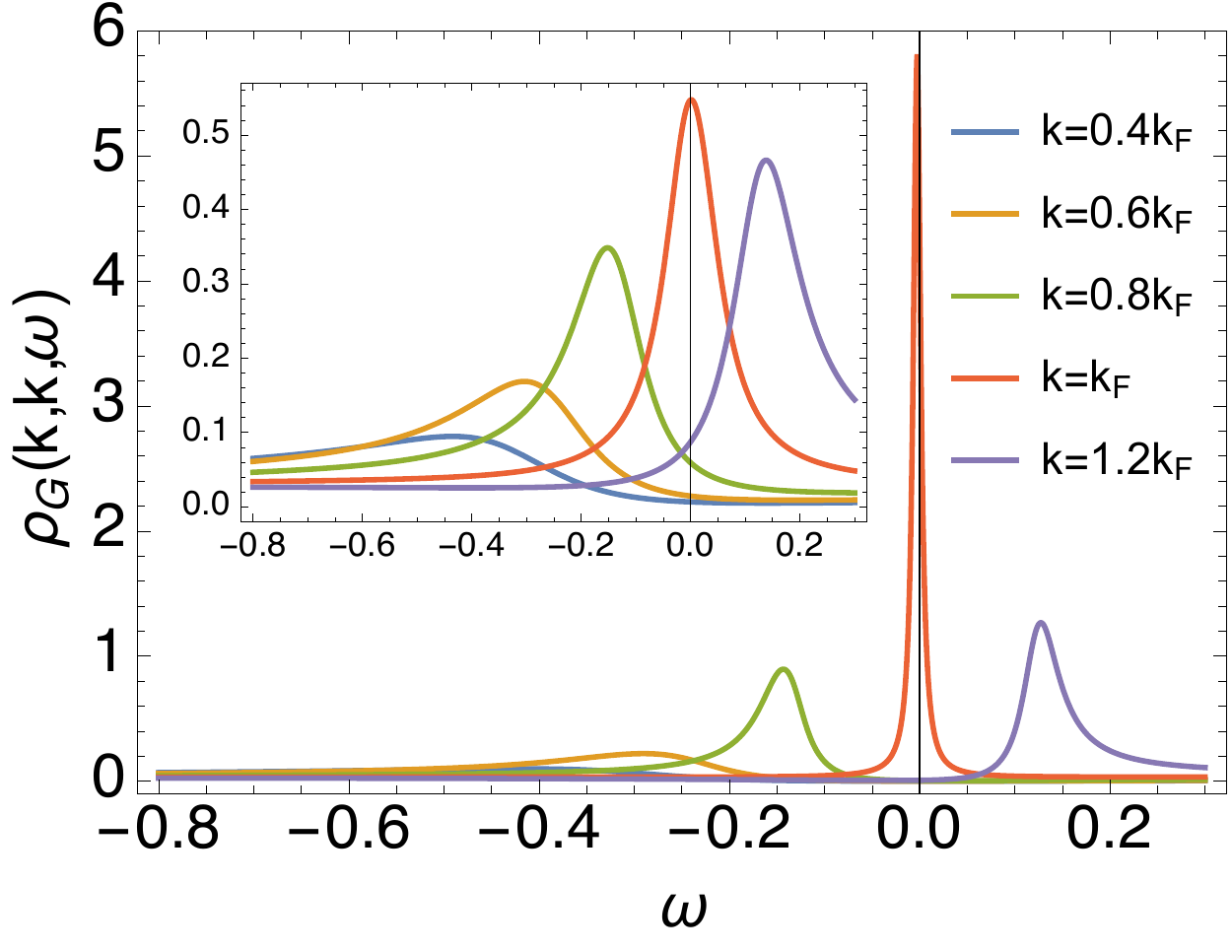}}
\subfigure[\;\; $t'=0.2$]{\includegraphics[width=.44\columnwidth]{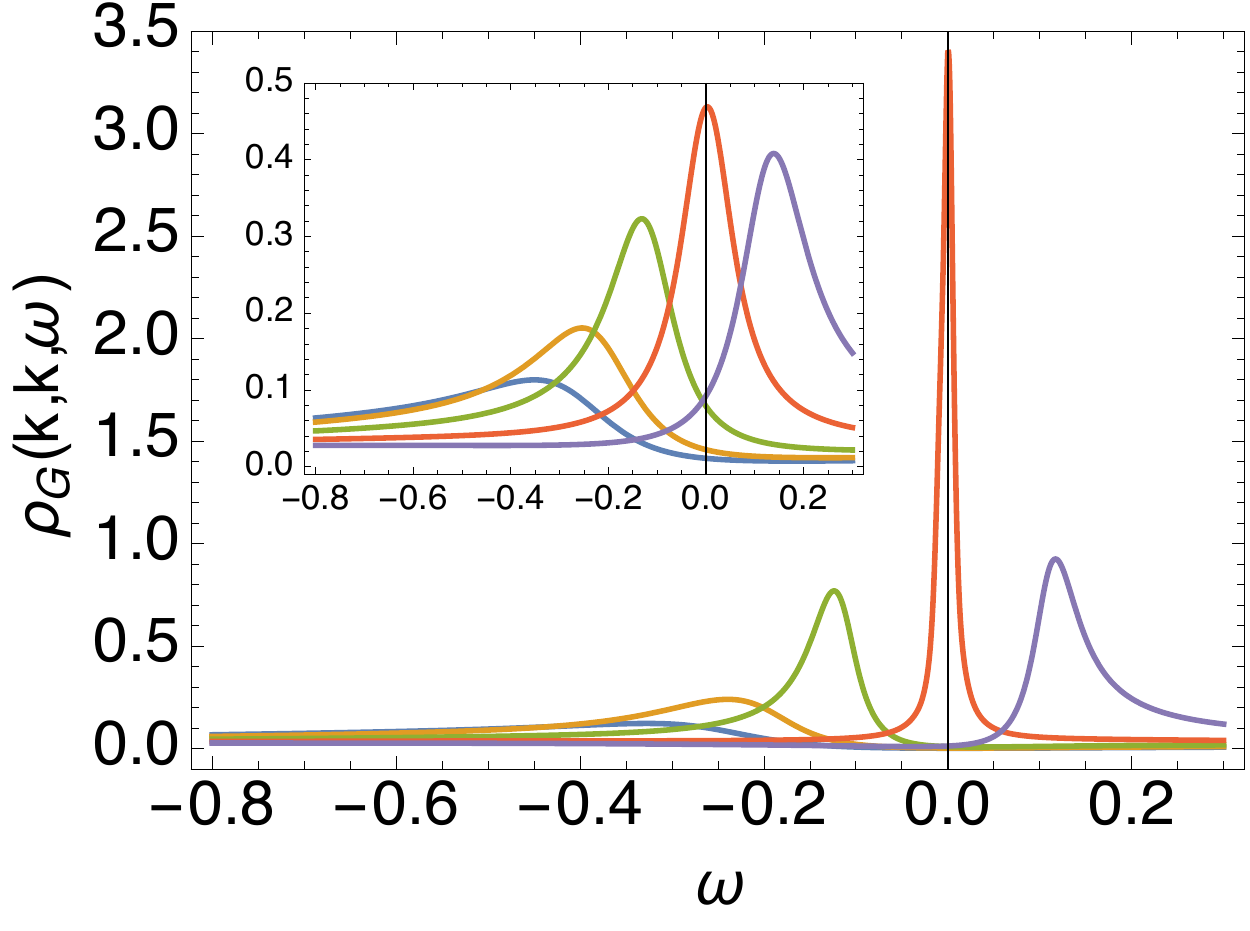}}
\subfigure[\;\; $t'=0$]{\includegraphics[width=.44\columnwidth]{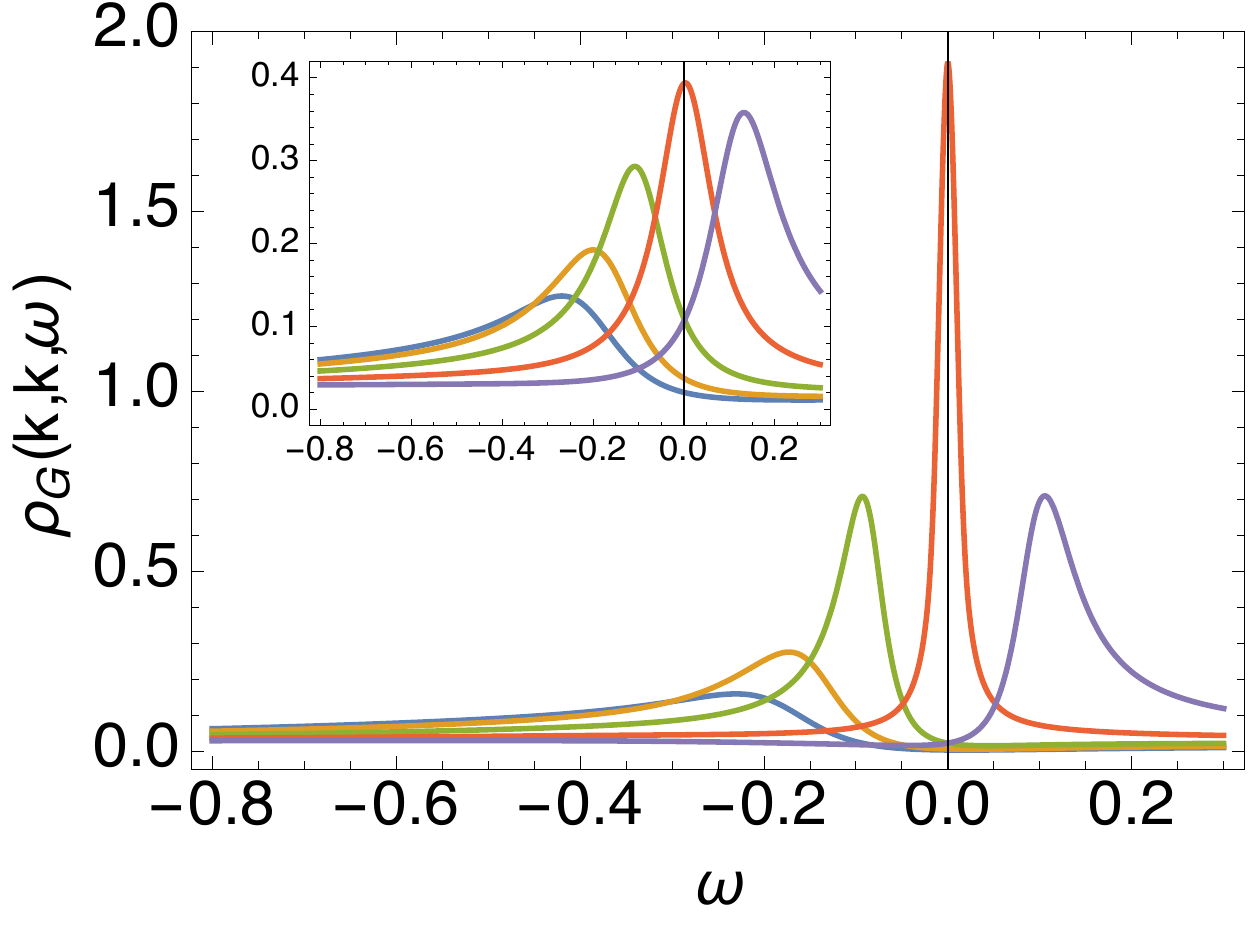}}
\subfigure[\;\; $t'=-0.2$]{\includegraphics[width=.44\columnwidth]{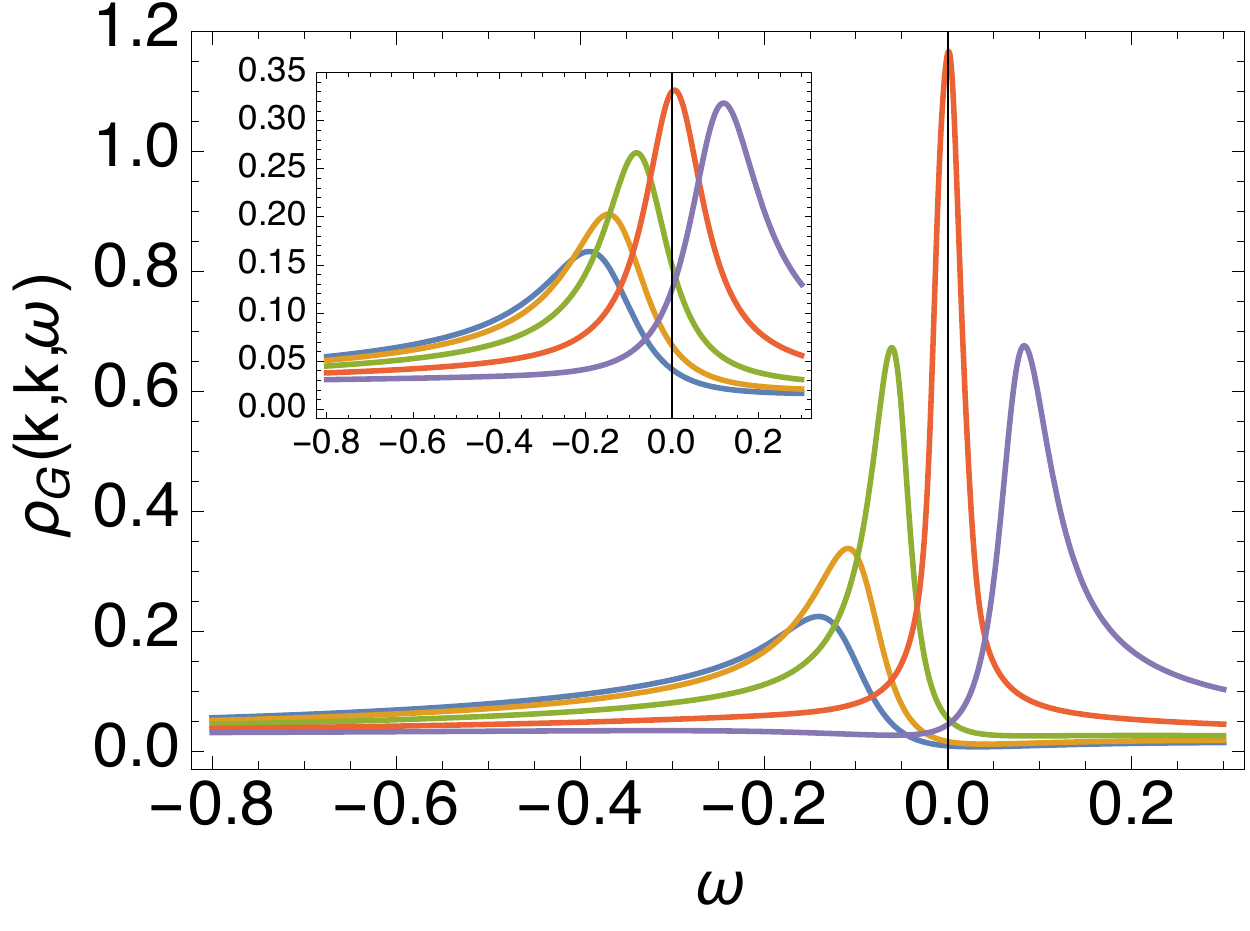}}
\subfigure[\;\; $t'=-0.4$]{\includegraphics[width=.44\columnwidth]{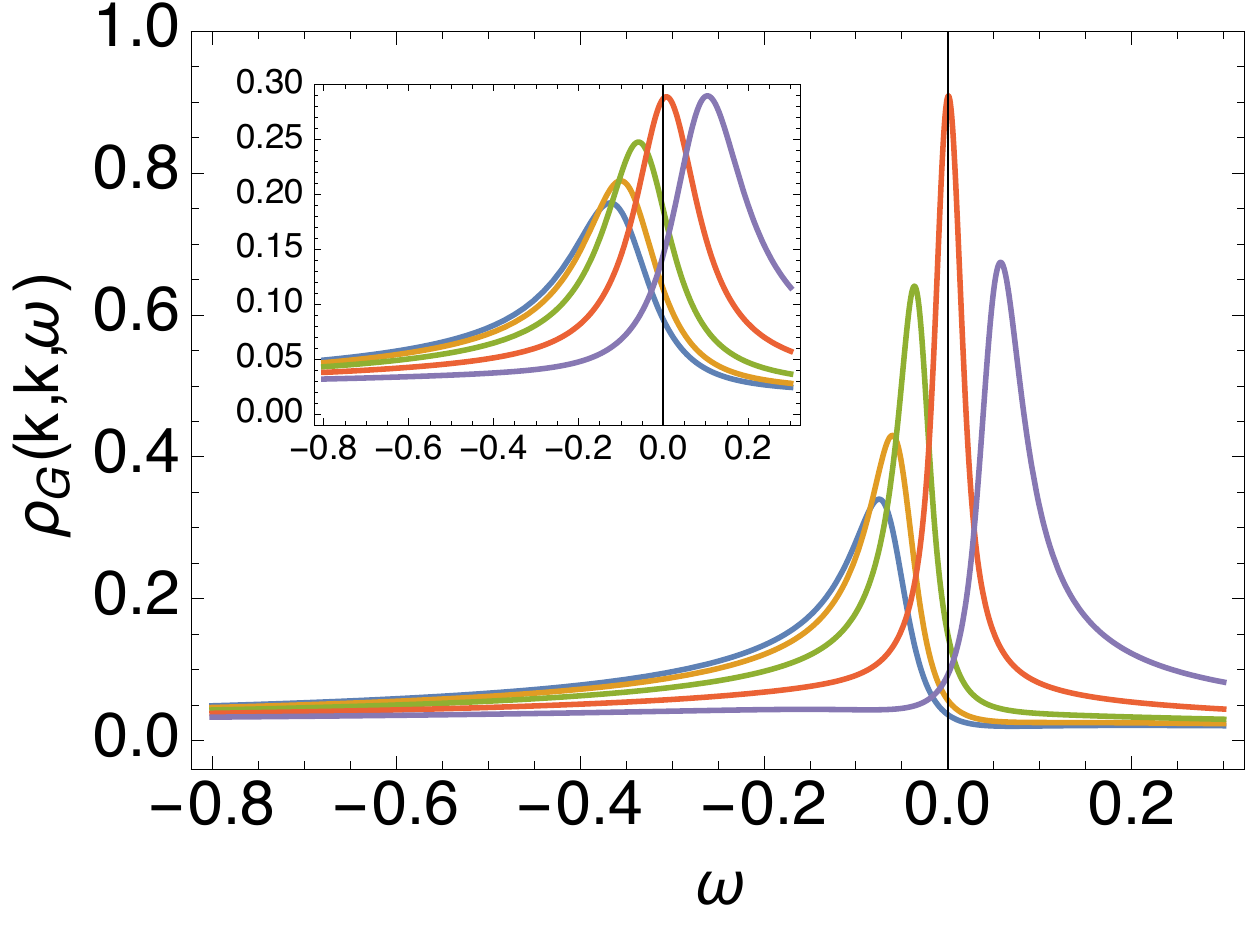}}
\caption{(Color online) EDC line shapes at different fixed values of momentum $k$ in nodal direction ($\Gamma\rightarrow X$). All figures including insets share the same legend. The parameters are set as $\delta=0.15$, $T=105K$ or $400K$ (inset) and $t'$ as specified.
The line peak and width in the vicinity of the Fermi surface  depends strongly on temperature. The peak magnitude at $\omega=0$  goes down as $t'$ decreases due to stronger correlation. 
}
\label{EDCnodal}
\end{figure}

\begin{figure}[!]
\subfigure[\;\; $t'=0.4$]{\includegraphics[width=.44\columnwidth]{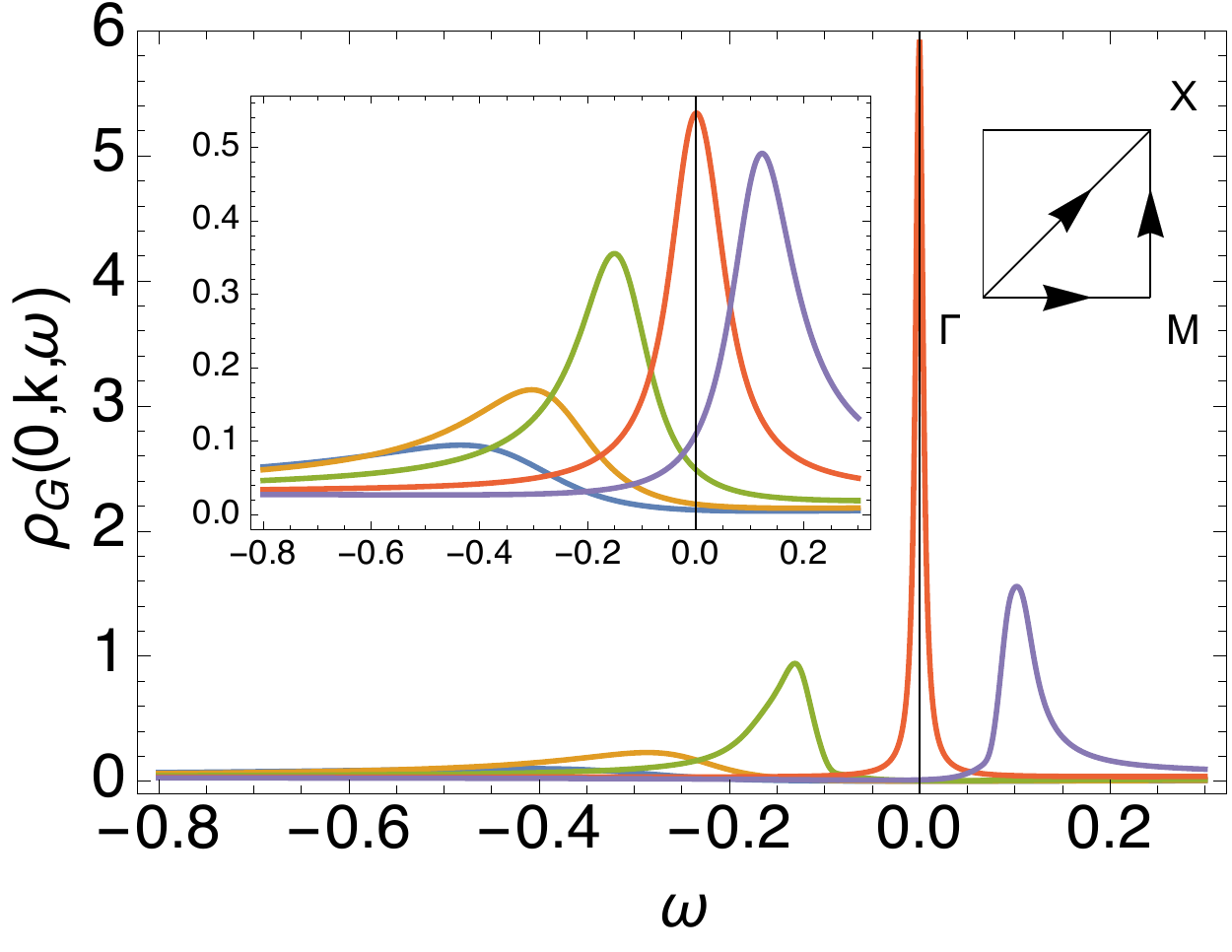}}
\subfigure[\;\; $t'=0.2$]{\includegraphics[width=.44\columnwidth]{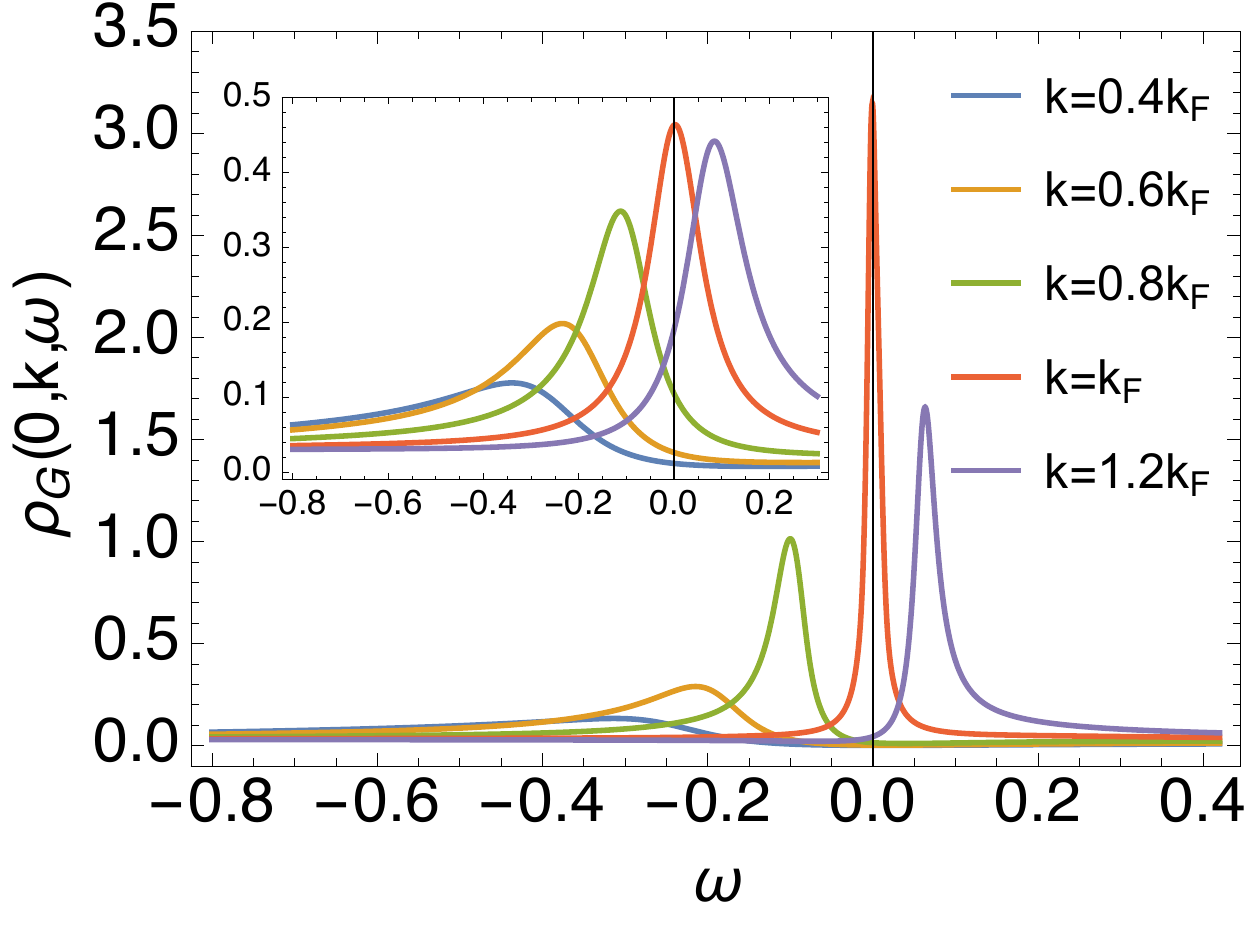}}
\subfigure[\;\; $t'=0$]{\includegraphics[width=.44\columnwidth]{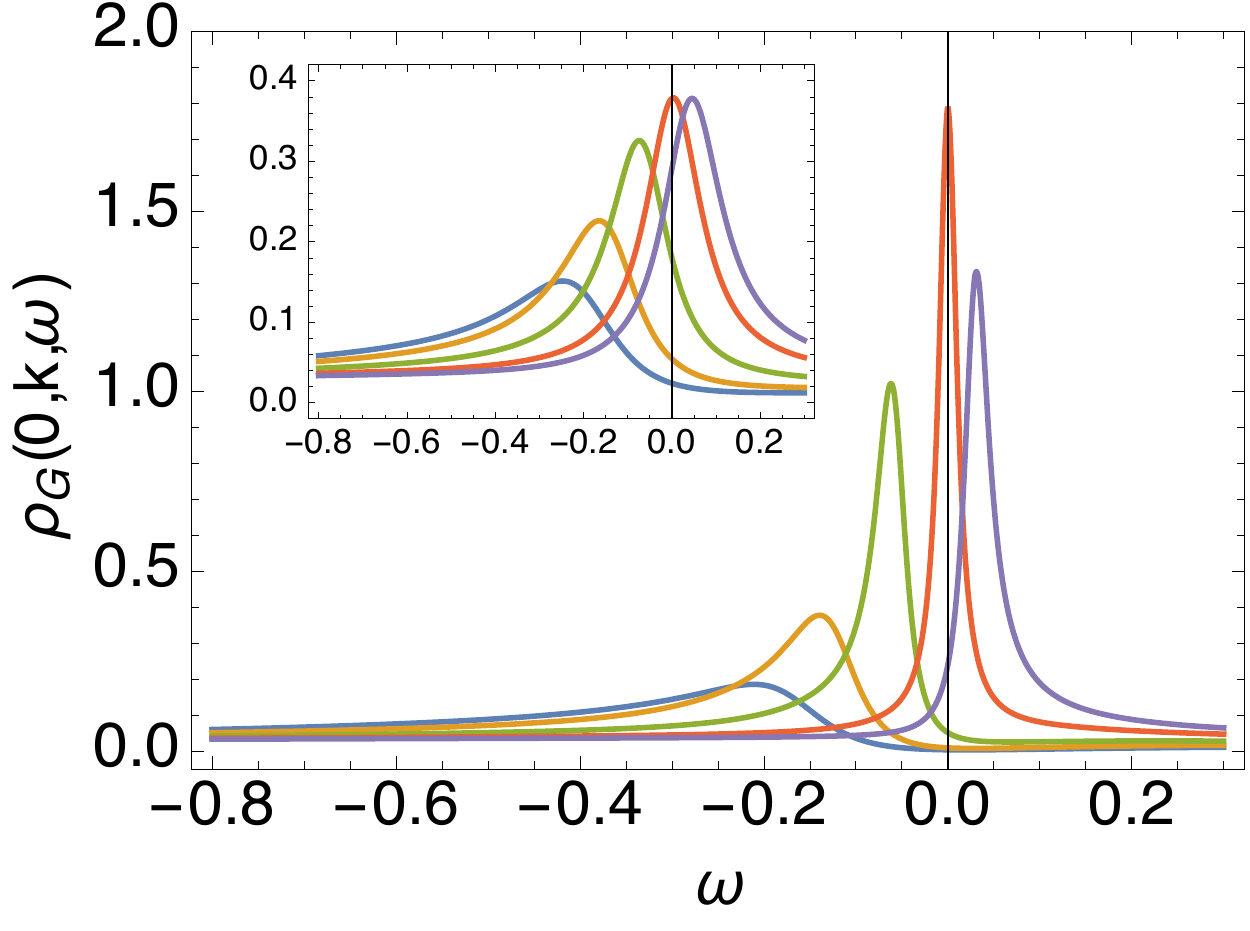}}
\caption{(Color online) EDC line shapes at different fixed values of momentum $k$ in antinodal direction ($\Gamma\rightarrow M$). All figures including insets share the same legend. The parameters are set as $\delta=0.15$, $T=105K$ or $400K$ (inset), and $t'$ as specified. The line peak and width in the vicinity of the Fermi surface depends strongly on temperature. The peak magnitude at $\omega=0$  goes down as $t'$ decreases due to stronger correlation. 
}
\label{EDCantinodal}
\end{figure}

\begin{figure}[!]
\subfigure[\;\; $t'=0.4$]{\includegraphics[width=.44\columnwidth]{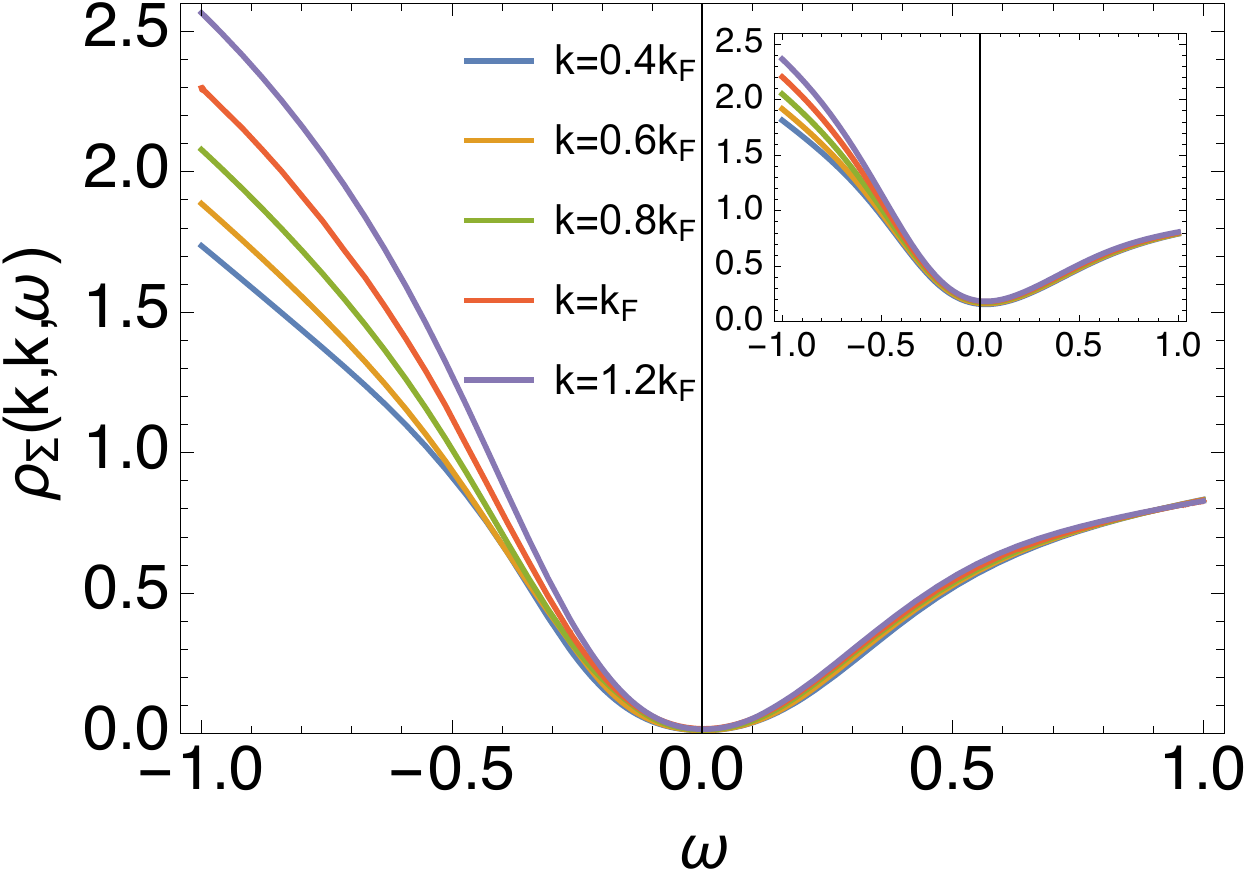}}
\subfigure[\;\; $t'=0.2$]{\includegraphics[width=.44\columnwidth]{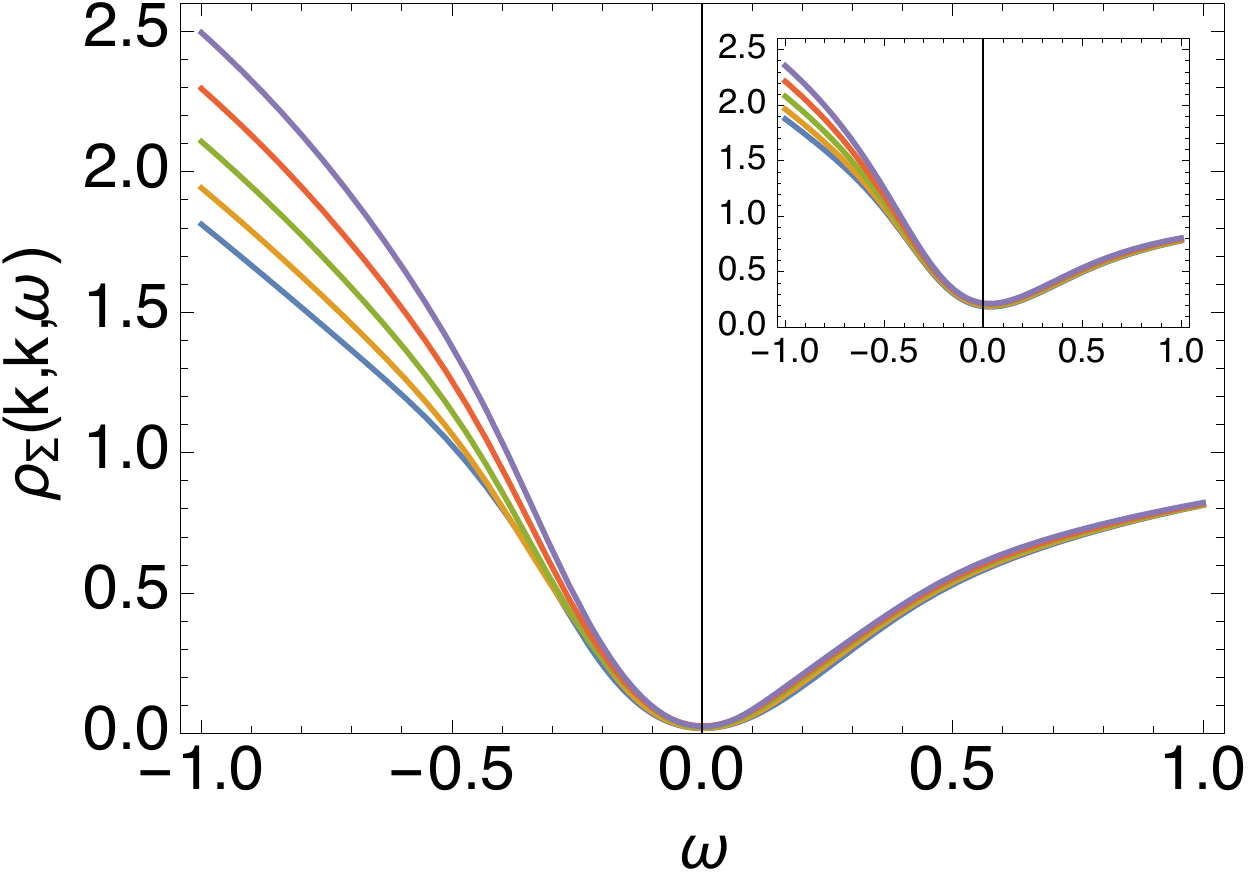}}
\subfigure[\;\; $t'=0$]{\includegraphics[width=.44\columnwidth]{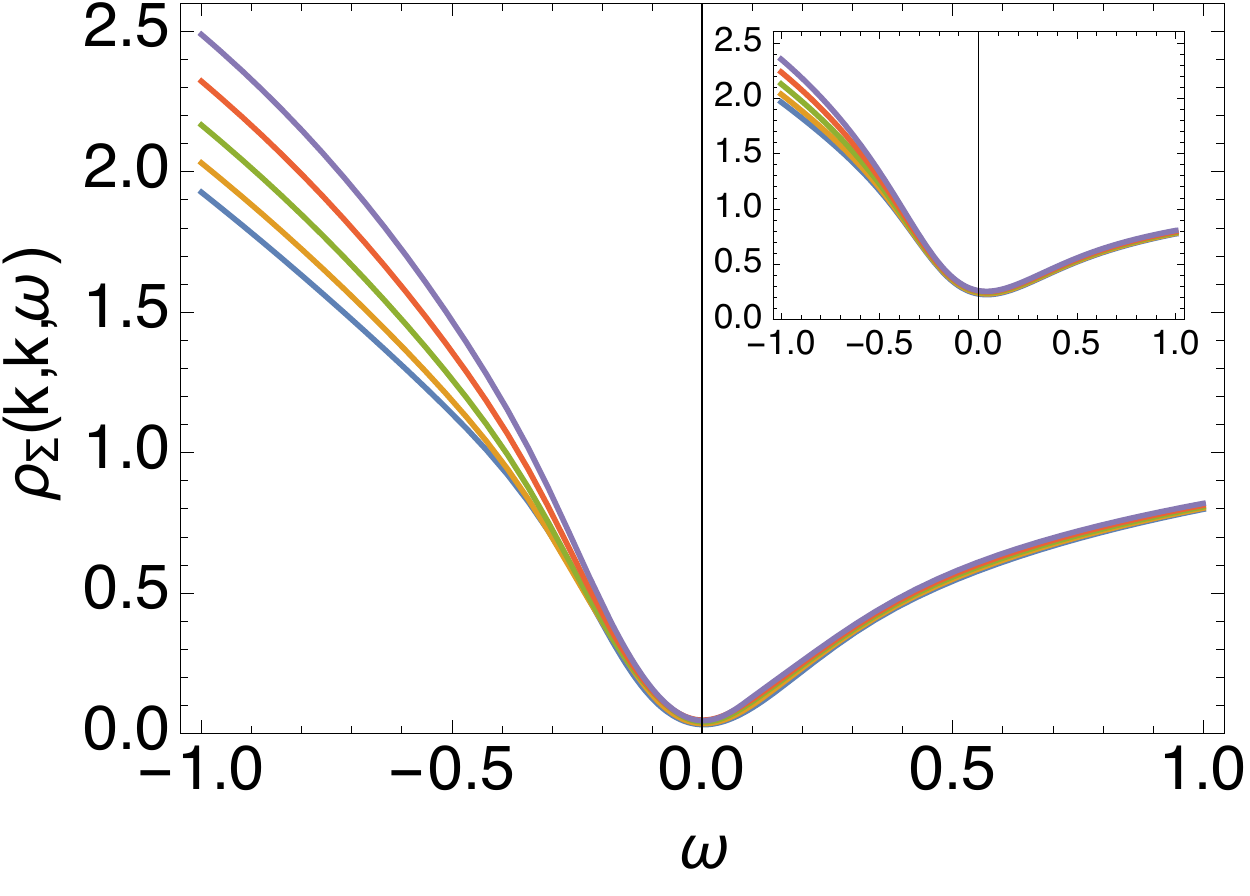}}
\subfigure[\;\; $t'=-0.2$]{\includegraphics[width=.44\columnwidth]{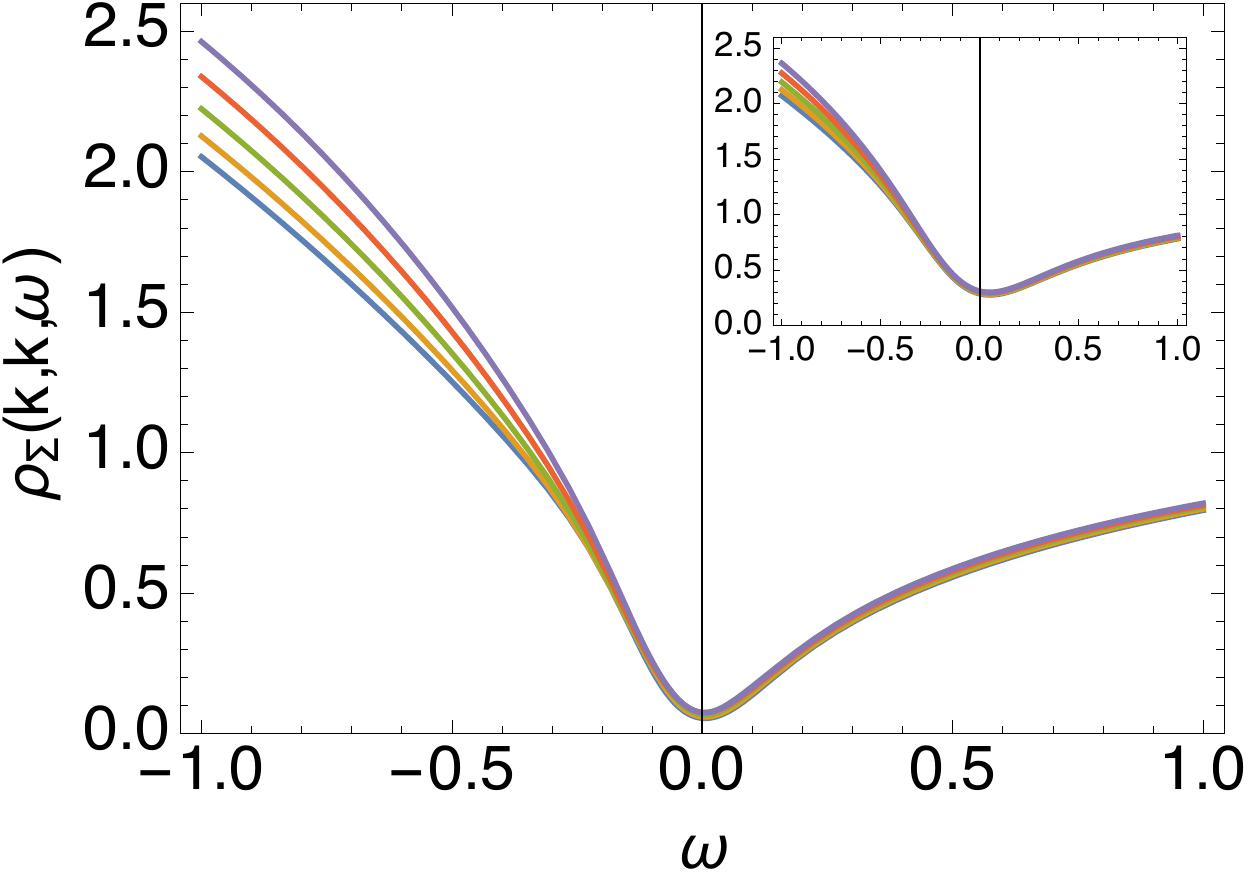}}
\subfigure[\;\; $t'=-0.4$]{\includegraphics[width=.44\columnwidth]{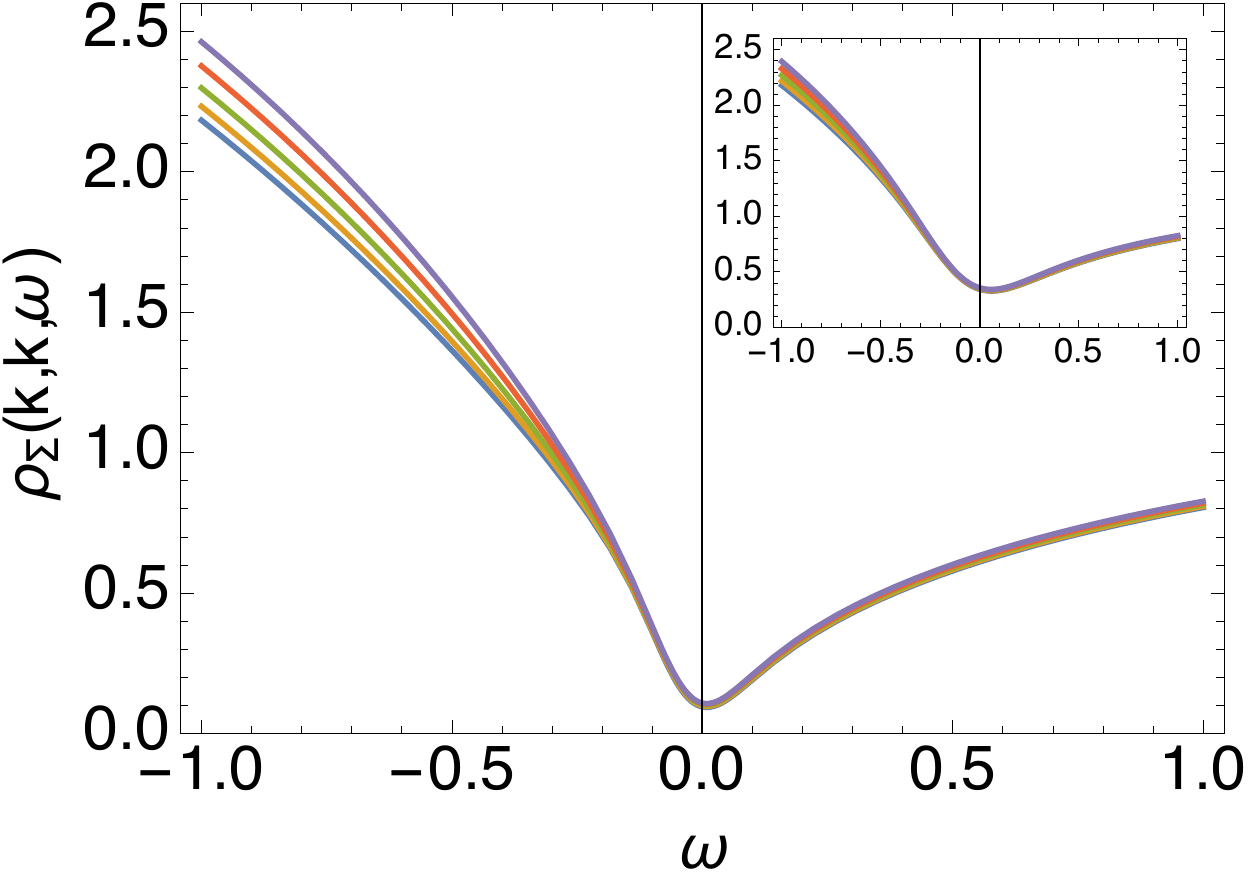}}
\subfigure[\;\; varying $t'$]{\includegraphics[width=.44\columnwidth]{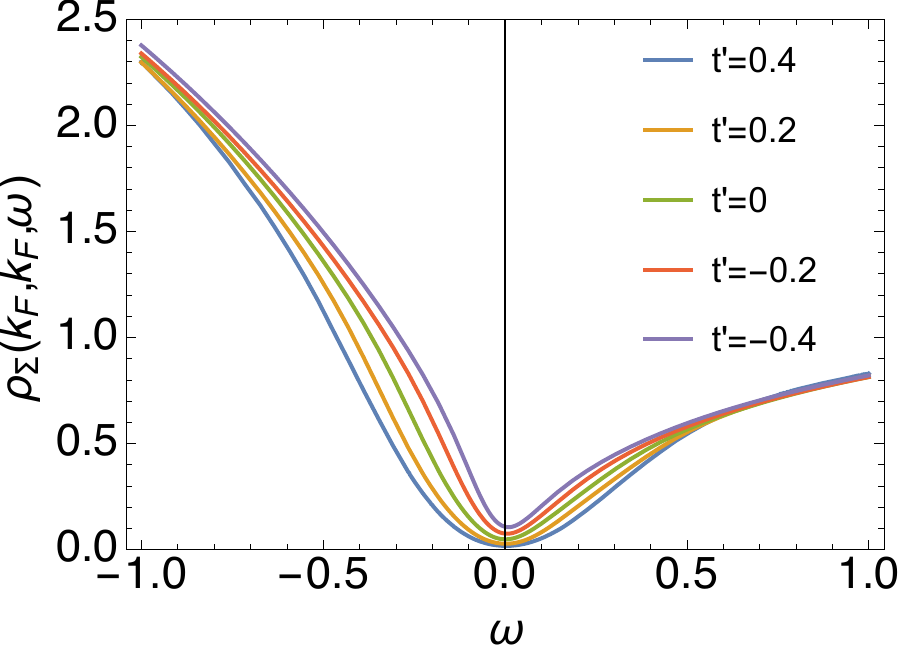}}
\caption{(Color online) (a)-(e):  The negative imaginary part of self energy $\rho_{\Sigma}$ at different $k$ in nodal ($\Gamma\rightarrow X$) direction with several $t'$. Here $\delta=0.15$, $T=105K$ and $400K$ (inset). In all cases, $\rho_{\Sigma}$ has a weak $k$-dependence and differs mostly at high energy on the unoccupied side. Increasing temperature raises the bottom of self-energy while leaving its high energy part almost unchanged.  (f): $\rho_{\Sigma}$ at fixed $k=k_F$ in nodal direction varying $t'$. Increasing $t'$ lowers the bottom of $\rho_{\Sigma}$ and makes its low energy part more rounded (Fermi-liquid like).}
\label{rhosigmaEDC}
\end{figure}

\begin{figure}[h]
\subfigure[\;\; $\rho_G(k_{Fnodal})$]{\includegraphics[width=.44\columnwidth]{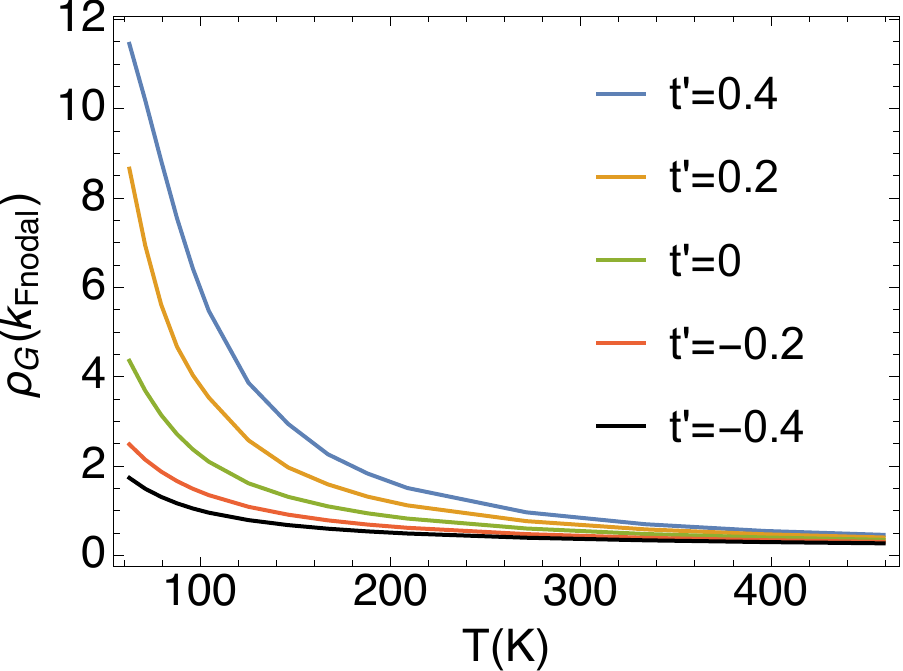}}
\subfigure[\;\; $\rho_G(k_{Fantinodal})$]{\includegraphics[width=.44\columnwidth]{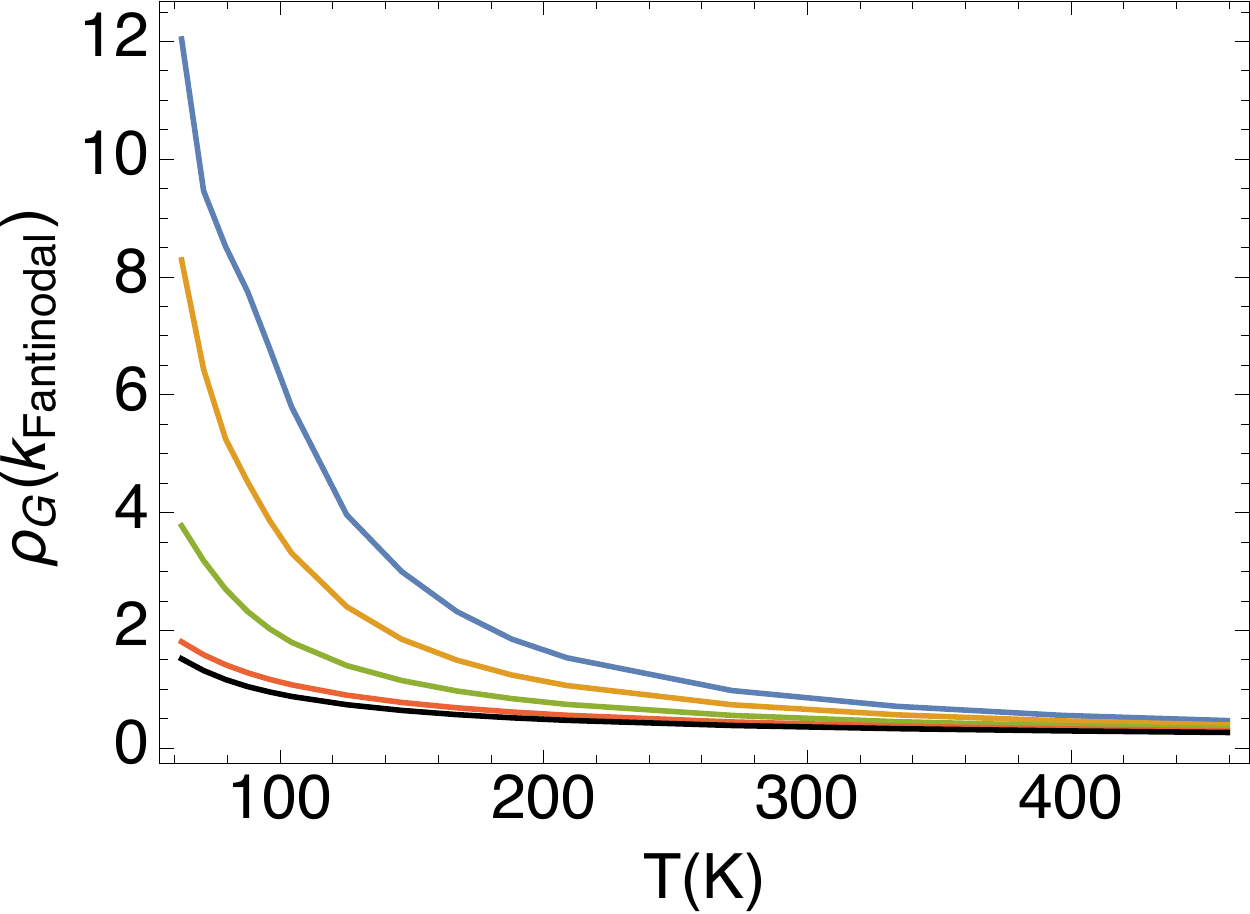}}
\subfigure[\;\; $\rho_\Sigma(k_{Fnodal})$]{\includegraphics[width=.44\columnwidth]{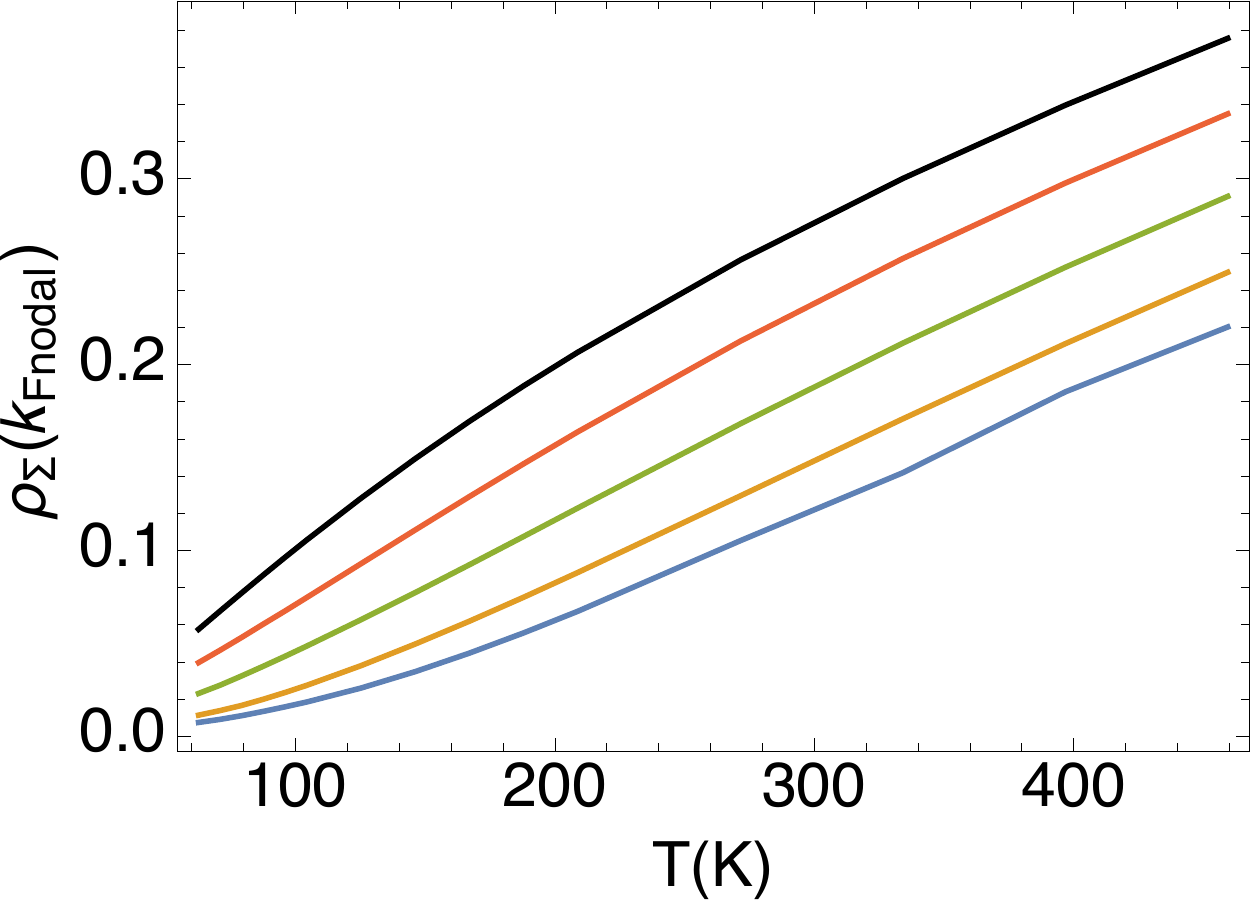}}
\subfigure[\;\; $\rho_\Sigma(k_{Fantinodal})$]{\includegraphics[width=.44\columnwidth]{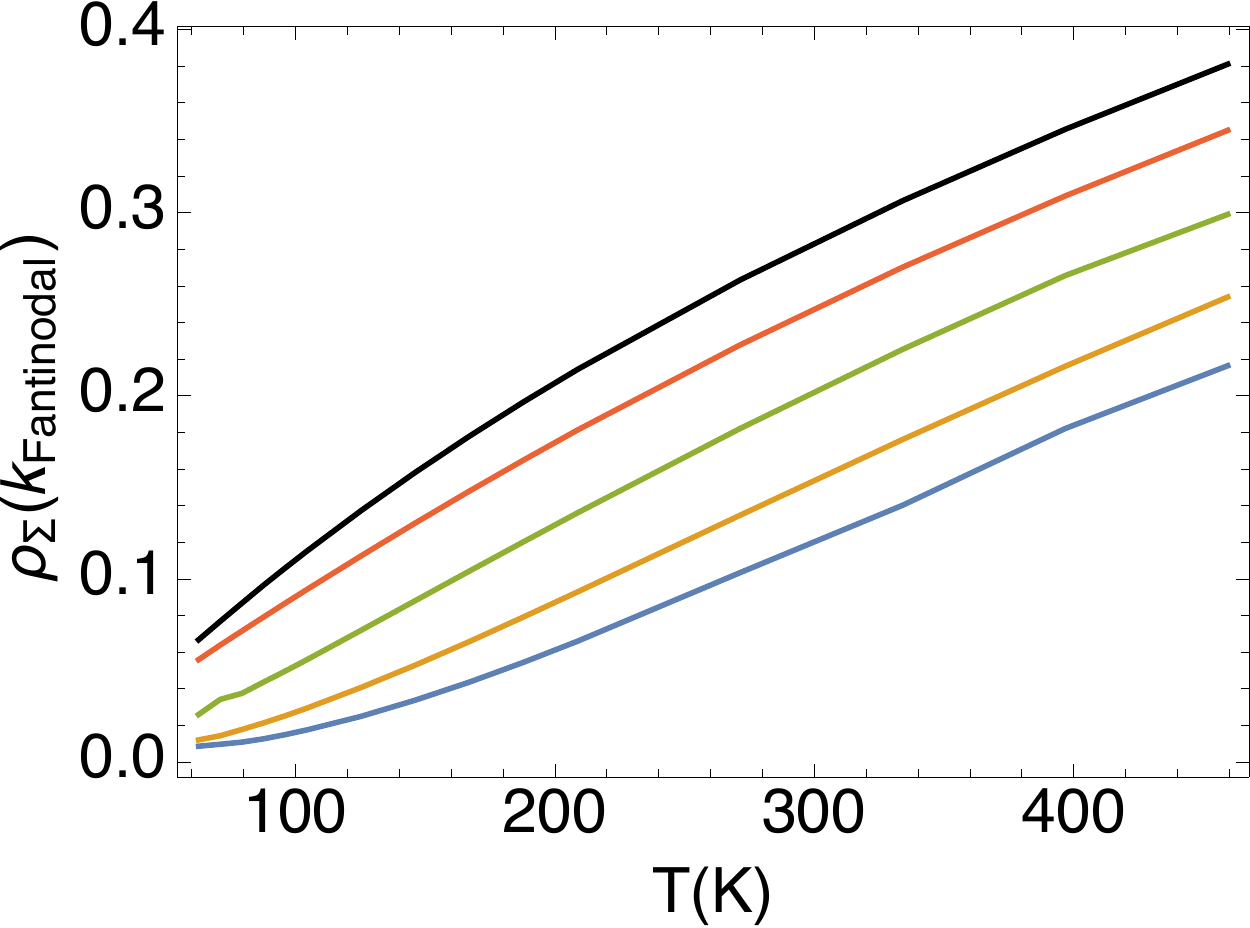}}
\caption{The spectral functions at  $\omega=0$: $\rho_G$ and $\rho_{\Sigma}$ at $k_F$ (in nodal and antinodal directions) vs $T$ with varying $t'$ at $\delta=0.15$: legend is the same for each figure.}
\label{kFT}
\end{figure}

\begin{figure}[!]
\subfigure[\;\; $t'=0.4$, nodal ($\Gamma\rightarrow X$)]{\includegraphics[width=.44\columnwidth]{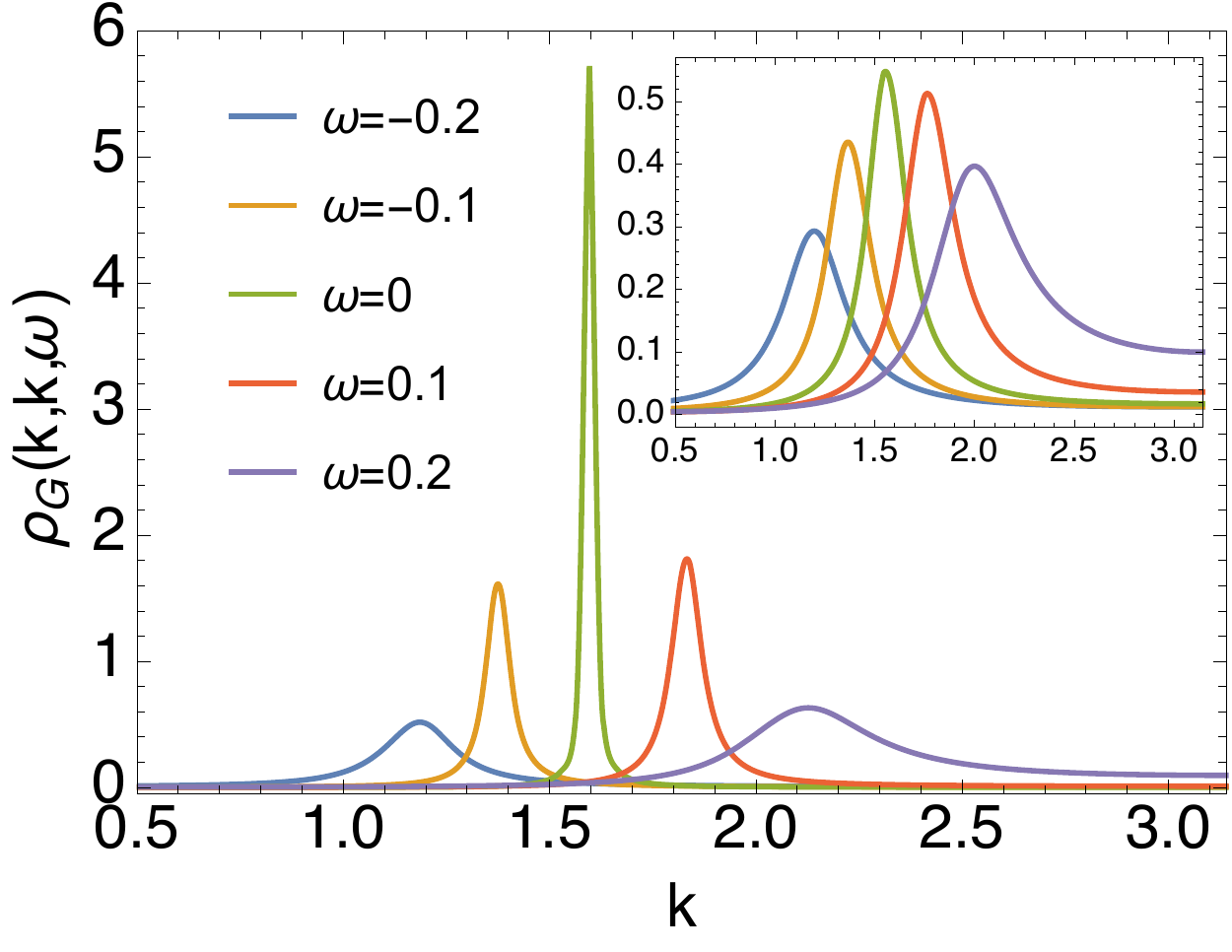}}
\subfigure[\;\; $t'=0.2$, nodal ($\Gamma\rightarrow X$)]{\includegraphics[width=.44\columnwidth]{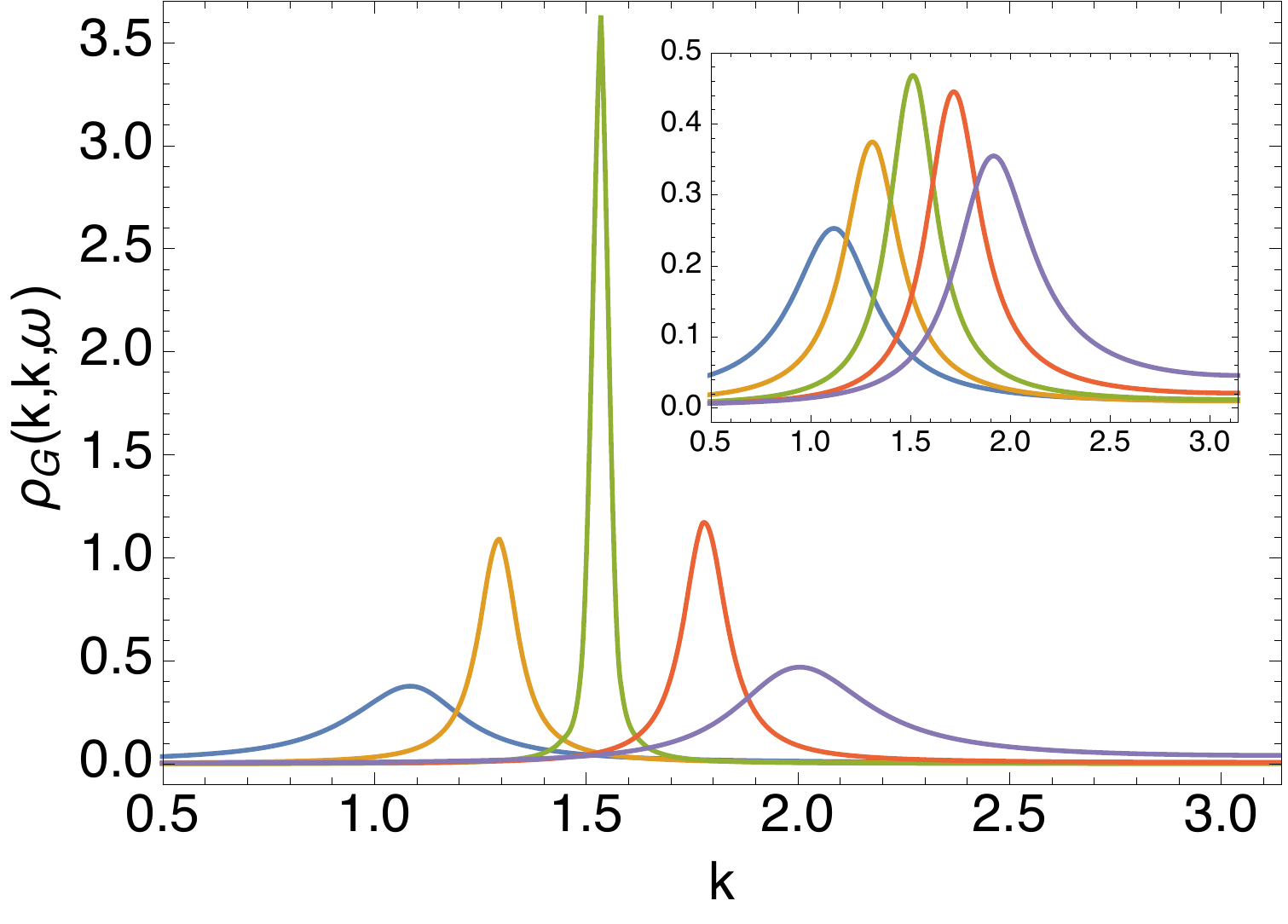}}
\subfigure[\;\; $t'=0$, nodal ($\Gamma\rightarrow X$)]{\includegraphics[width=.44\columnwidth]{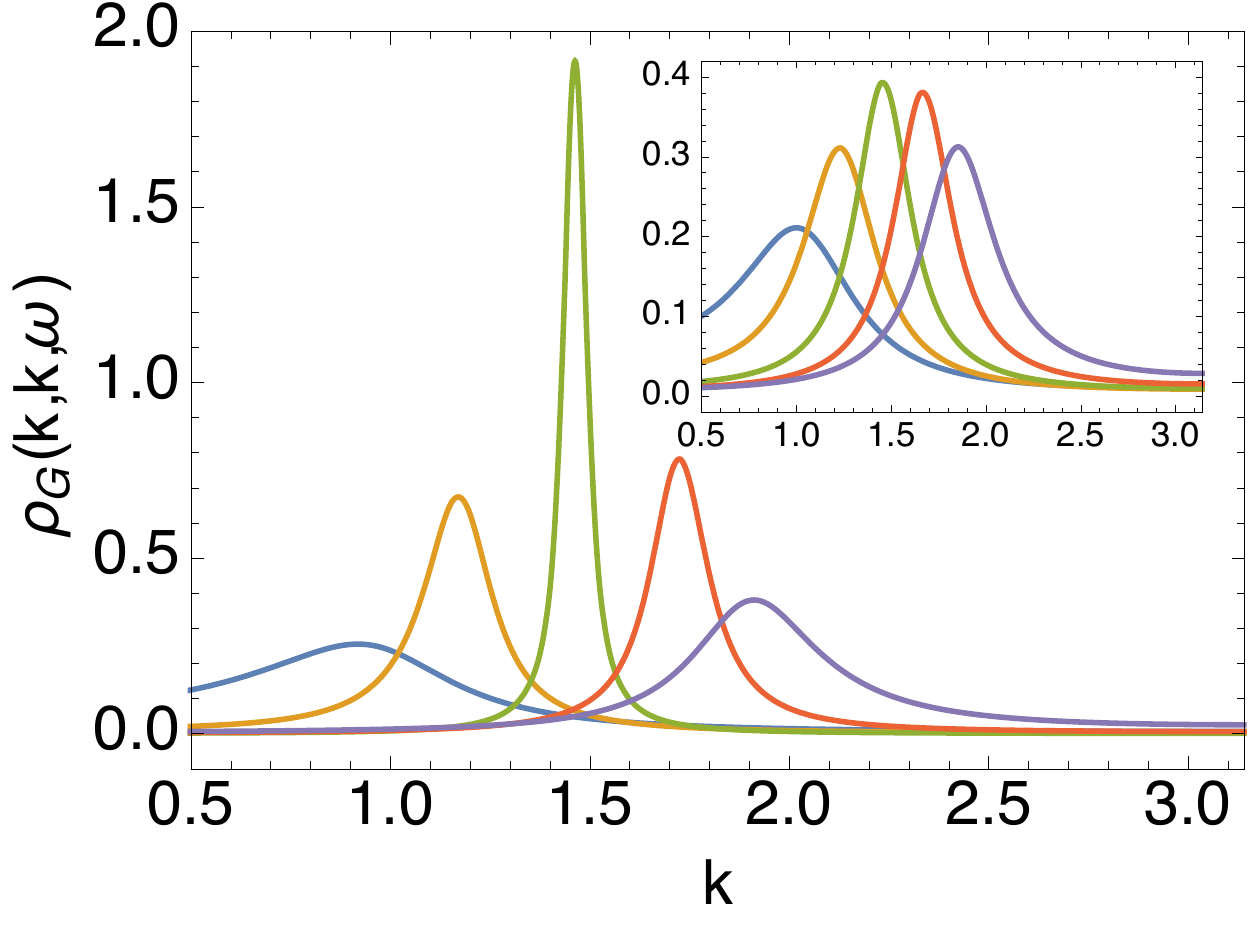}}
\subfigure[\;\; $t'=-0.2$, nodal ($\Gamma\rightarrow X$)]{\includegraphics[width=.44\columnwidth]{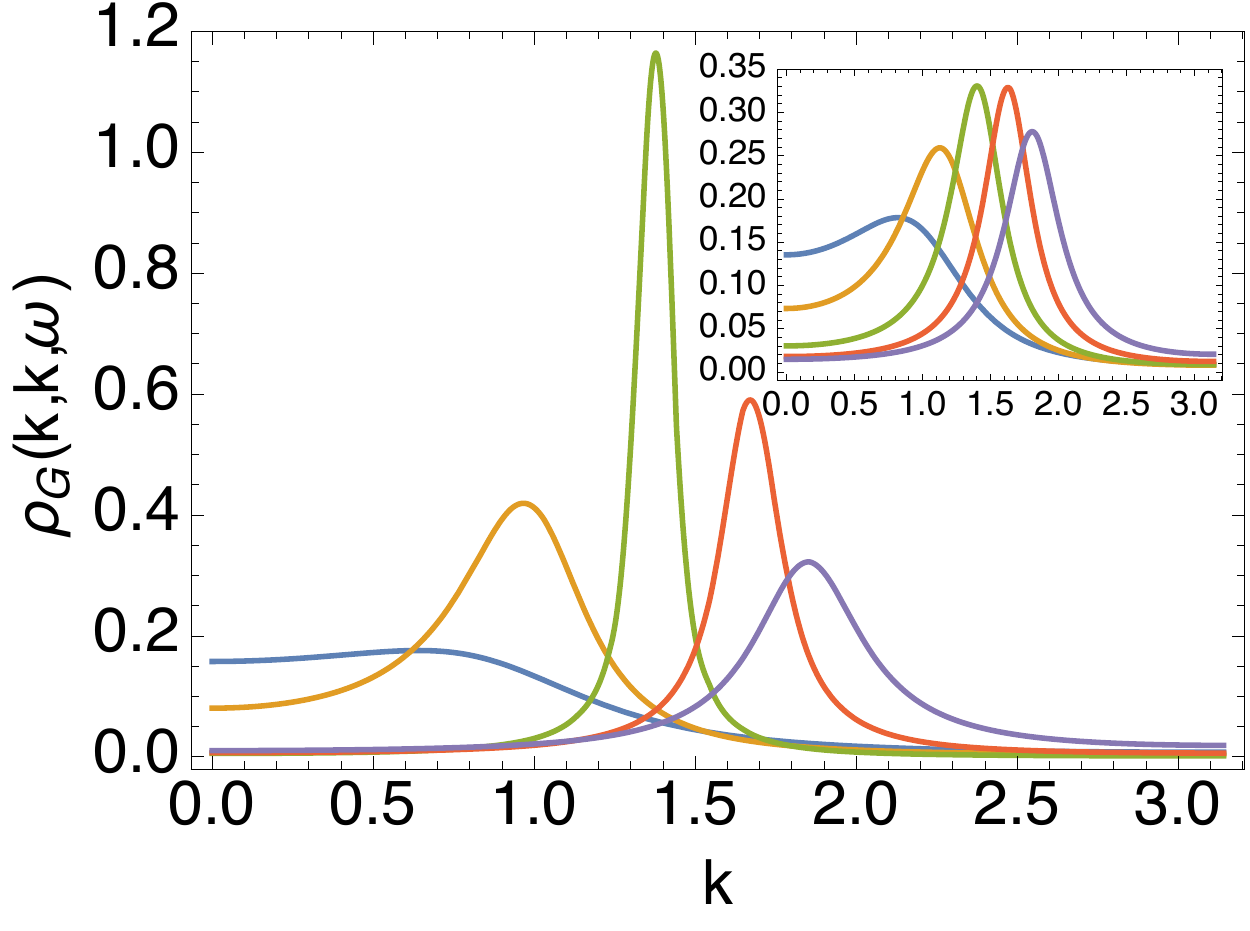}}
\subfigure[\;\; $t'=-0.4$, nodal ($\Gamma\rightarrow X$)]{\includegraphics[width=.44\columnwidth]{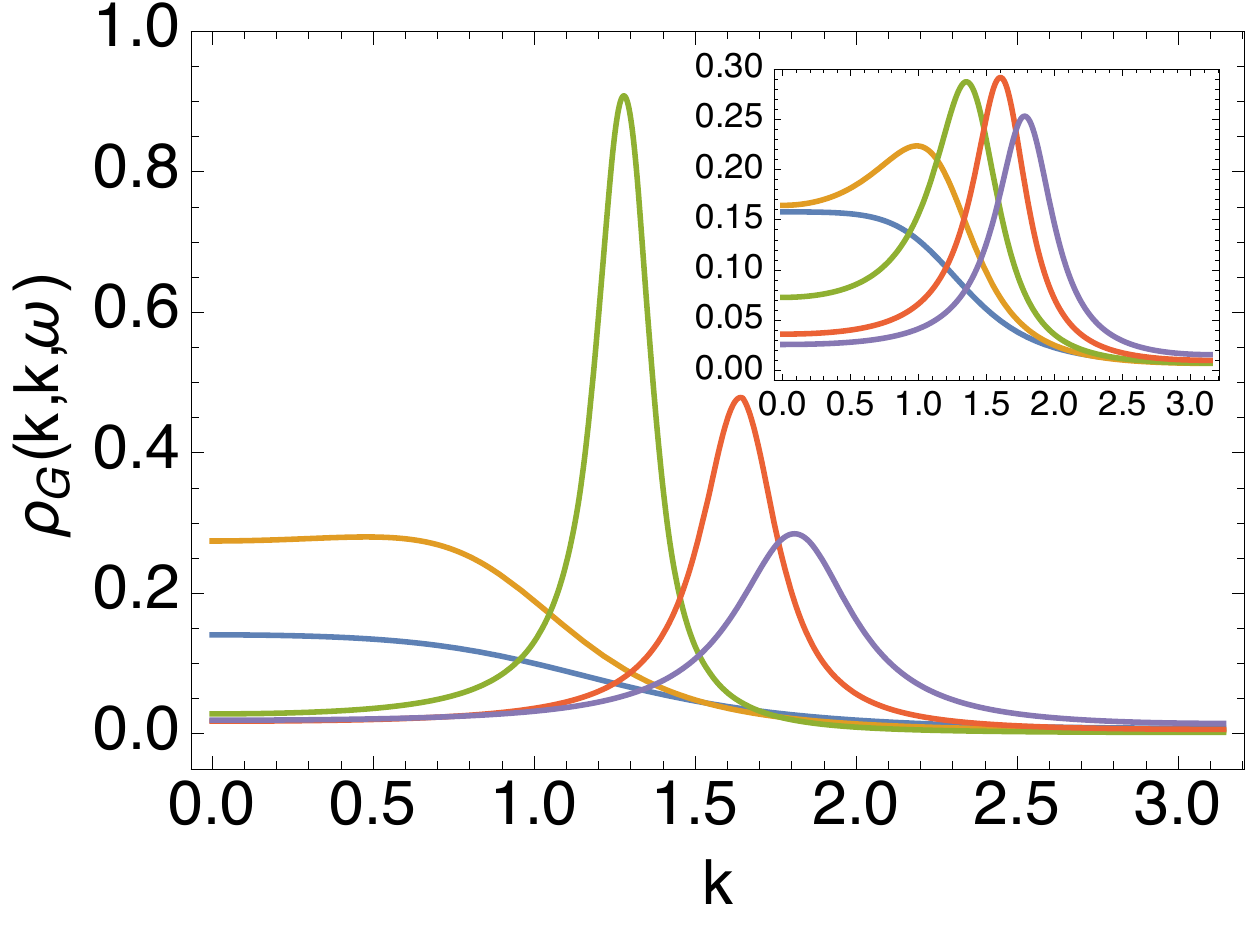}}
\caption{(Color online) MDC line shapes at different fixed values of frequency $\omega$ in each curve. All figures including insets share the  legend. Here the parameters are set as $\delta=0.15$, $T=105K$ and $400K$ (inset). $k$ is scanned along the nodal ($\Gamma\rightarrow X$) direction. In all cases, they have a highest peak with a symmetric shape at $\omega=0$. Consistently, the peak height decreases with smaller $t'$, or stronger correlation. 
}
\label{MDCnodal}
\end{figure}

\begin{figure}[!]
\subfigure[\;\; $t'=0.4$, antinodal ($\Gamma\rightarrow M$)]{\includegraphics[width=.44\columnwidth]{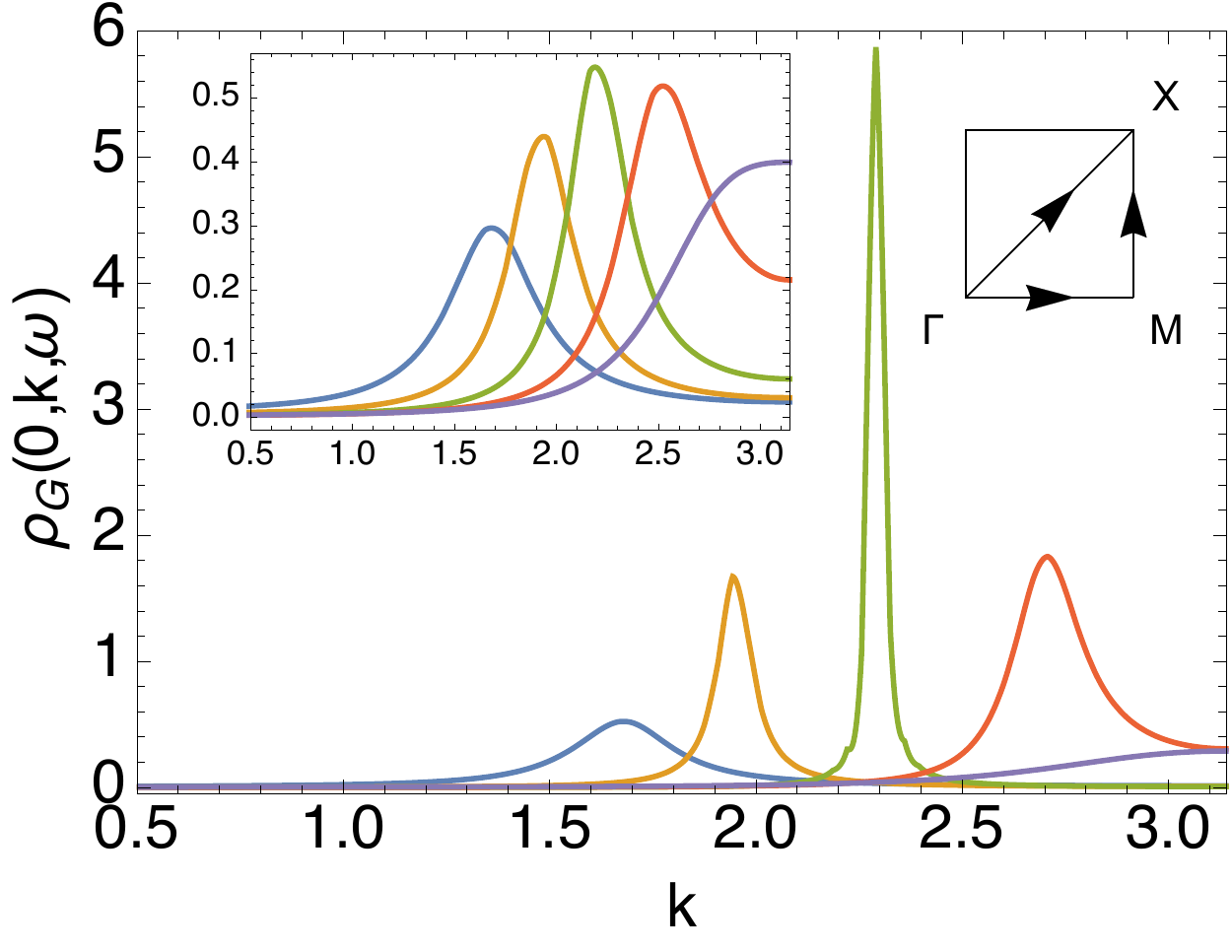}}
\subfigure[\;\; $t'=0.2$, antinodal ($\Gamma\rightarrow M$)]{\includegraphics[width=.44\columnwidth]{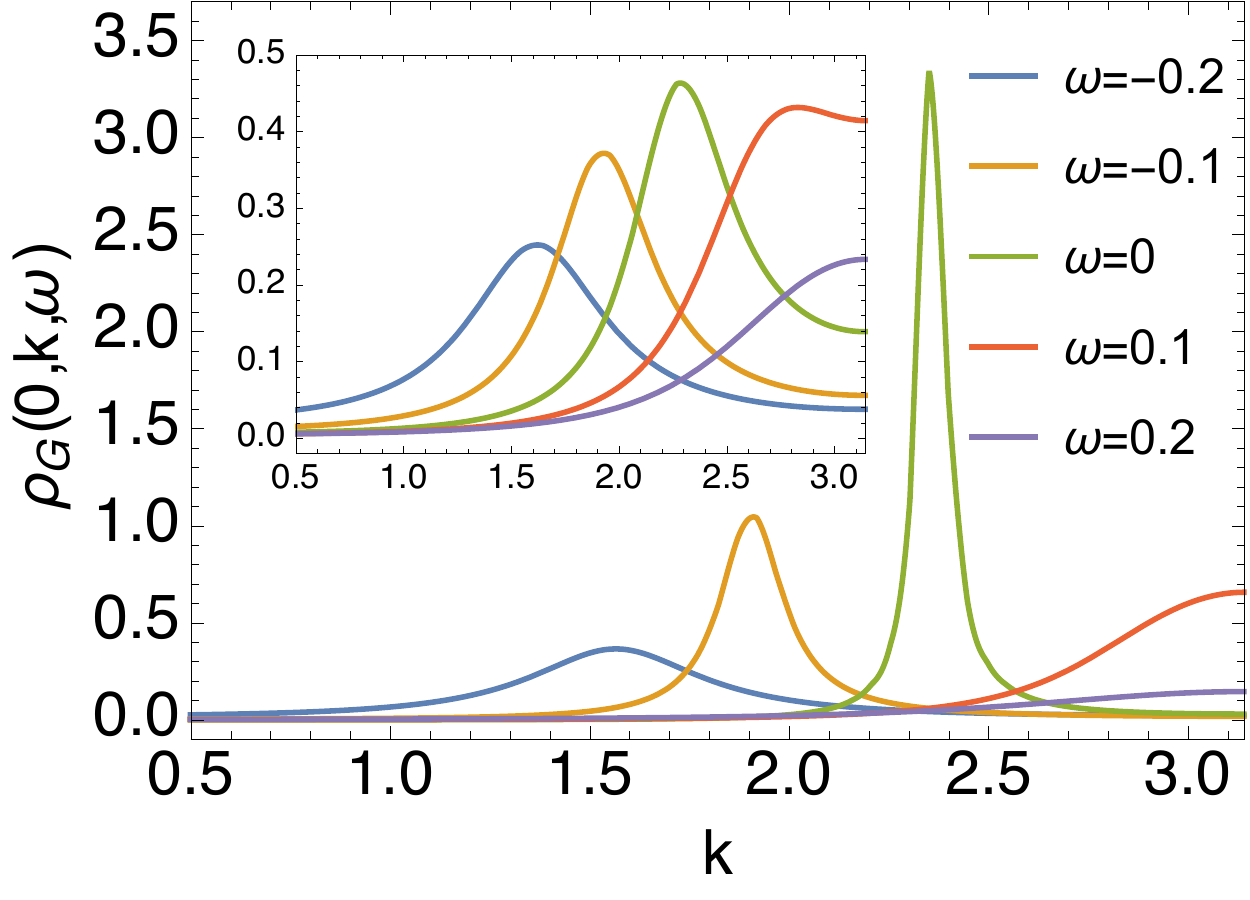}}
\subfigure[\;\; $t'=0$, antinodal ($\Gamma\rightarrow M$)]{\includegraphics[width=.44\columnwidth]{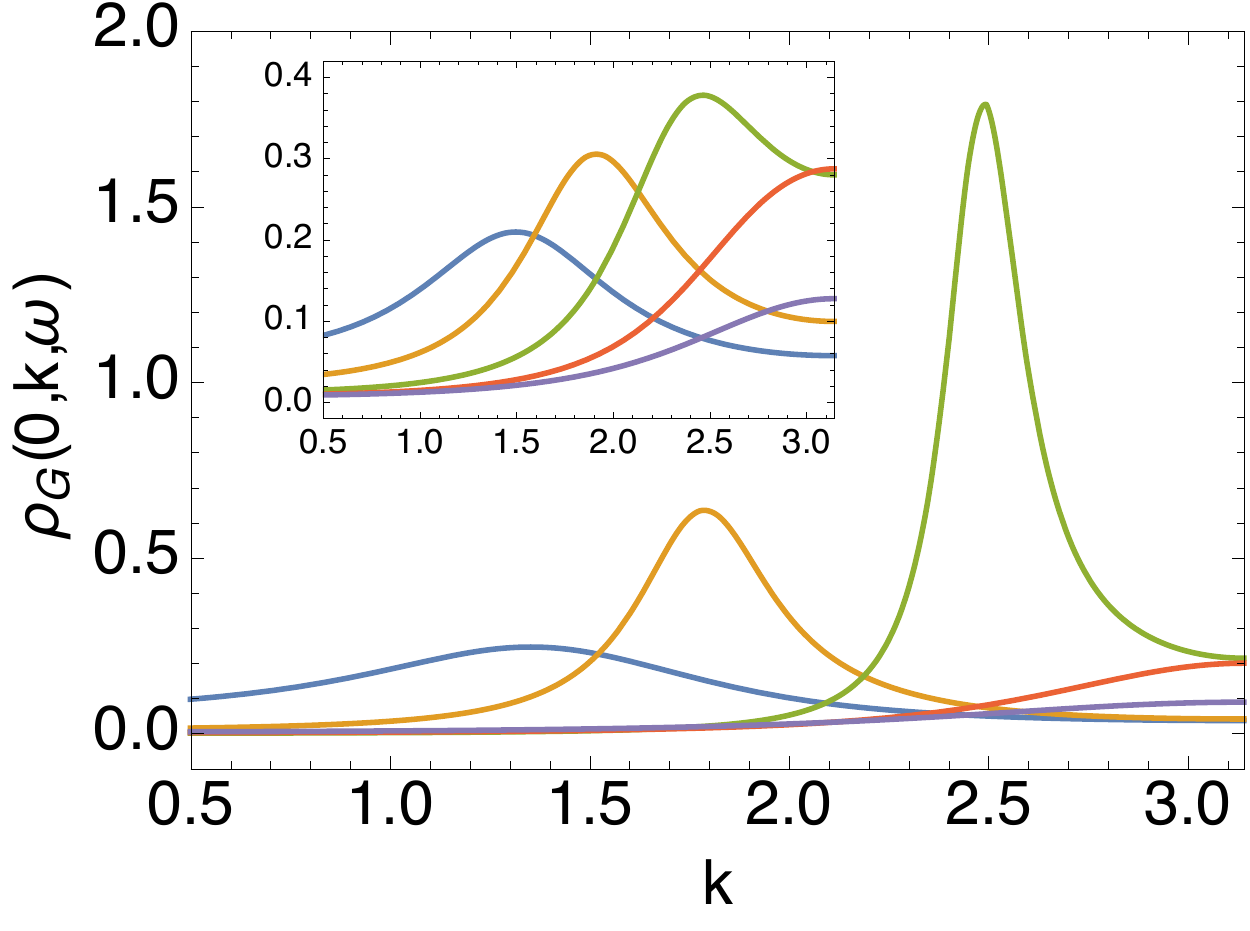}}
\subfigure[\;\; $t'=-0.2$, antinodal ($M\rightarrow X$)]{\includegraphics[width=.44\columnwidth]{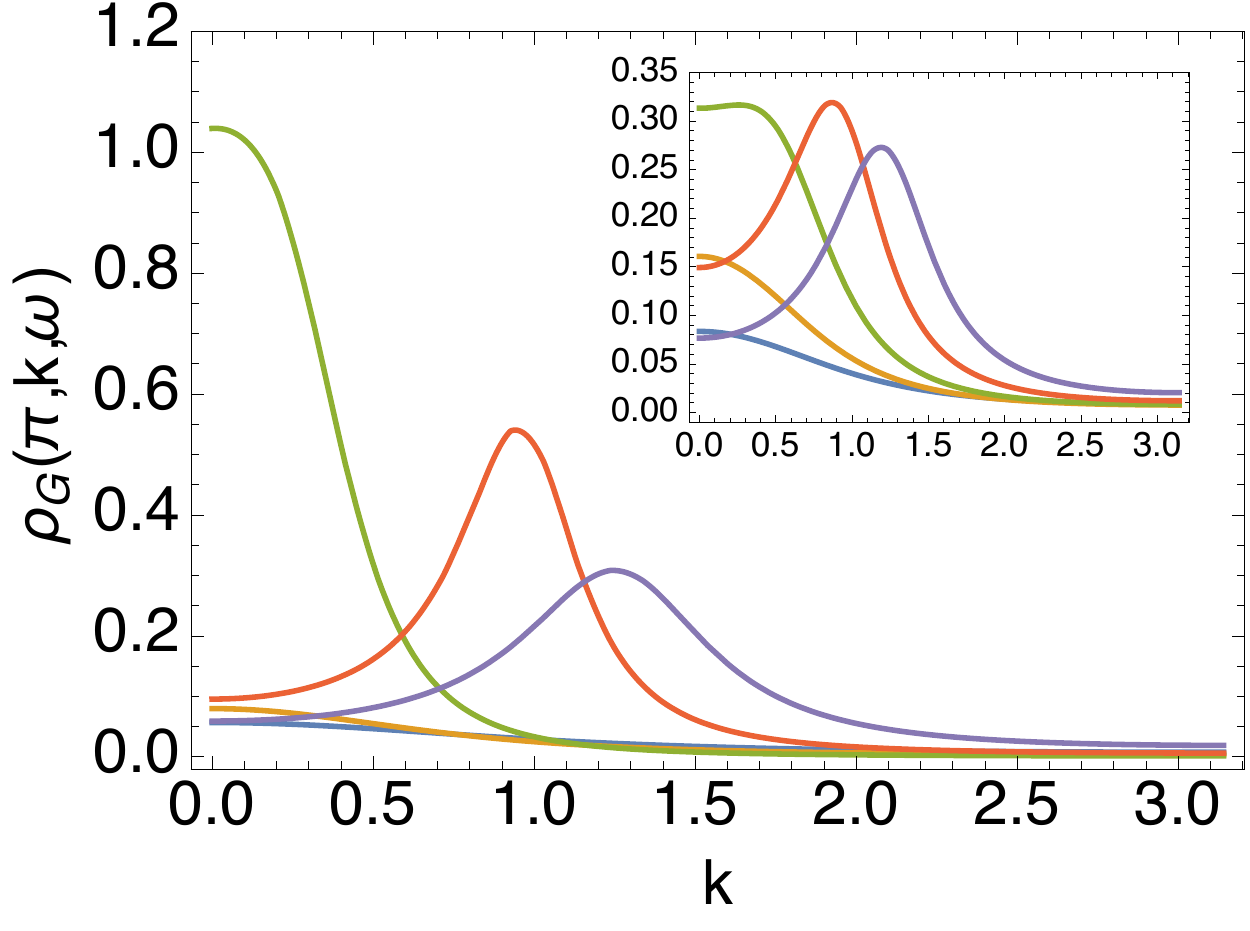}}
\subfigure[\;\; $t'=-0.4$, antinodal ($M\rightarrow X$)]{\includegraphics[width=.44\columnwidth]{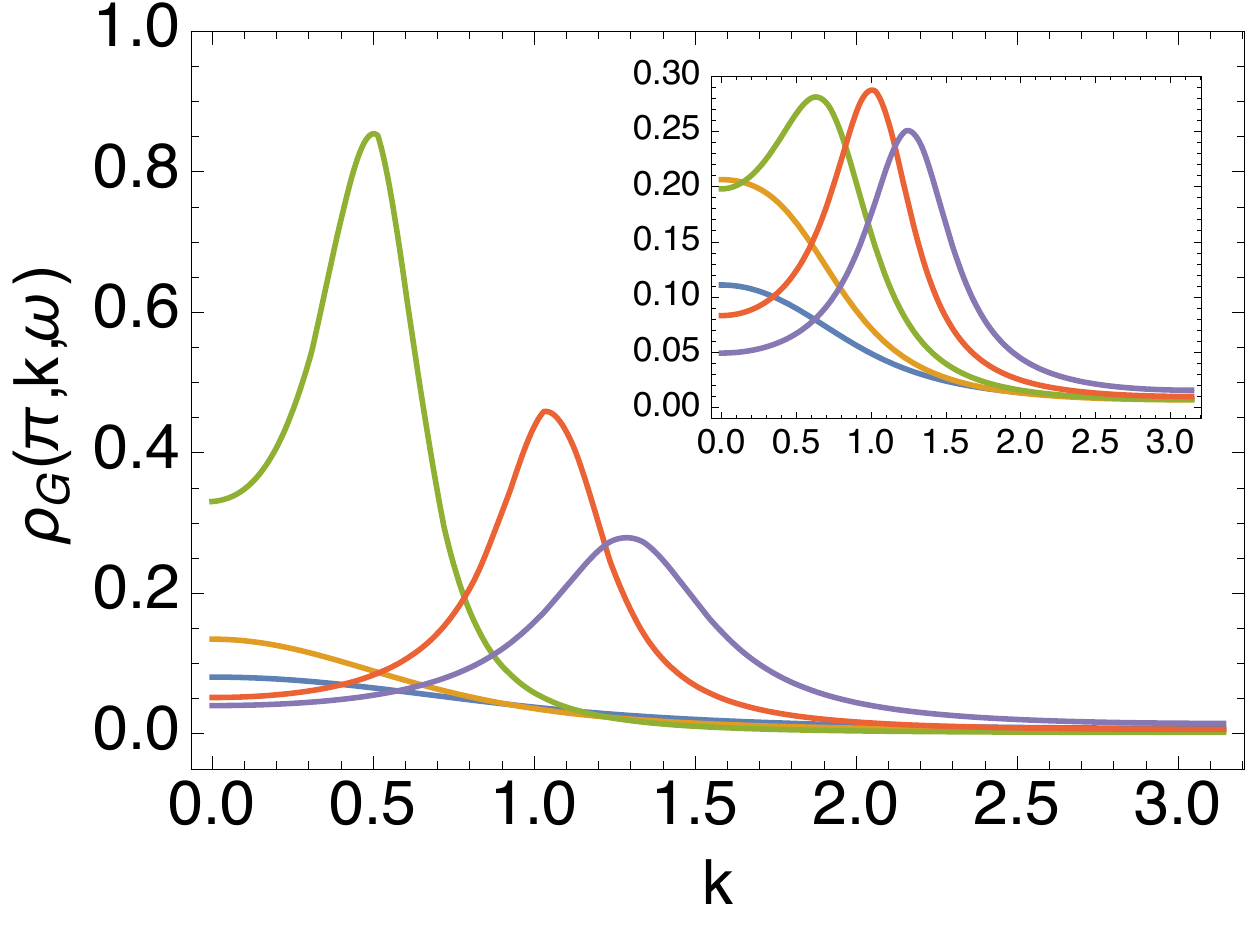}}
\caption{(Color online) MDC line shapes at different fixed values of frequency $\omega$ in each curve. All figures including insets share the same legend. Here the parameters are set as $\delta=0.15$, $T=105K$ and $400K$ (inset). $k$ is scanned along the antinodal ($\Gamma\rightarrow M$ for $t'\geq0$ or $M\rightarrow X$ for $t'<0$) directions. 
}
\label{MDCantinodal}
\end{figure}

\section{Results}\label{results}

\subsection{Spectral properties}\label{spectralproperties}

\subsubsection{Spectral Function and Self-energy}

In earlier studies \cite{Gweon}, the  ECFL spectral function obtained  phenomenologically\cite{ECFL,Gweon,Anatomy} has been compared  with experimental data measured by the angle-resolved photoemission spectroscopy (ARPES) at optimal doping, leading to very good fits.  Later we calculated the spectral function from the raw second order ECFL equations in the symmetrized model \cite{Hansen} but it is only valid for doping $\delta\gssim 0.25$. Here we present the  result at optimal doping $\delta=0.15$ from a microscopic calculation of ECFL by numerically solving the improved set of second order equations\cite{SP, Sriram-Edward}. 

We display the energy distribution curves (EDCs) in \figdisp{EDCnodal} and \figdisp{EDCantinodal}, obtained by fixing $k$ and scanning $\omega$ 
at optimal doping and various $t'$. These quantities can be measured in ARPES experiment. \figdisp{EDCnodal} shows the EDCs for several constant $k$ along nodal ($\Gamma\rightarrow X$) and  \figdisp{EDCantinodal} for the antinodal direction ($\Gamma\rightarrow M$ for $t'>0$). Note that the value of $k_F$ depends on $t'$ and direction in $k$ space. The fixed value of $k$ is given in terms of $k_F$ based on the specific $t'$ and direction. The antinodal ($M\rightarrow X$) $k_F$ for $t'\leq -0.2$ is close to zero. The corresponding EDCs are too close to resolve clearly; hence those ones are not presented. 

We observe that at low temperatures the EDC peak  gets sharper as $k$ approaches the Fermi surface.The insets show that a small heating ($\Delta T\sim 0.06 t$) strongly suppresses the region around the Fermi surface $k\sim k_F$ while it leaves the region away from Fermi surface almost unchanged. As a result, a weaker $k$-dependence of peak height can be viewed in the higher temperature. It also shows that the EDC line shape is asymmetric for $k<k_F$, consistent to ARPES experiment. As $t'$ decreases from positive (electron doped) to negative (hole doped), the correlation becomes stronger, and therefore the spectral peak gets lower. Slight anisotropy is found for $t'\leq 0.2$ in that the peak at the Fermi surface is a bit higher in the nodal direction than in the antinodal direction, indicating a weak $k$-dependence of self-energy. 

The spectral function of the Dyson self-energy is defined as 
\beq
\rho_{\Sigma}(\vec{k},\omega)=-\frac{1}{\pi}\Im m \, \Sigma(\vec{k},\omega).
\eeq
It is calculated from the spectral function obtained from solving the set of ECFL equations~(1-6). 
\beq
\rho_{\Sigma}(\vec{k},\omega)=\frac{\rho_G(\vec{k},\omega)}{\pi^2\rho_G^2(\vec{k},\omega)+\{\Re e G(\vec{k},\omega)\}^2}.
\eeq
where $\Re e G$ is calculated through Hilbert transform of $\rho_G$. As observed in \figdisp{rhosigmaEDC} (a-e), the self-energy shows asymmetry from intermediate frequencies at essentially  all  values of $t'$ and $k$, which is consistent with previous studies \cite{Hansen,Sriram-Edward}, unlike the symmetric curves in standard Fermi liquid theory. Further they all appear to  depend weakly on $k$. This is qualitatively different from the strong $k$-dependence of the low energy behaviors of the self-energy in one dimension \cite{PSS}. This weak $k$-dependence supports our approximation of resistivity formula ignoring vertex correction in the next section.  The inset indicates that the heating makes the most difference in the low energy region by lifting the bottom. In \figdisp{rhosigmaEDC} (f), $\rho_\Sigma$ at $k_F$ for different $t'$ are put together. As $t'$ increase from negative to positive, its minimum goes down, indicating a lower decay rate, and the bottom region becomes rounded and more Fermi-liquid like.

We also study the temperature-dependent $\rho_G(k_F)$ and $\rho_\Sigma(k_F)$ at $\omega=0$ for $k_F$ in the nodal and antinodal direction in \figdisp{kFT}. Also, (a) and (b) shows that the spectral function peak is very sensitive to temperature changes. A sharp drop happens over a small temperature region ($<1\%$ bare bandwidth), wiping out the quasiparticle peak for $T>400$K in either direction. Another angle to observe this phenomenon is through the self-energy, $\rho_\Sigma(k_F)=1/(\pi^2\rho_G(k_F))$, describing the decay rate of a quasiparticle. The huge increase of $\rho_\Sigma(k_F)$ upon small warming shows a rapid drop in the lifetime of a quasiparticle. Note that the $\rho_\Sigma$ curvature dependence on $t'$ is similar to that of the plane resistivity in Fig. (4) of \refdisp{SP}.

\begin{figure}[!]
\subfigure[\;\; $\delta=0.11$]{\includegraphics[width=.32\columnwidth]{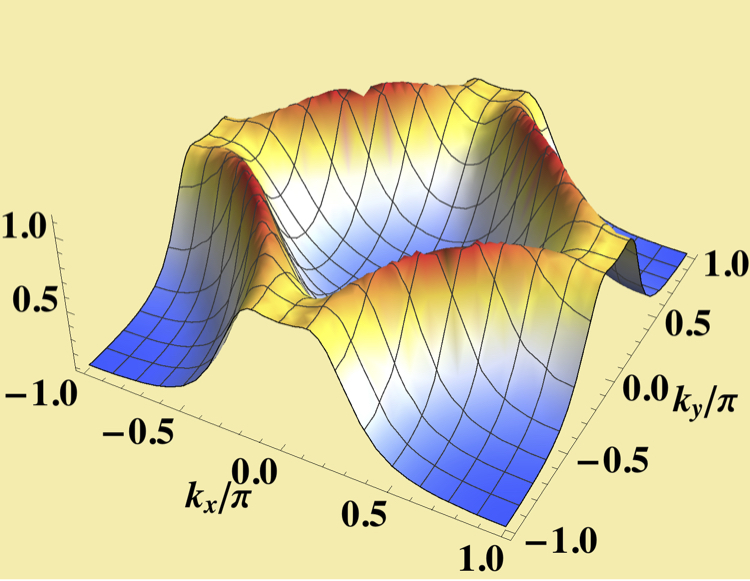}}
\subfigure[\;\; $\delta=0.14$]{\includegraphics[width=.32\columnwidth]{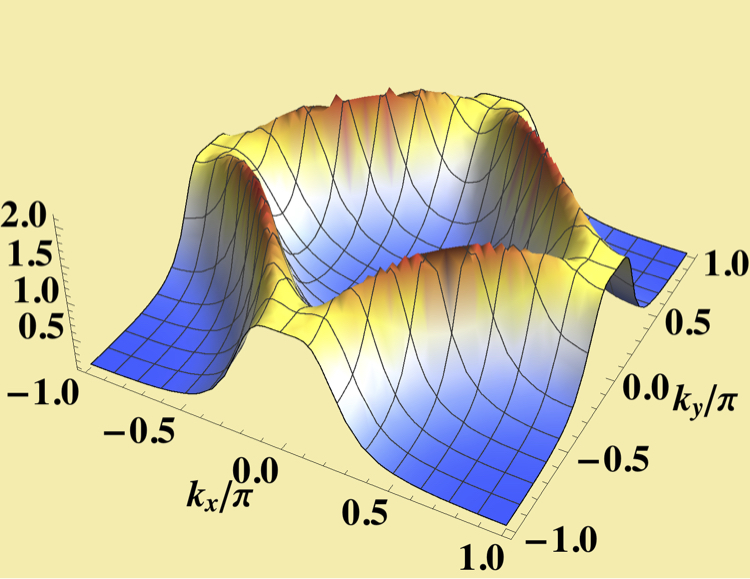}}
\subfigure[\;\; $\delta=0.17$]{\includegraphics[width=.32\columnwidth]{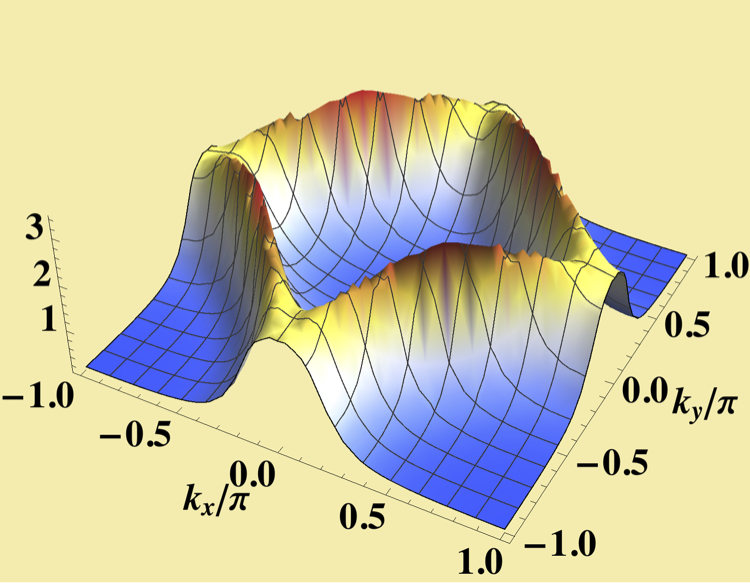}}
\subfigure[\;\; $\delta=0.2$]{\includegraphics[width=.32\columnwidth]{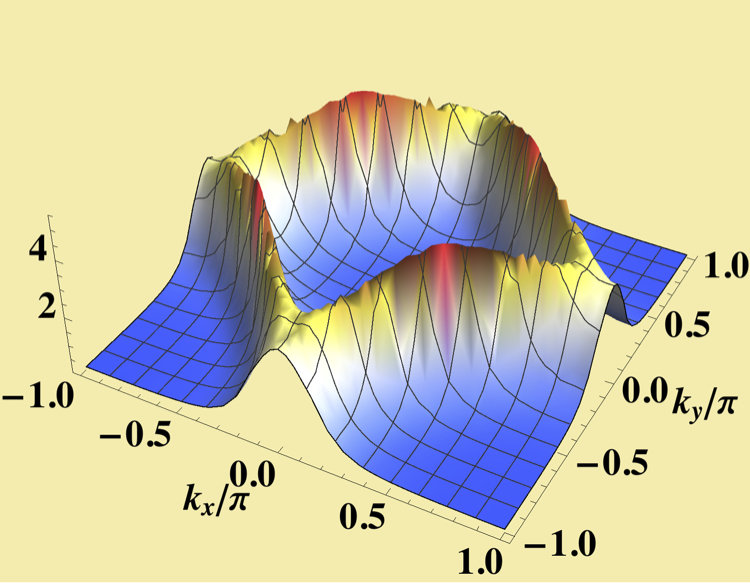}}
\subfigure[\;\; $\delta=0.23$]{\includegraphics[width=.32\columnwidth]{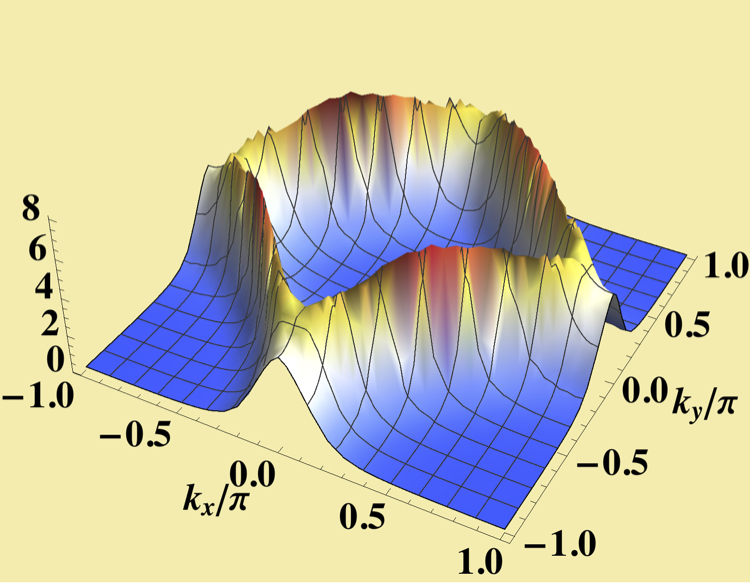}}
\caption{(Color online) The 3D plot of The spectral function peak height at several dopings at $t'=-0.2$, $T=63K$. The ridge in the spectral function peak tracks the Fermi surface. As $\delta$ increases, we find that the Fermi surface changes from open (hole-like) to close (electron-like), with the critical $\delta\approx 0.17$. The ridge height increases generally as $\delta$ goes up, showing decreasing correlation strength.}
\label{FS3D}
\end{figure}

\begin{figure}[!]
\subfigure[\;\; $\rho_G(\pi,k_y)$, $M \to X$]{\includegraphics[width=.32\columnwidth]{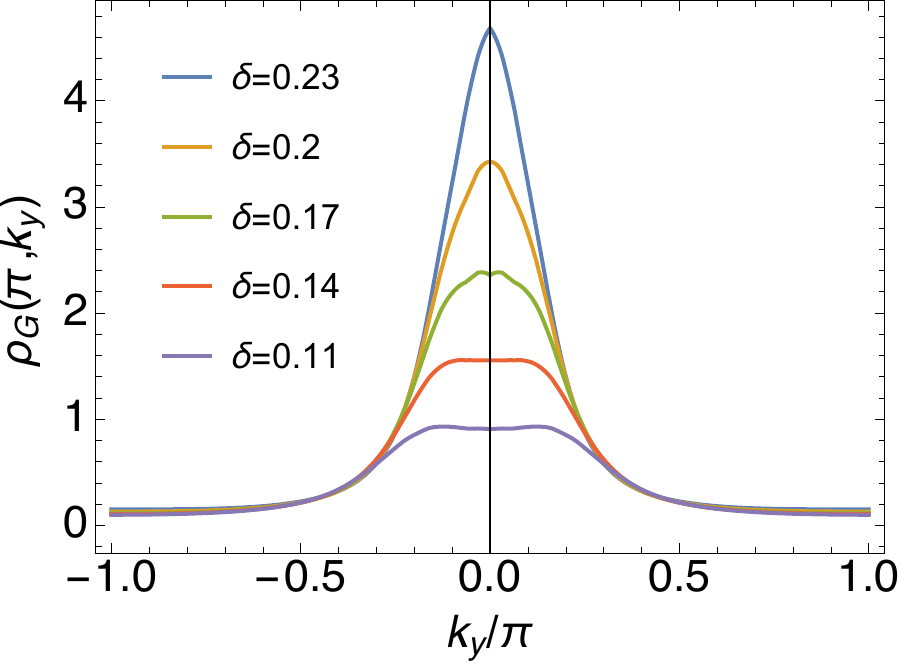}}
\subfigure[\;\; $\rho_G(\pi/2,k_y)$]{\includegraphics[width=.32\columnwidth]{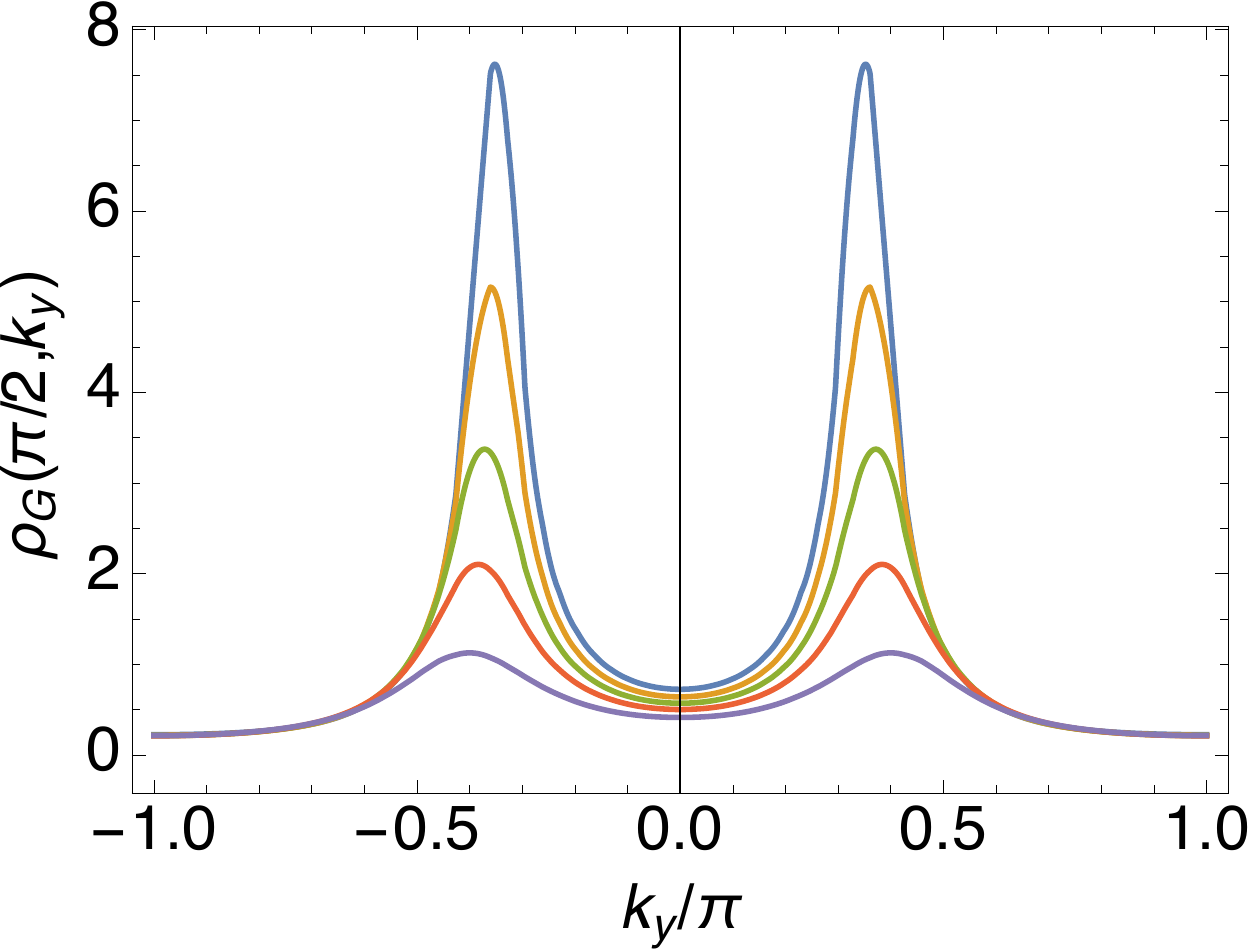}}
\subfigure[\;\; $\rho_G(0,k_y)$, $\Gamma\to M$]{\includegraphics[width=.32\columnwidth]{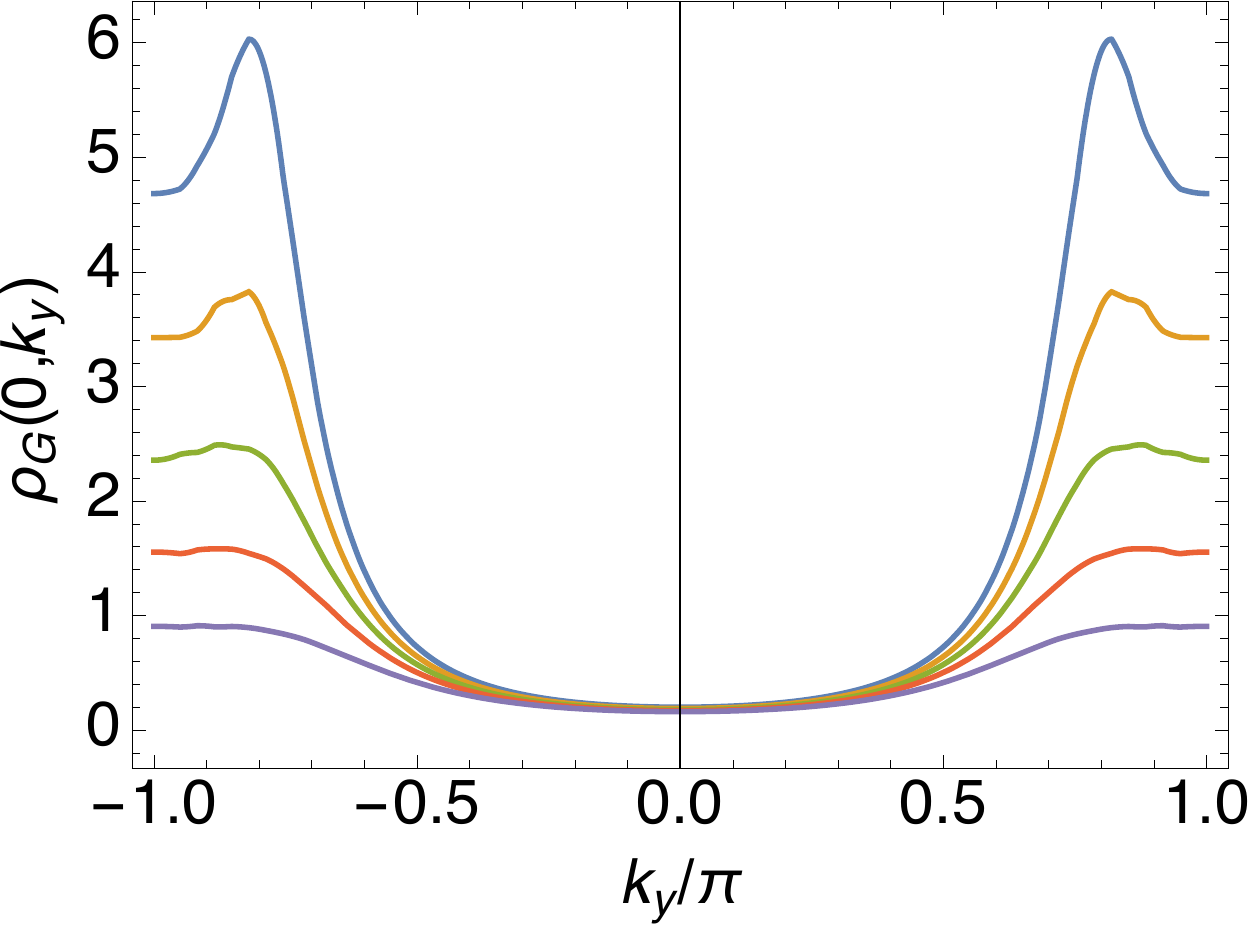}}
\subfigure[\;\; $\rho_G(k,k)$, nodal ($\Gamma\to X$)]{\includegraphics[width=.32\columnwidth]{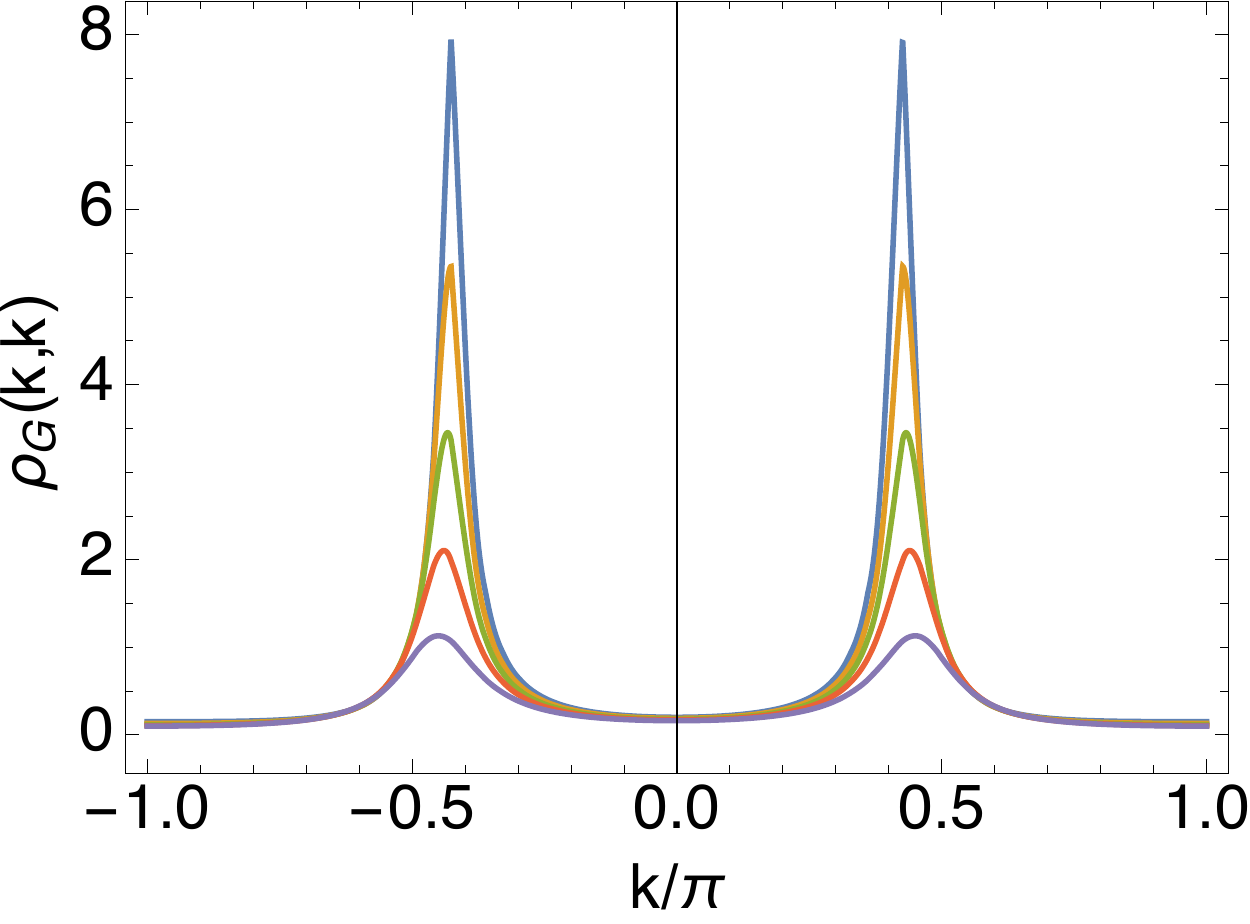}}
\subfigure[\;\; Tight-binding model]{\includegraphics[width=.32\columnwidth]{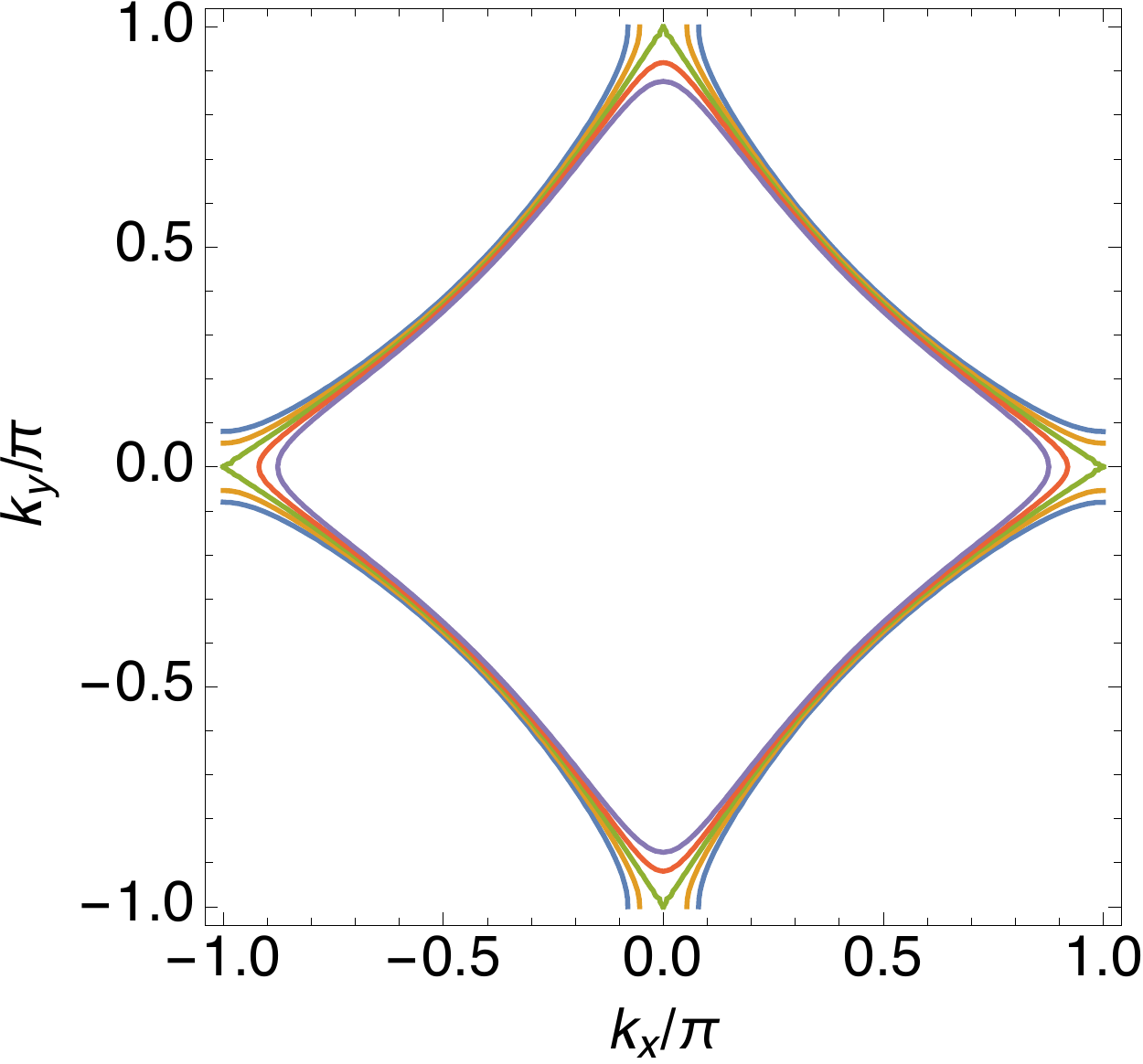}}
\caption{(Color online) The spectral function peak height in typical directions of momentum space at several dopings at $t'=-0.2$ and $T=63K$. All panels share the same legend. Panel (a) shows evidence of Lifshitz transition (Fermi surface changed from opened to closed) at $\delta\approx 0.17$, similar to the tight binding model case shown in panel (e). Panel (b), (c), (d) provide other angles to observe this transition, in complimentary with the 3D plots in \figdisp{FS3D}. }
\label{FS}
\end{figure}

The momentum distribution curves (MDCs) are plotted in \figdisp{MDCnodal} and \figdisp{MDCantinodal}, obtained by fixing $\omega$ and scanning $k$ in nodal and antinodal directions respectively, at optimal doping and various $t'$. As expected from the EDC case, the MDC peak is highest at the Fermi surface $\omega=0$, which gets broadened the most upon warming. However, unlike the EDC case, the MDC peaks that are far away from $k=0$ or $\pi$ look more symmetric. This difference is consistent with experimental findings. 
The spectral function in the early phenomenological versions of ECFL \refdisp{Gweon,Anatomy} lead to a somewhat exaggerated asymmetry in MDC curves, and has been the subject of further phenomenological adjustments in \refdisp{Gweon-Kazue}, to reconcile with experiments.  The present microscopic results show that the greater symmetry of the MDC spectral lines comes about naturally, without the need for any adjustment of the parameters.

\subsubsection{Fermi Surface}
The Fermi surface (FS) structure can be observed in the momentum distribution of spectral function peak height. We present the case for $t'=-0.2$, which is  roughly the parameter describing  the $LSCO$  cuprate material \cite{LSCOFS}, and vary the doping $\delta$ in \figdisp{FS3D}. The FS is hole-like (open) for low doping (a and b) and becomes electron-like (closed) for high doping in (d and e). The transition point $\delta\approx 0.17$ can be explicitly seen in \figdisp{FS}a which is close to the non-interacting case with tight-binding model in \figdisp{FS}e, consistent to experimental findings\cite{LSCO1,LSCO2,LSCOFS}. At higher (hole) doping which leads to a weaker effective correlation \cite{SP}, the quasiparticle peak height increases and becomes more Fermi-liquid like. 

The FS is only well-defined at zero temperature. Following \refdisp{Shastry-Gamma}  we can define a pseudo-FS at finite temperature, by examining  a specifically weighted first moment of the energy:
\beq
\gamma_{k \sigma}(\mu,T)=-{\int\rho_G(k,\omega)  \frac{ {d\omega} \, \omega} {\cosh(\beta \omega/2)}}\bigg/ \int \rho_G(k,\omega)  \frac{{d\omega}}{\cosh(\beta \omega/2)}
\eeq
We  {\em define} a pseudo-FS as the surface in $\vec{k}$ space where $\gamma_{k \sigma}$ changes sign from positive to negative. Shastry has recently shown \cite{Shastry-Gamma} that 
at $T=0$, the pseudo-FS  becomes the exact Luttinger-Ward FS. It is further suggested that it is  useful to study  a $T$ dependent effective carrier density
\beq
N_{eff}=\sum_{k \sigma} \Theta(\gamma_{k \sigma}(\mu,T)),
\eeq
where $\Theta$ is the Heaviside step function, such that  $N_{eff}=N$  at zero temperature. At finite temperatures we expect  that $N_{eff} \neq N$, and the difference between the two gives  insights into the different T scales at play. This is especially applicable in strongly correlated materials, where it is well known \cite{badmetal,HFL,WXD} that Gutzwiller correlations result in the Fermi liquid regime, the strange metal regime and the bad metal regime, followed by a high T regime, with three crossover temperatures. 
\begin{figure}[ht]
\subfigure{\includegraphics[width=.5\columnwidth]{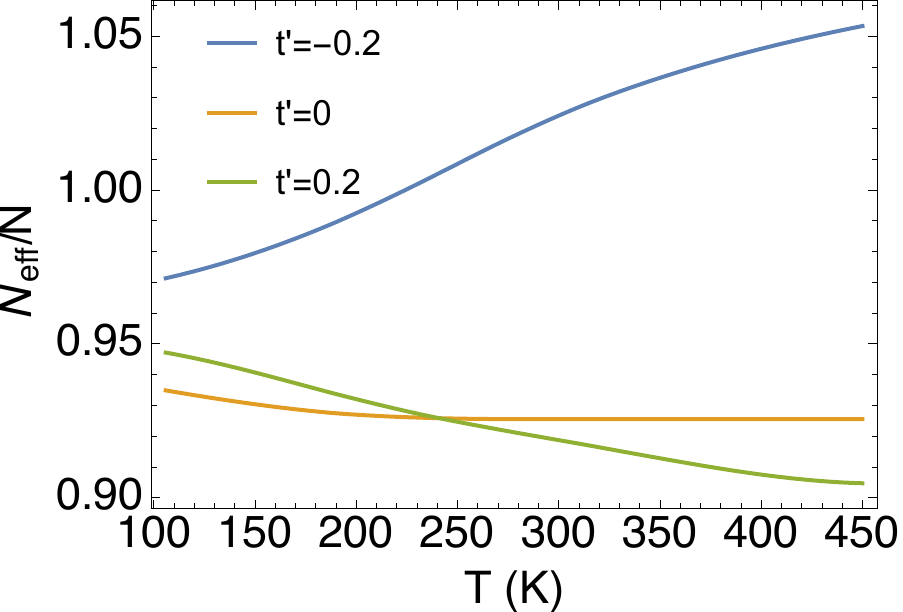}}
\caption{(Color online) $N_{eff}/N$ vs $T$ at $\delta=0.15$ and various $t'$. For electron-doped ($t'\geq0$) case, $N_{eff}$ increases as one lowers the temperature, while in the hole-doped ($t'=-0.2$) case, $N_{eff}$ decreases upon cooling down. At lower temperature, one expects that $N_{eff}$  equals  $N$. } 
\label{Neff}
\end{figure}
In \figdisp{Neff}, we show how $N_{eff}/N$ changes with temperature for different $t'$. For $t'\geq0$, $N_{eff}$ increases monotonically toward $N$ as $T$ goes down. And for $t'<0$, $N_{eff}$ decreases from larger to smaller than $N$ upon cooling. With further lowering $T$ one  expects that $N_{eff}$  equals  $N$.

\begin{figure}[!]
\subfigure[\;\;]{\includegraphics[width=.32\columnwidth]{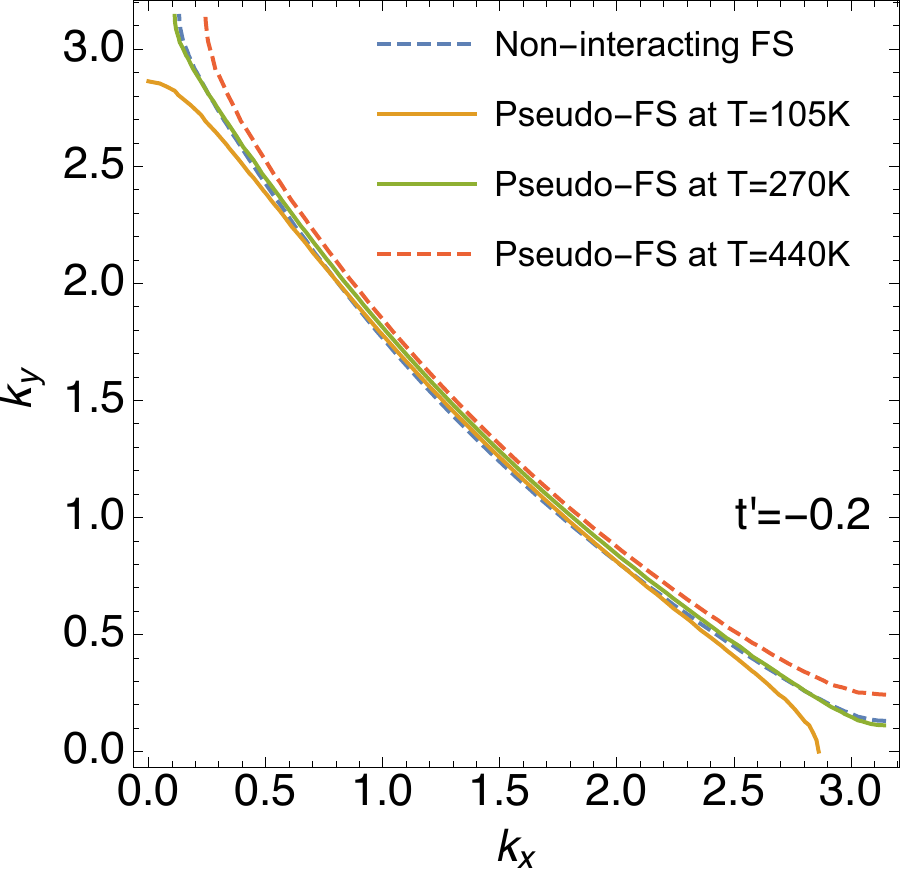}}
\subfigure[\;\;]{\includegraphics[width=.32\columnwidth]{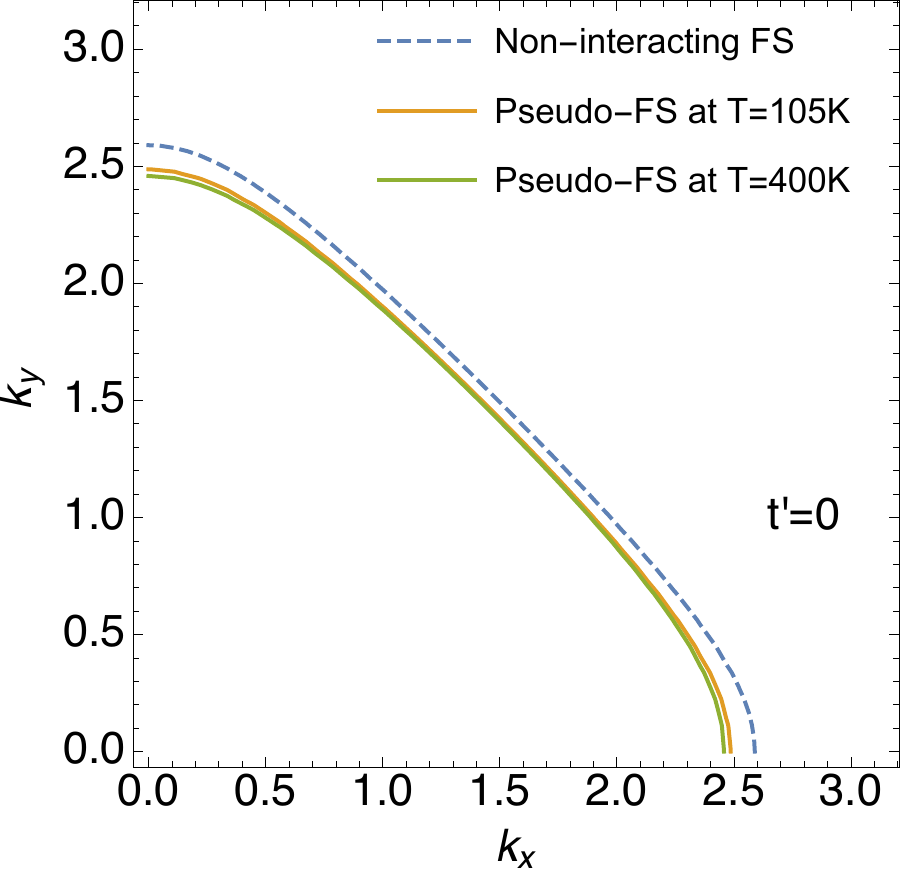}}
\subfigure[\;\;]{\includegraphics[width=.32\columnwidth]{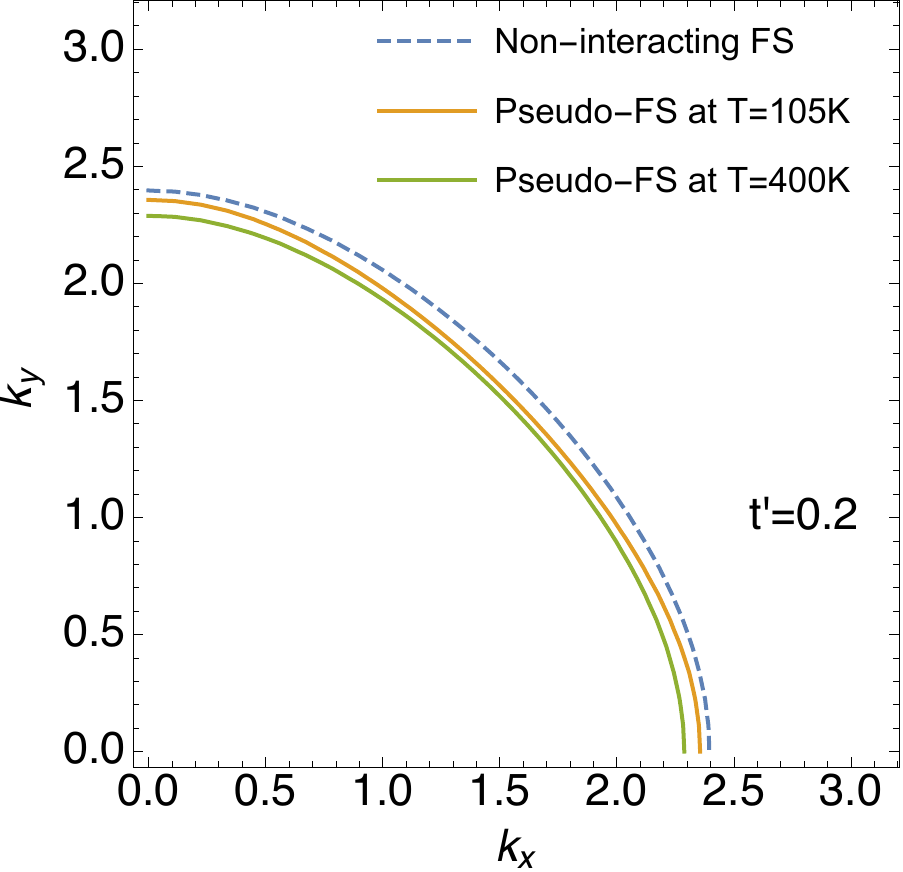}}
\caption{(Color online) Comparison between the non-interacting FS and pseudo-FS at low and high temperature. Here we fix $\delta=0.15$ and vary $t'$. Generally, as we cool down the system, the pseudo-FS  approaches the non-interacting system or FS from the right ($t'=-0.2$) or left ($t'=0, 0.2$) side. The exception is that at $T=105K$ and $t'=-0.2$ the pseudo-FS turns out to be closed (electron-like). This delicate effect is a consequence of the redistribution of weight in the spectral function, and its thermal sensitivity is presumably related to the nearby Lifshitz transition point for the choice of $t'=-.2$. We cannot access very low $T$ for our system sizes, but it is expected that the pseudo-FS flips back to being hole like at a low $T$.   }
\label{FScom}
\end{figure}

\begin{figure}[!]
\subfigure[\;\; $t'=0.2$, $T=400K$]{\includegraphics[width=.24\columnwidth]{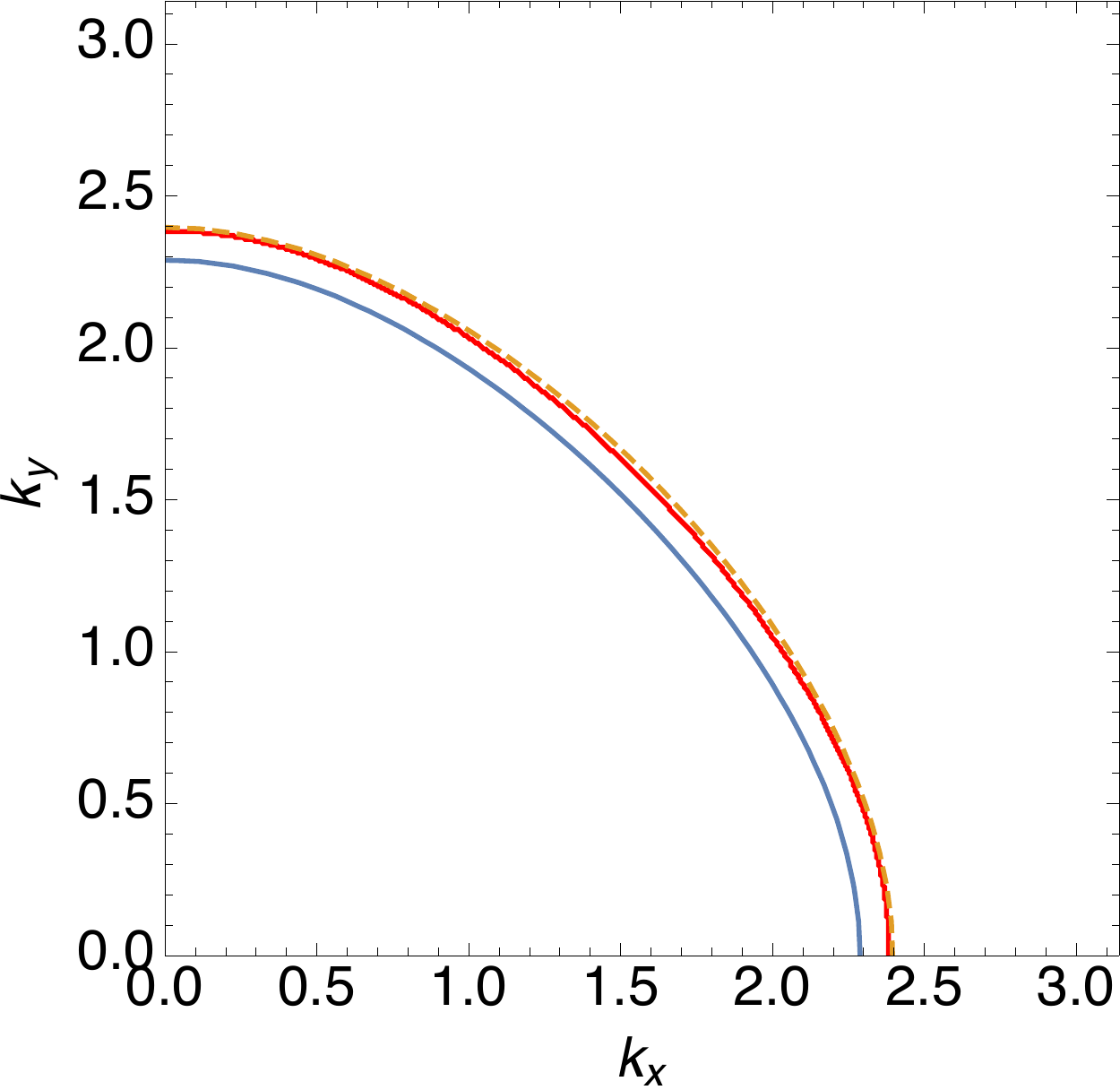}}
\subfigure[\;\; $t'=0$, $T=400K$]{\includegraphics[width=.24\columnwidth]{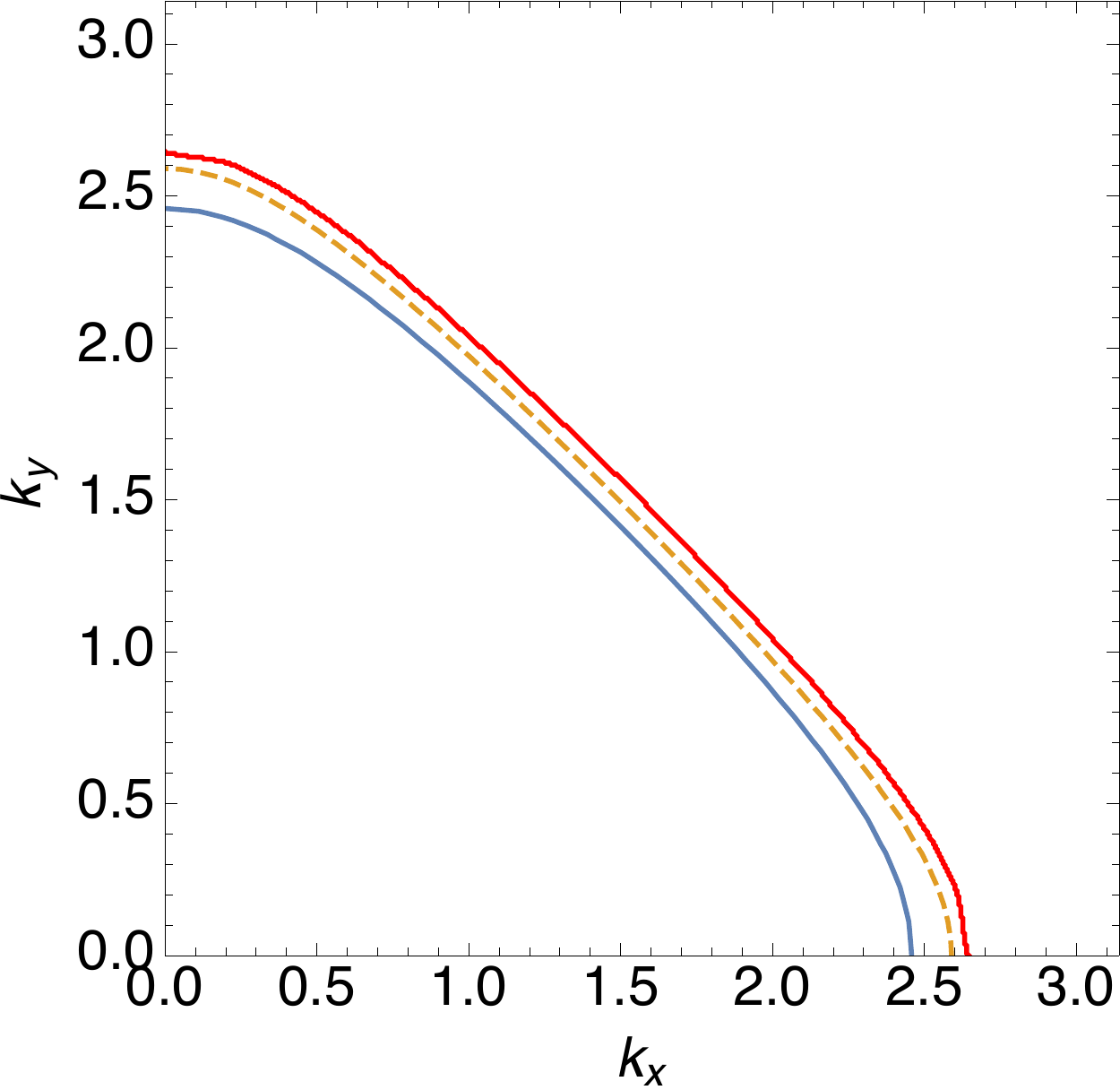}}
\subfigure[\;\; $t'=-0.2$, $T=440K$]{\includegraphics[width=.24\columnwidth]{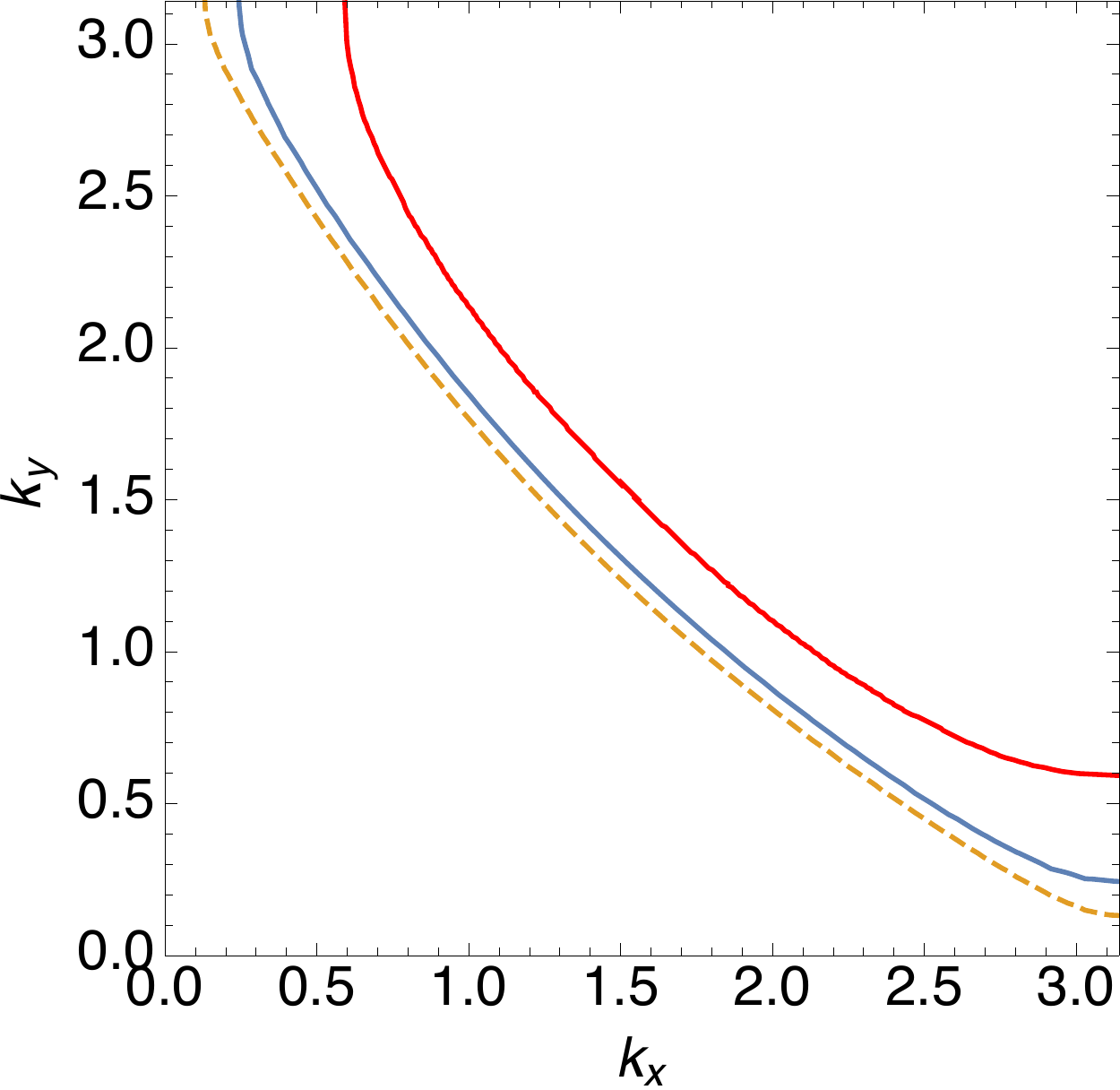}}
\subfigure[\;\;$t'=-0.4$, $T=420K$]{\includegraphics[width=.24\columnwidth]{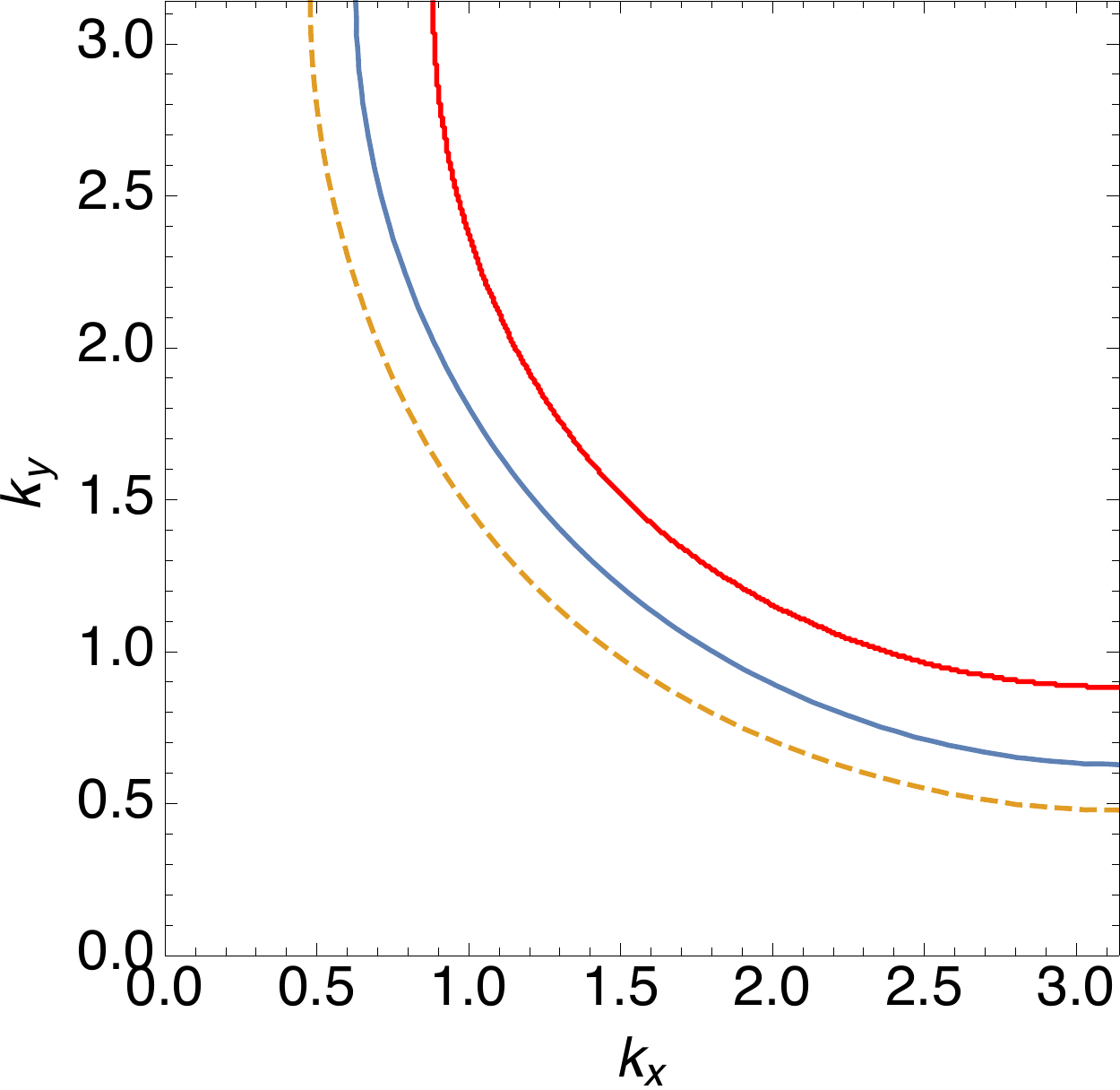}}
\subfigure[\;\; $t'=0.2$, $T=105K$]{\includegraphics[width=.24\columnwidth]{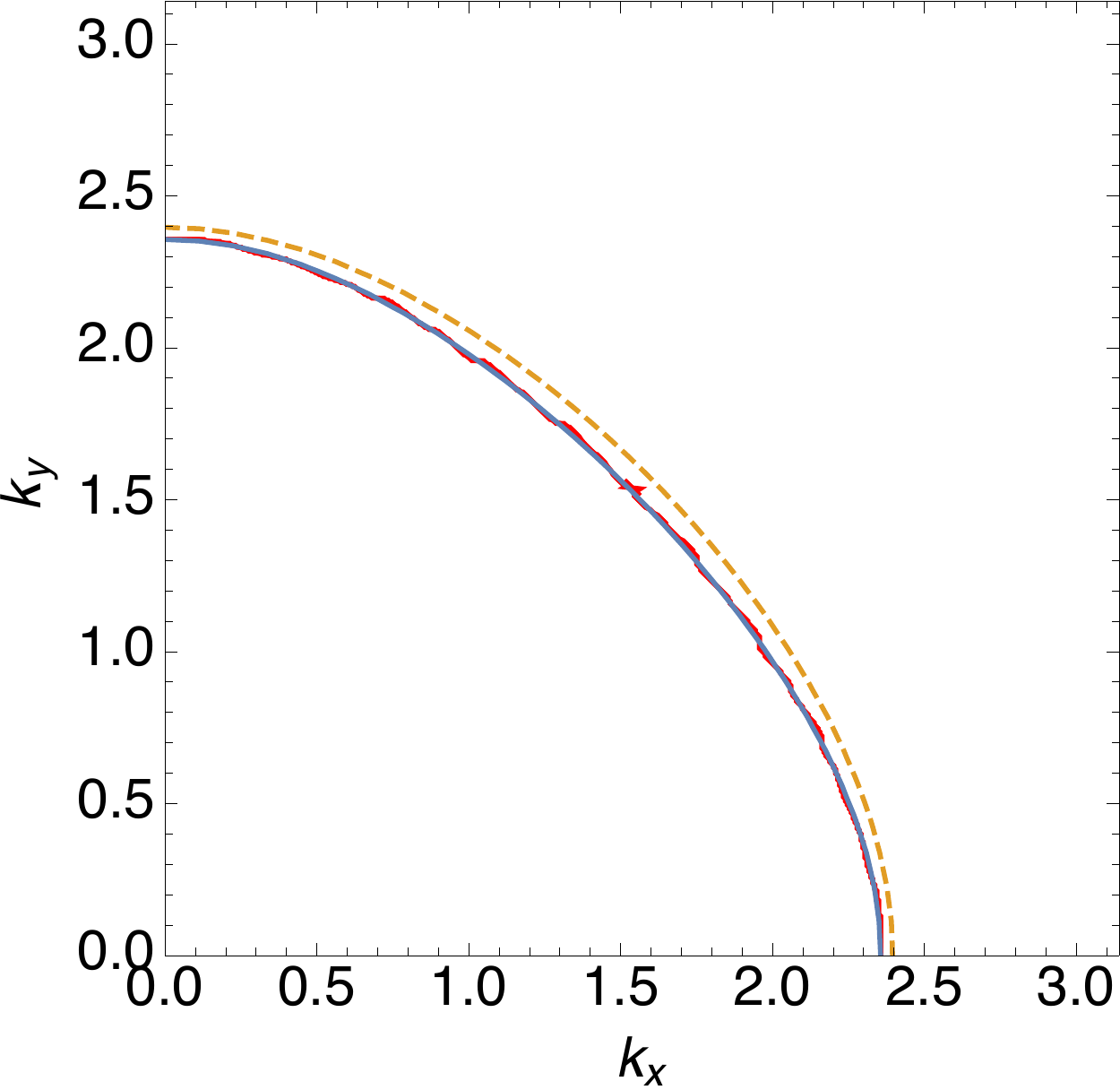}}
\subfigure[\;\; $t'=0$, $T=105K$]{\includegraphics[width=.24\columnwidth]{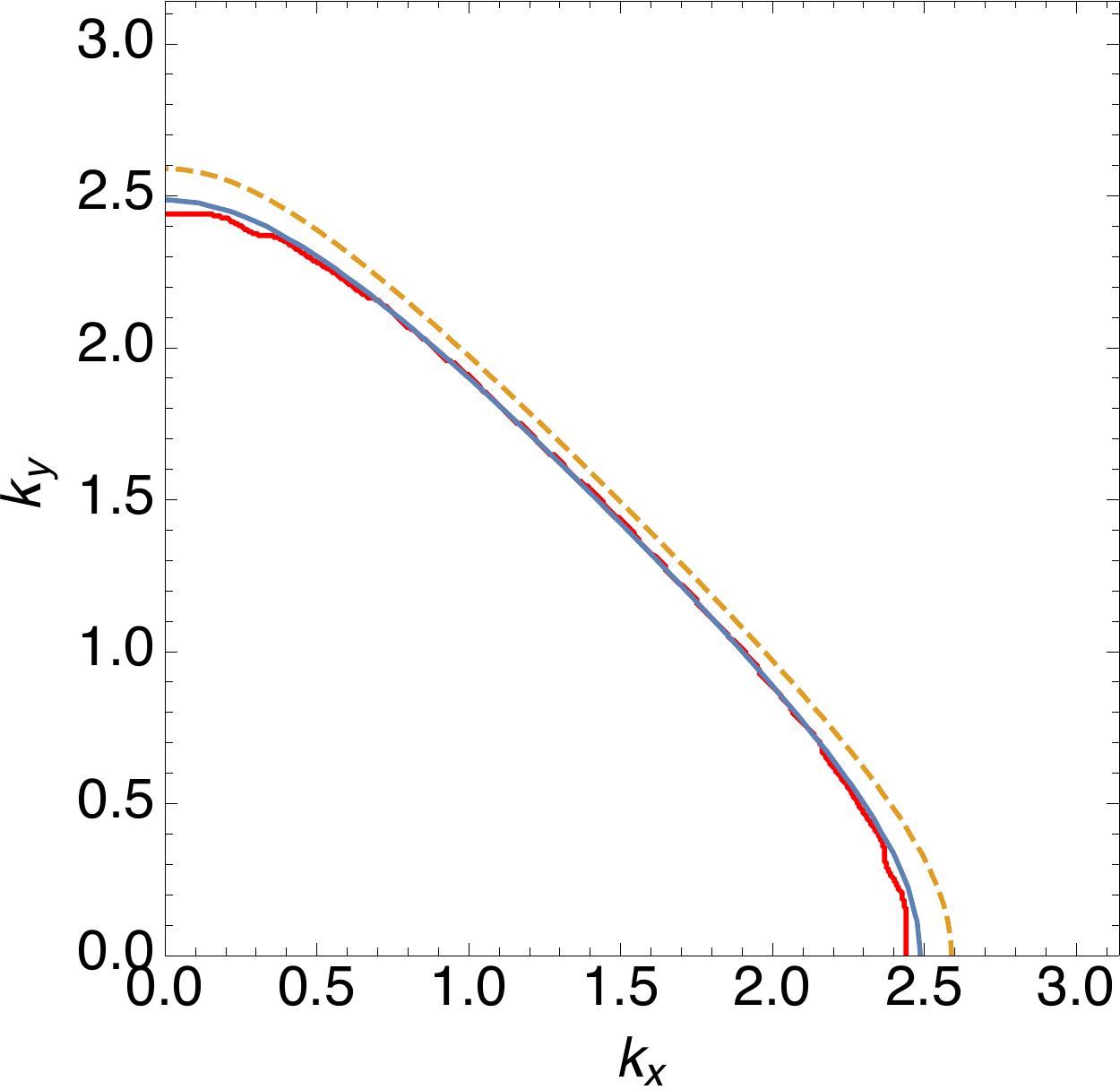}}
\subfigure[\;\; $t'=-0.2$, $T=270K$]{\includegraphics[width=.24\columnwidth]{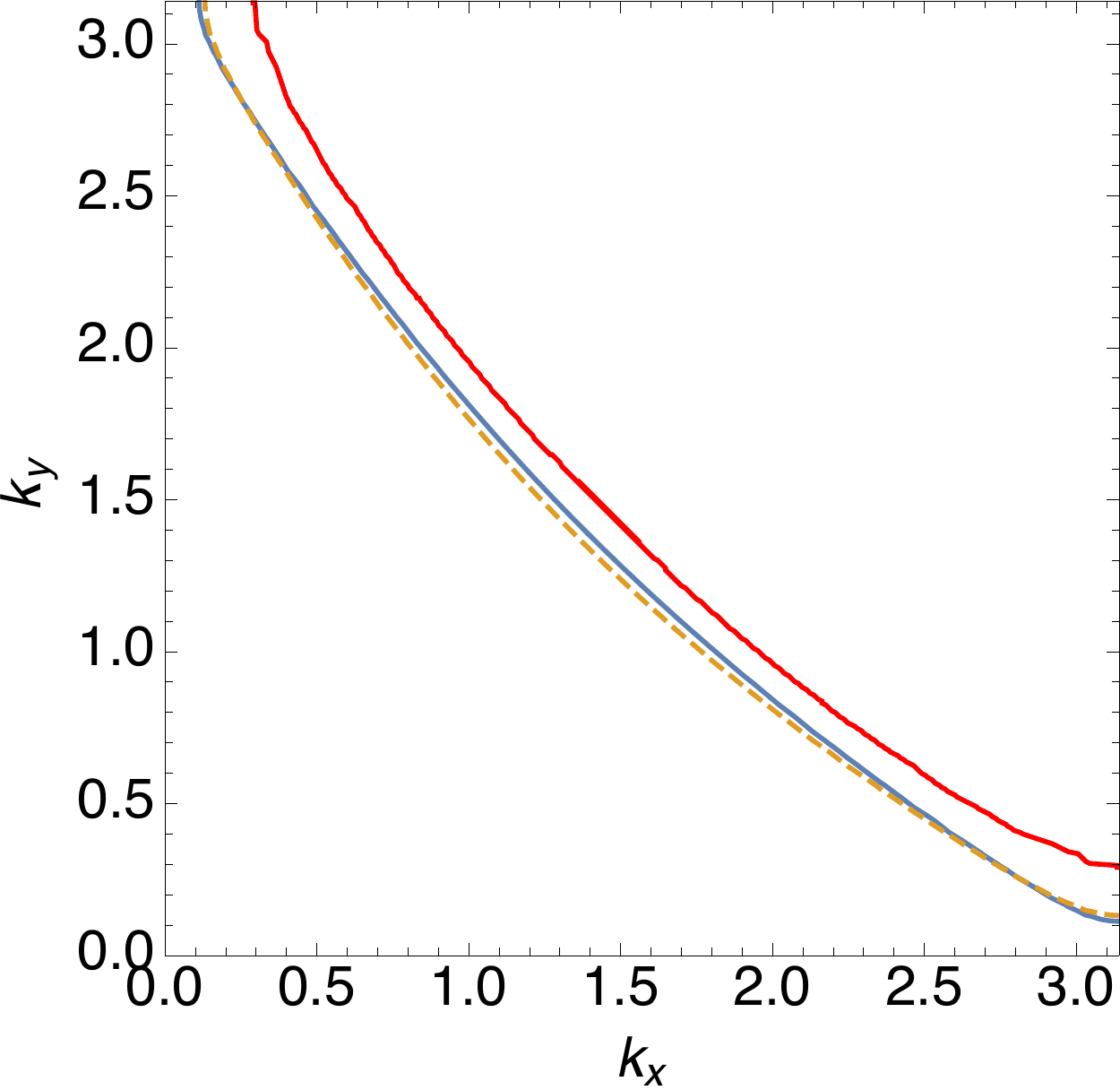}}
\subfigure[\;\;$t'=-0.4$, $T=105K$]{\includegraphics[width=.24\columnwidth]{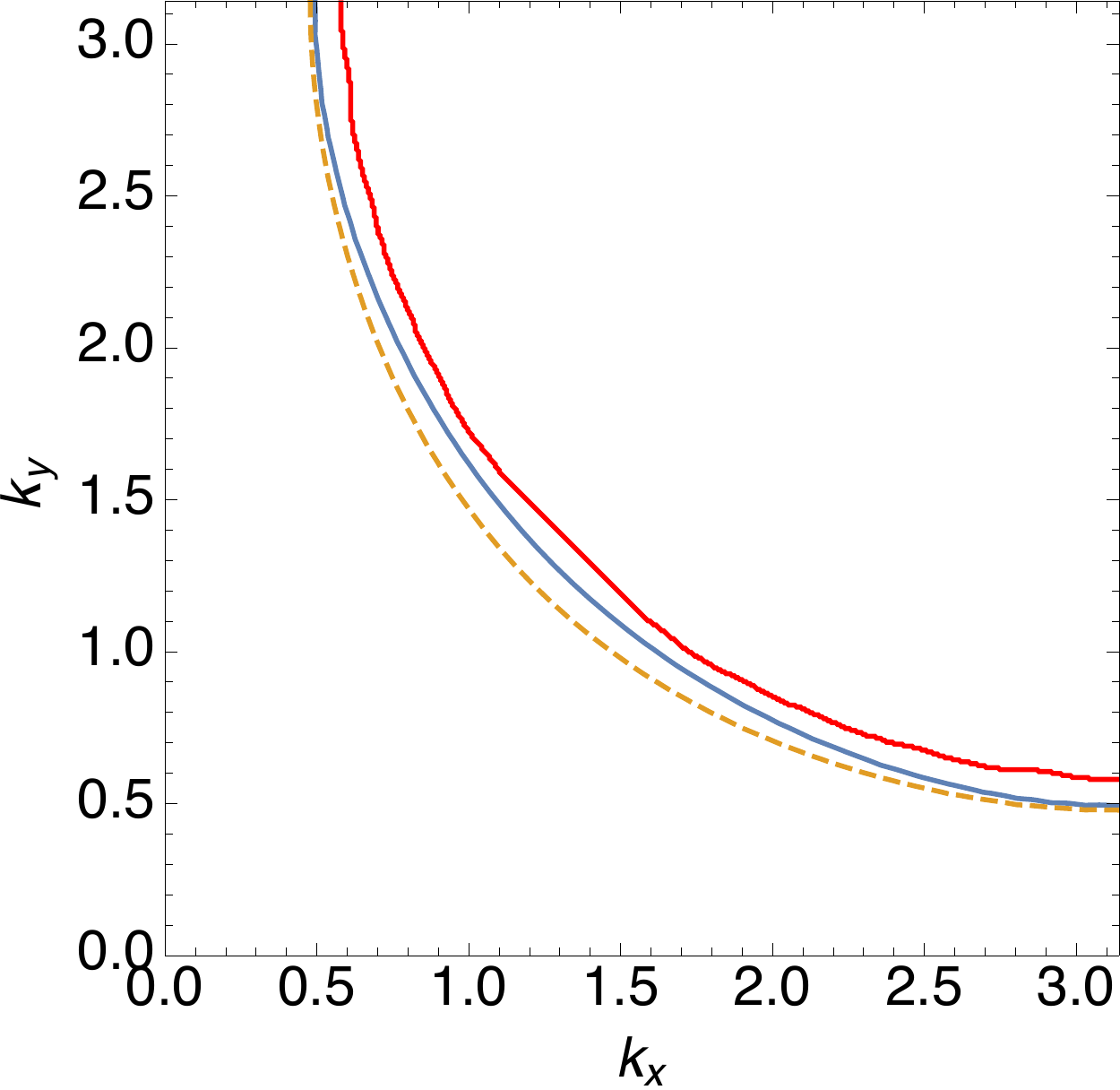}}
\caption{(Color online)  Comparison between the pseudo-FS from $\gamma_k$ (blue), the spectral peak (red) and the non-interacting FS (dashed) at various $t'$ and fixed $\delta=0.15$. Note that the spectral peak (location) curve and the pseudo-FS are not exactly the same, but deviate from the non-interacting FS in the same direction.
As $T$ decreases, the difference between them gets smaller.}
\label{PFScom}
\end{figure}

At low temperatures ($T\ll t$), we find that the roots of $\gamma_k$ are close to the location of the ridge of spectral peak height shown in \figdisp{PFScom}, and hence it can be taken as an approximate or a pseudo finite-temperature FS. \figdisp{FScom} shows that the pseudo-FS is getting close to the true FS at zero temperature as $T$ goes down for both electron-doped and hole-doped systems.  

\begin{figure}[thb]
\begin{center}
\subfigure{\includegraphics[width=.5\columnwidth]{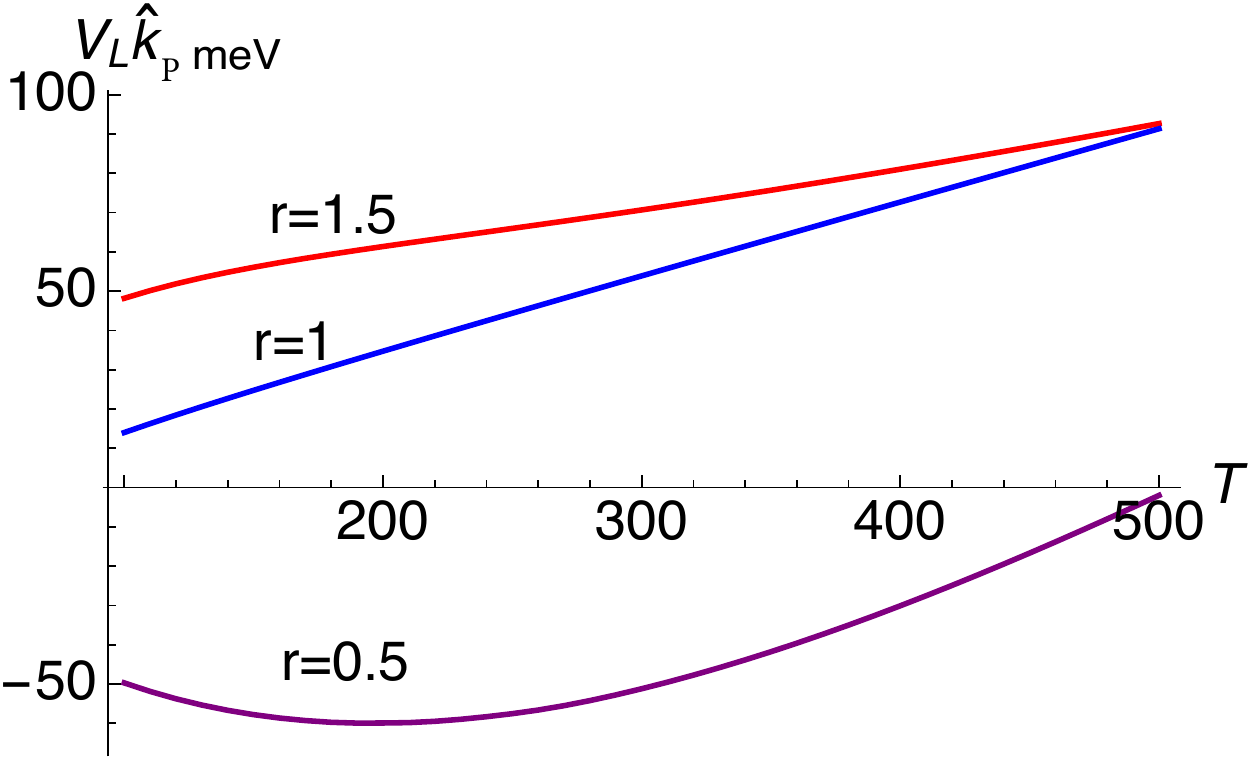}}
\caption{The  location of the peak of the spectral function $A(\hat{k},\omega)$ in \disp{ak-model} in  units of  $\hat{k}_P V_L$ versus T,  at three values  of $r$. The model spectral function, \disp{ak-model}, is from \refdisp{Matsuyama}. It  is obtained by a low energy expansion of the two ECFL self energies $\Psi$ and $\Phi$ (equivalently $\chi$) in  \disp{mu2} and \disp{eq2}.
As $T\to0$ all the curves move towards $\hat{k}=0$ as one expects, but the approach from finite $T$ display significant differences  depending on the value of $r$.
 The values of the parameters used here are $\eta=.01, \Delta_0=50,\Omega_\Phi=5000$ (in meV), and $c_\alpha=10$. An estimated \cite{Matsuyama}  $V_L\sim 2$ eV$A^0$ gives the shift in wavevector $\Delta \hat{k} \sim .05\,A^0$, at 500K for $r=1.5$. } 
\label{model-akw}
\end{center}
\end{figure}

To understand better the deviations at finite T  seen in  \figdisp{Neff}, \figdisp{FScom} and  \figdisp{PFScom}, it is helpful to recall a phenomenological spectral-function\cite{Matsuyama} (see Eq.~(9) and Eq.~(SI-20,21) in \refdisp{Matsuyama}). This function is obtained by expanding the two self energies in \disp{mu2} and \disp{eq2} at low energies in a power series. It
captures many features of the ECFL  calculations in terms of a few parameters, and is given as 
\beq
A(\hat{k},\omega)= \frac{z_0}{\pi} \frac{\Gamma(\omega)}{\Gamma(\omega)^2+ (\omega- V_L \hat{k})^2} \times (1- \frac{\xi}{\sqrt{1+  c_\alpha \xi^2}}), \label{ak-model}
\eeq
where $\hat{k}$ is the component of $\vec{k}$ normal to the FS; $\xi= \frac{1}{\Delta_0} (\omega- r\;  V_L  \hat{k})$; $\Gamma(\omega)= \eta + \frac{\pi}{\Omega_\Phi}(\omega^2+ \pi^2 k_B^2 T^2)$; $\Delta_0$ and $\Omega_\Phi$  are the low and high  energy scales; $V_L$ is the Fermi velocity,   $z_0,r$ and $c_\alpha$ are numerical constants. The  important variable $r\in[0,2]$  determines the location of a  feature in the dispersion known as the ``kink". It is analyzed using this model spectral function in \refdisp{Matsuyama}. Here $r=1$ is at the border of two regimes $r<1$ with kinks in the unoccupied side, and $r>1$ with kinks in the occupied side of the distribution.    In \figdisp{model-akw} we plot the location of the peak in the spectral function  \disp{ak-model} against T, for three values  $r=0.5,1$ and $1.5$. From this we see that these regimes display either a shrinking or an enlargement of the FS with increasing $T$. This corresponds to the types of behavior  seen in the \figdisp{FScom} and \figdisp{PFScom}.

\begin{figure}[h!]
\subfigure[\;\; $T=105K$]{\includegraphics[width=.49\columnwidth]{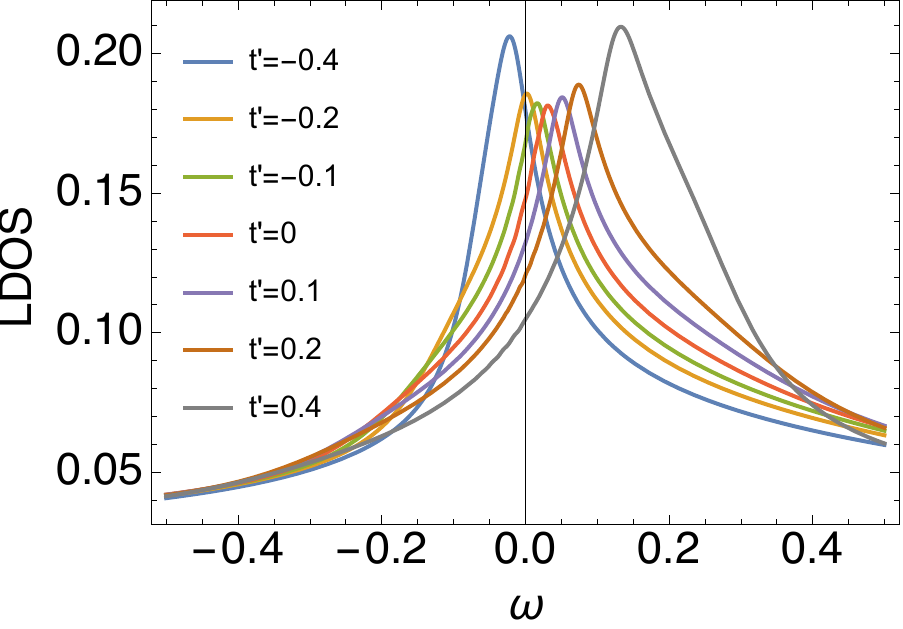}}
\subfigure[\;\; $T=400K$]{\includegraphics[width=.49\columnwidth]{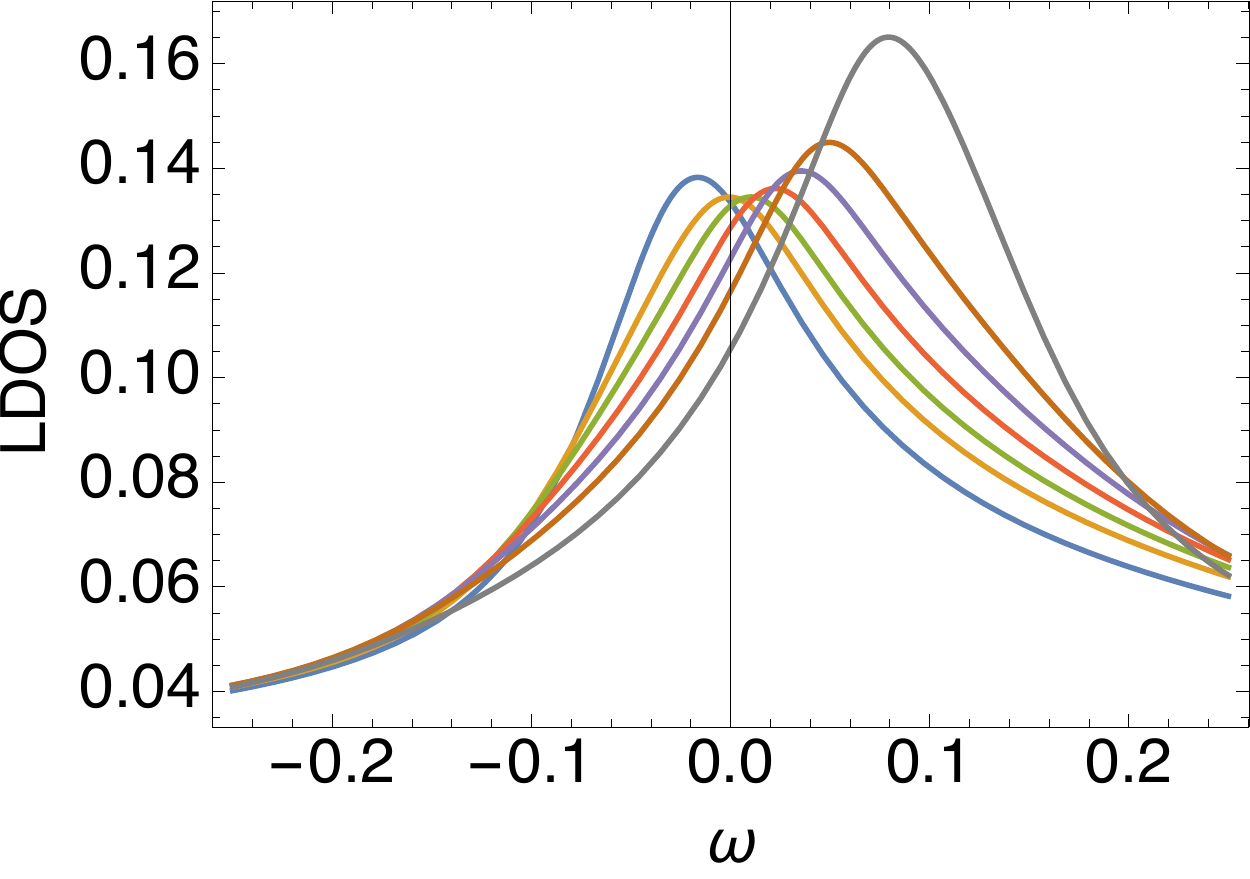}}
\subfigure[\;\; Tight-binding model for reference, $T=0$]{\includegraphics[width=.49\columnwidth]{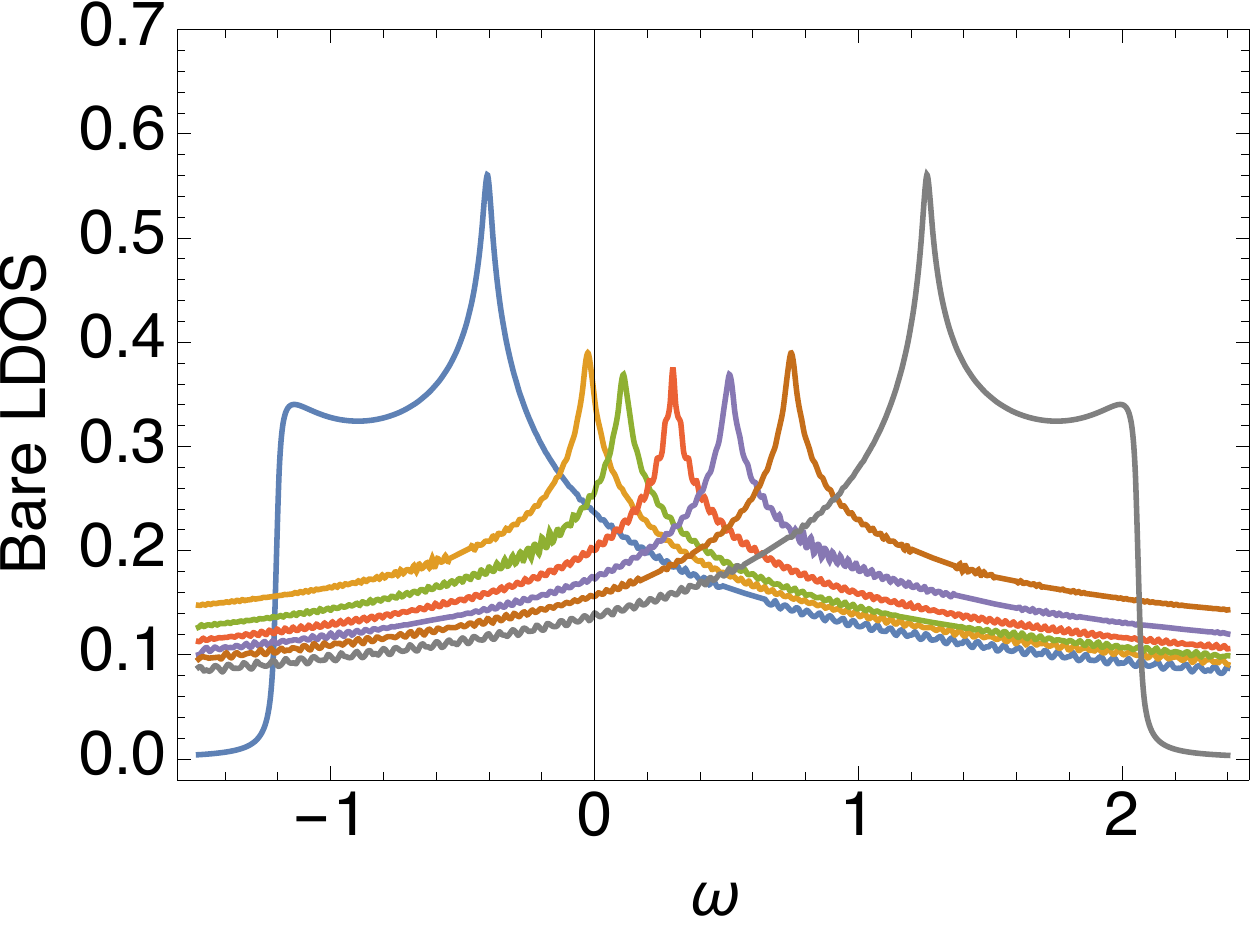}}
\caption{(Color online) Local density of states with varying $t'$ while fixing $\delta=0.15$, at $T=105K$ and $400K$ from ECFL and at $T=0$ from the bare case. All figures share the same legend.}
\label{LDOSvarytp}
\end{figure}

\begin{figure}[!]
\subfigure[\;\; $t'=0$, $T=105K$]{\includegraphics[width=.49\columnwidth]{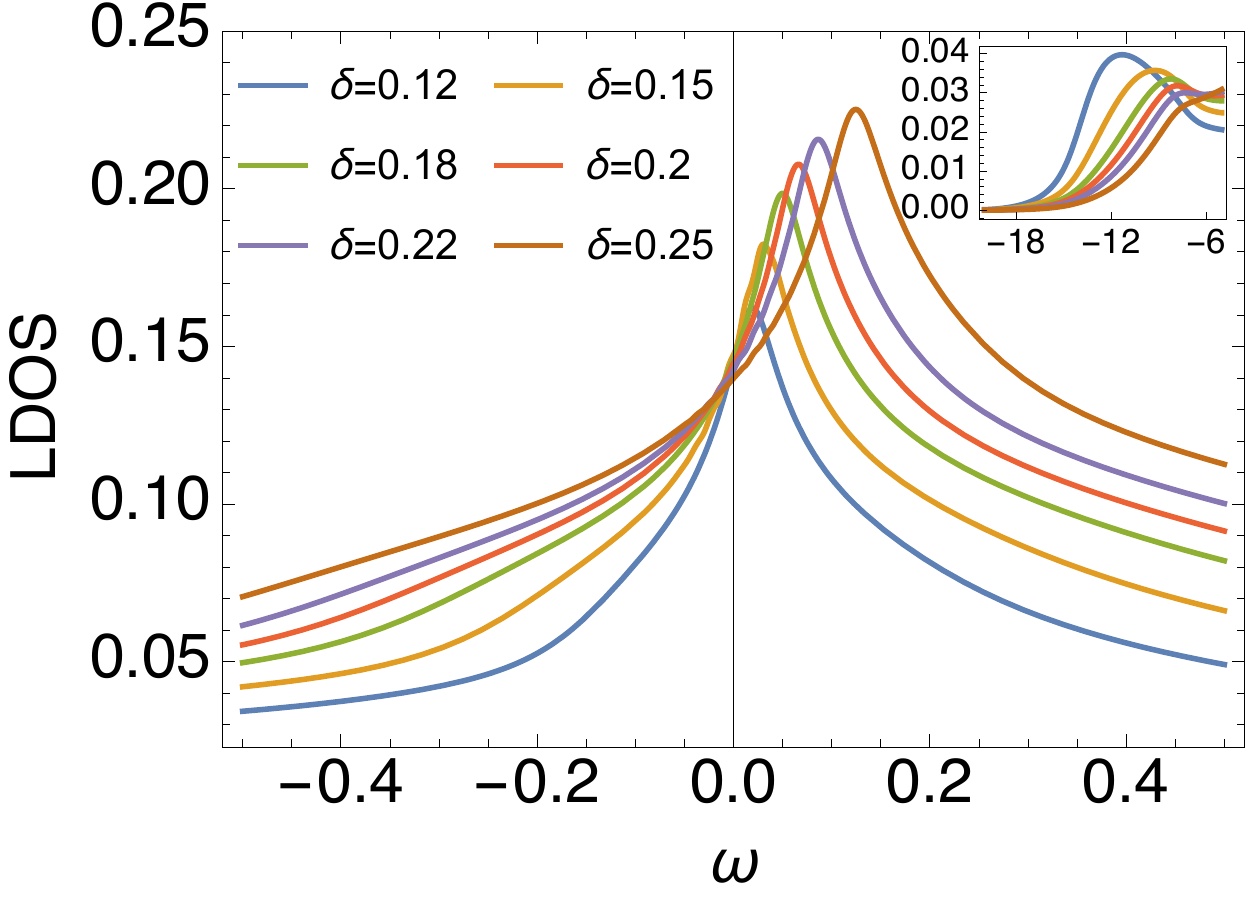}}
\subfigure[\;\; $t'=-0.4$, $T=105K$]{\includegraphics[width=.49\columnwidth]{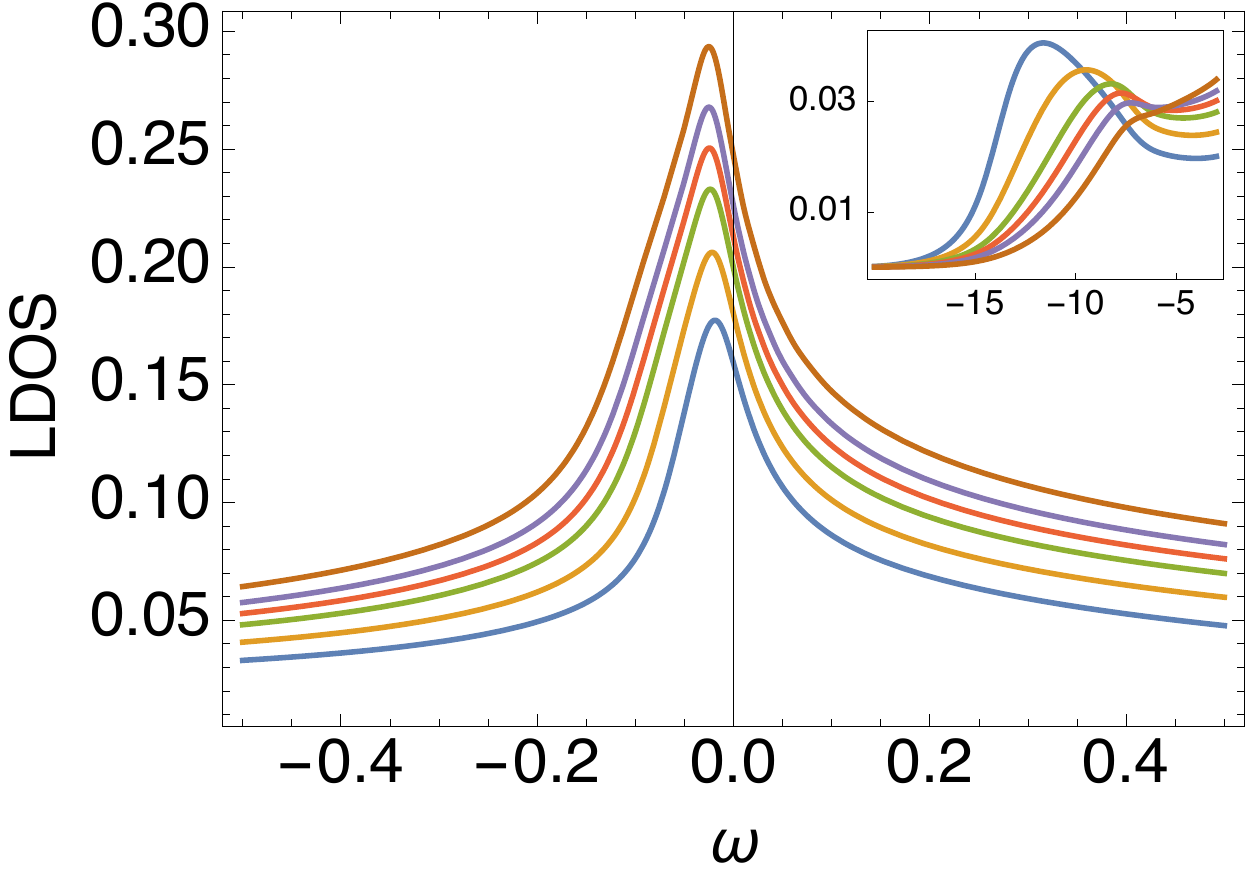}}
\subfigure[\;\; $t'=0$, $T=400K$]{\includegraphics[width=.49\columnwidth]{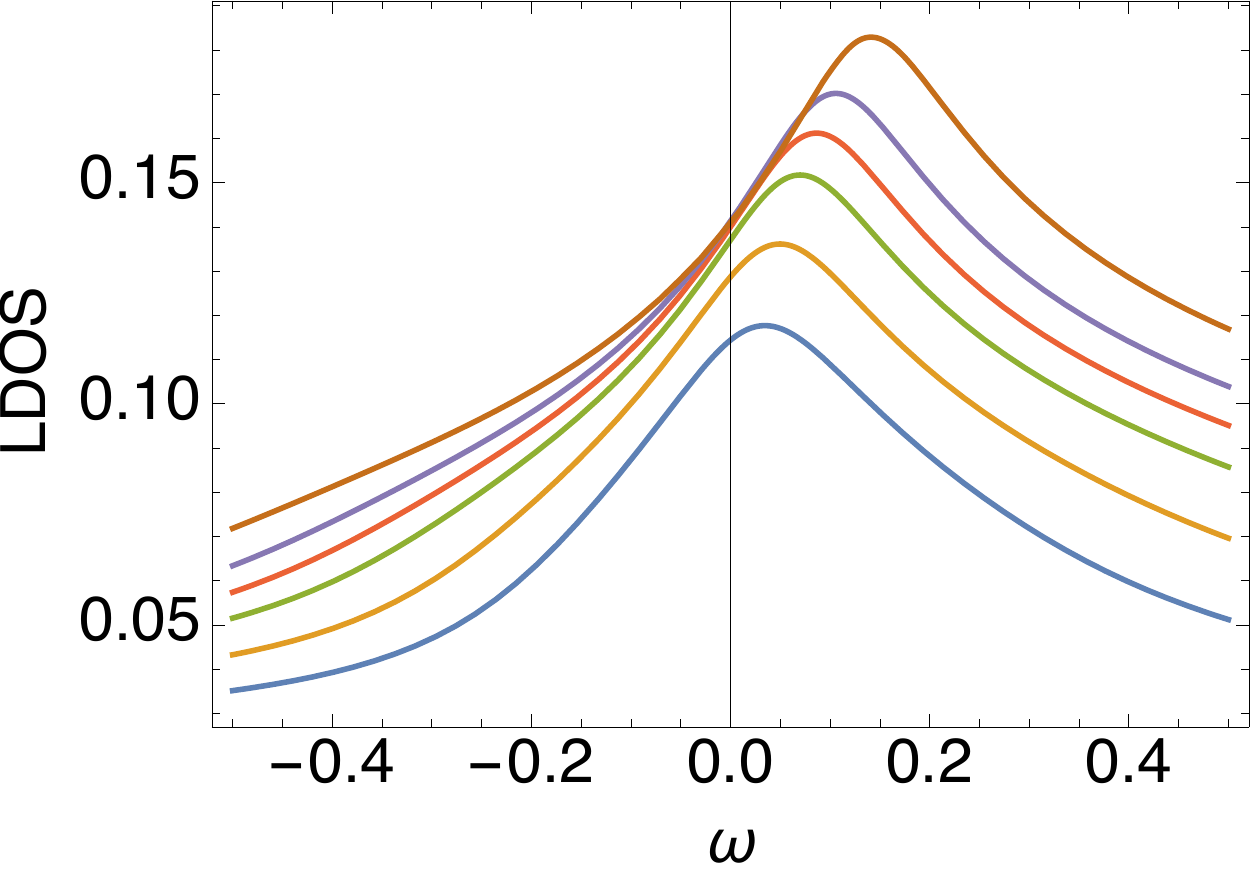}}
\subfigure[\;\; $t'=-0.4$, $T=400K$]{\includegraphics[width=.49\columnwidth]{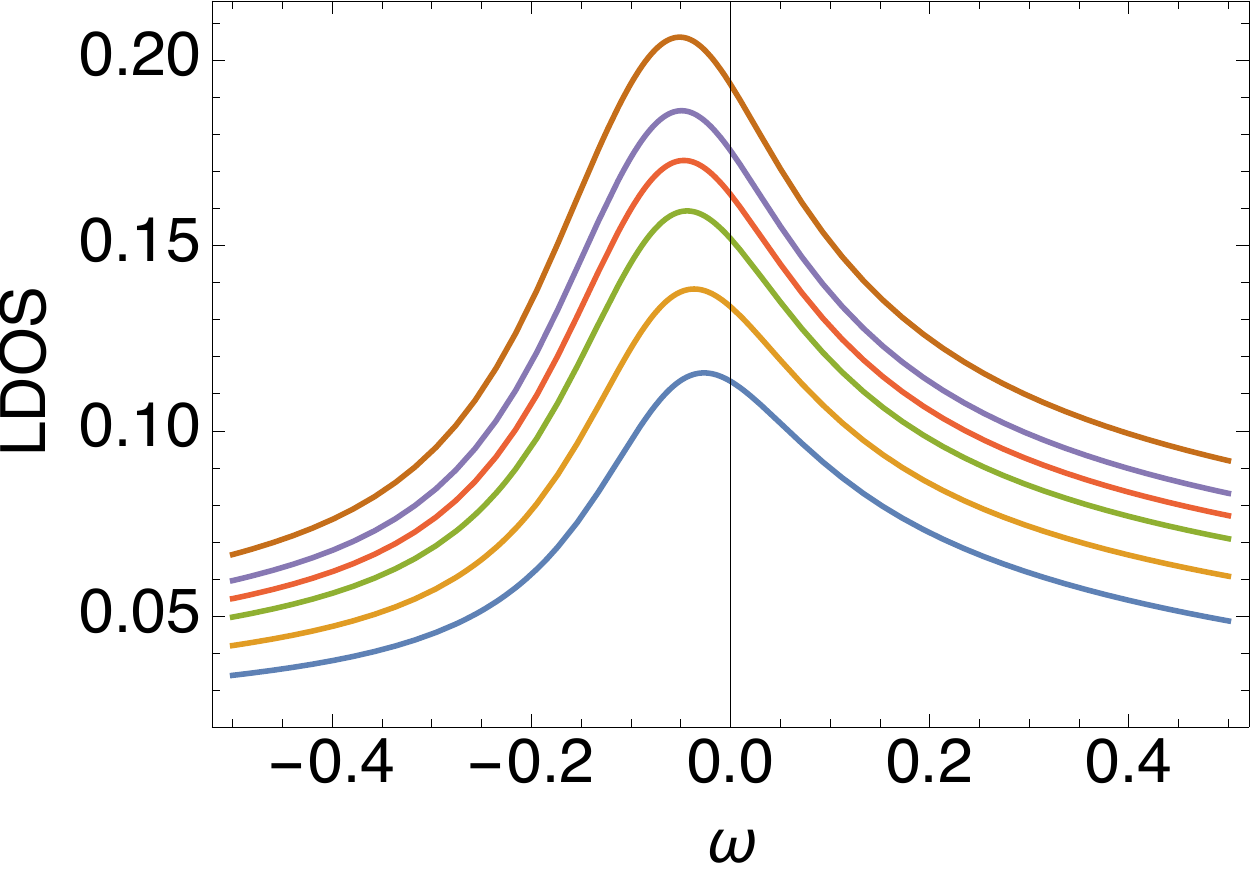}}
\subfigure[\;\; $t'=0$, Tight-binding model for reference, $T=0$]{\includegraphics[width=.49\columnwidth]{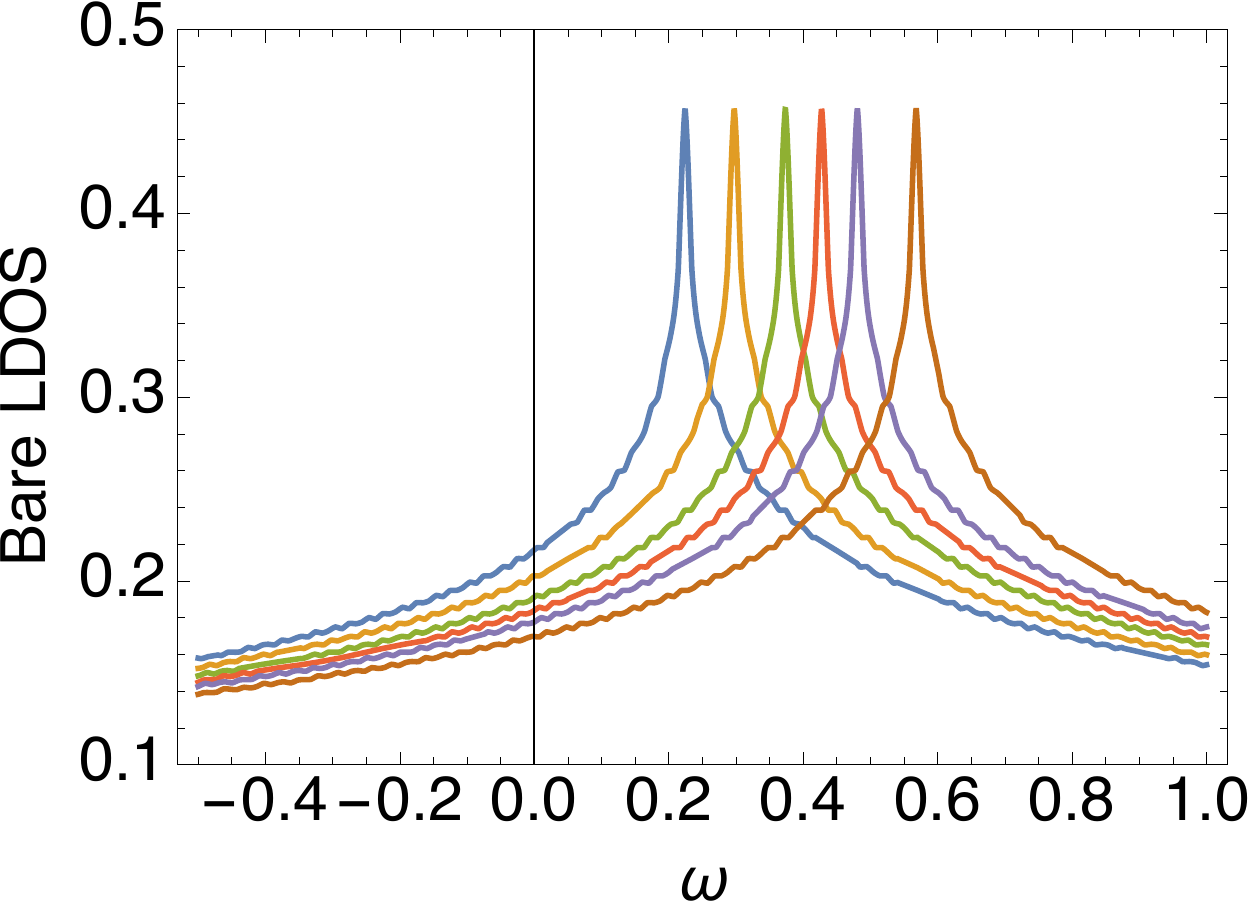}}
\subfigure[\;\; $t'=-0.4$, Tight-binding model for reference, $T=0$]{\includegraphics[width=.49\columnwidth]{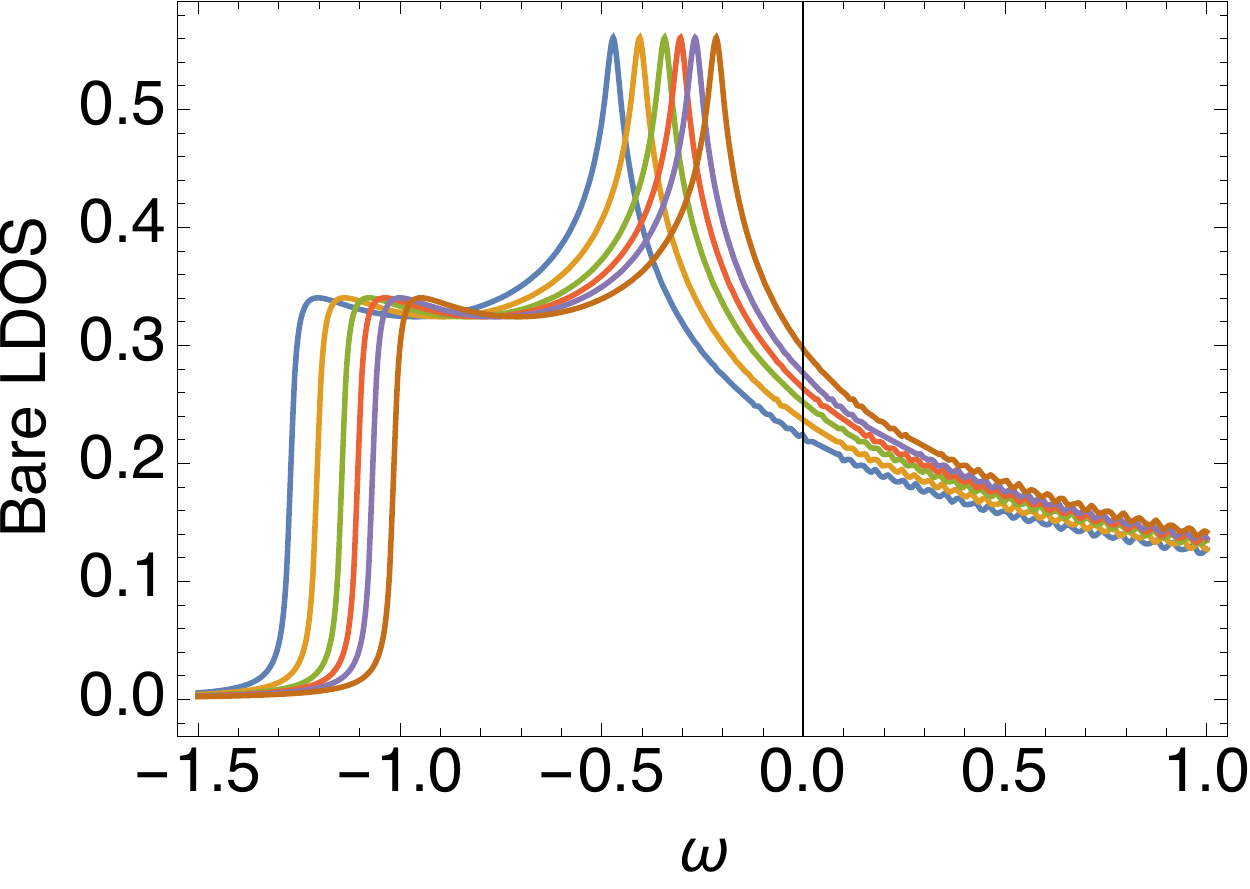}}
\caption{(Color online) Local density of states with varying $\delta$ while fixing $t'=0$ and $-0.4$, at $T=105K$ and $400K$ from ECFL and at $T=0$ from the bare case. All figures share the same legend.}
\label{LDOSvarydelta}
\end{figure}

\subsubsection{Local density of states}

The local density of states (LDOS) is calculated by $\sum_{\vec{k}}(1/N_s)\rho_G(\vec{k},\omega)$ and plotted in \figdisp{LDOSvarytp} and (\ref{LDOSvarydelta}), varying $t'$ with fixed $\delta=0.15$ and varying $\delta$ with fixed $t'=0, -0.4$ respectively. This quantity can be measured by Scanning Tunneling Microscopy \cite{STM1,STM2,STM3,STM4,STM5}.

In \figdisp{LDOSvarytp}, comparing panel (a) and (c), we observe that the LDOS peak gets smoothened and also broadened by the electron-electron interaction. Although the relative position for different $t'$ remains unchanged after turning on interaction, the strong correlation brings them closer by renormalizing the bare band into the effective one, as shown in the inset of \figdisp{dispersionJvar}. From panel (a) to (b), raising temperature tends to have a stronger suppression on the peak with lower $t'$. It means the system with higher $t'$ has a higher Fermi-liquid temperature scale, and therefore it is more robust to heating, which is consistent to the previous findings of the spectral function.
 
In \figdisp{LDOSvarydelta}, from the electron-like panels (a, c, e) to the strongly hole-like panels (b, d, f), the LDOS peaks shifts from $\omega>0$ to $\omega<0$. In contrast to the noninteracting tight binding model in (e) and (f) where the peak height is independent of doping, (a) - (d) have smaller peaks in general and show that the height decreases at smaller doping with more weight in the lower Hubbard band (insets). This is again a feature of strong correlation. As the system approaches the half-filling limit ($\delta\rightarrow0$), the correlation enhances and further suppresses the quasiparticle peak, which contributes to the central peak of LDOS. We also observe that (a) is similar to the density-dependence of the location of Kondo or Abrikosov-Suhl resonance in Anderson impurity problem\cite{Sriram-Edward}. It can be understood as a generic characteristic in strongly correlated matter given the relation between density and the effective interaction.

\subsection{Resistivity} \label{re}
\begin{figure}[!]
\subfigure[\;\; $t'=-0.2$]{\includegraphics[width=.49\columnwidth]{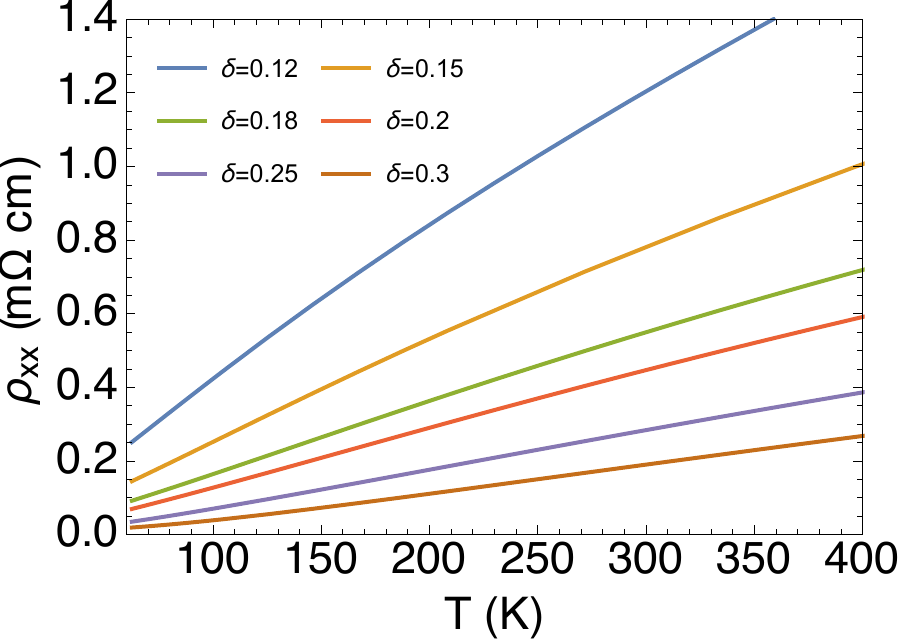}}
\subfigure[\;\; $t'=-0.1$]{\includegraphics[width=.49\columnwidth]{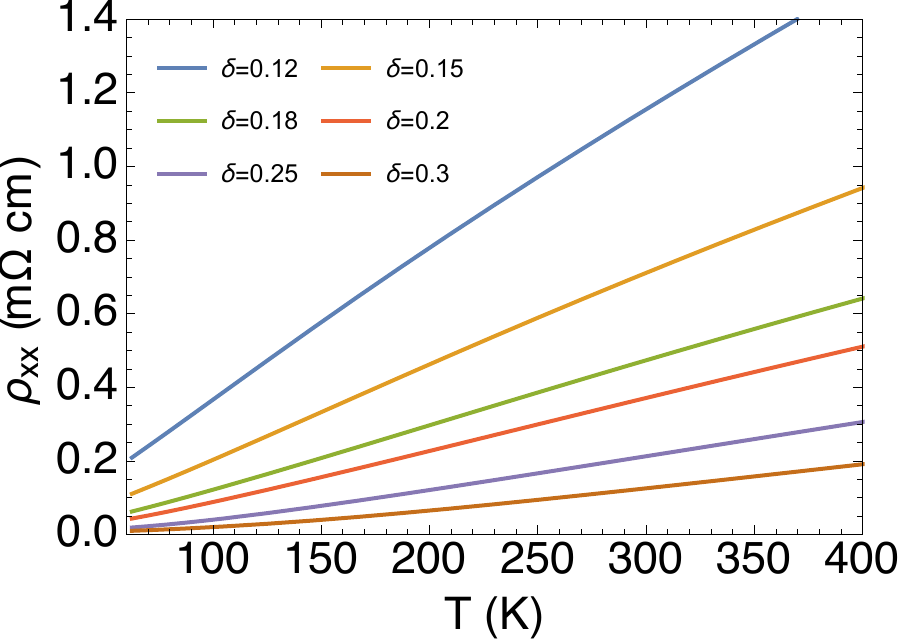}}
\subfigure[\;\; $t'=0$]{\includegraphics[width=.49\columnwidth]{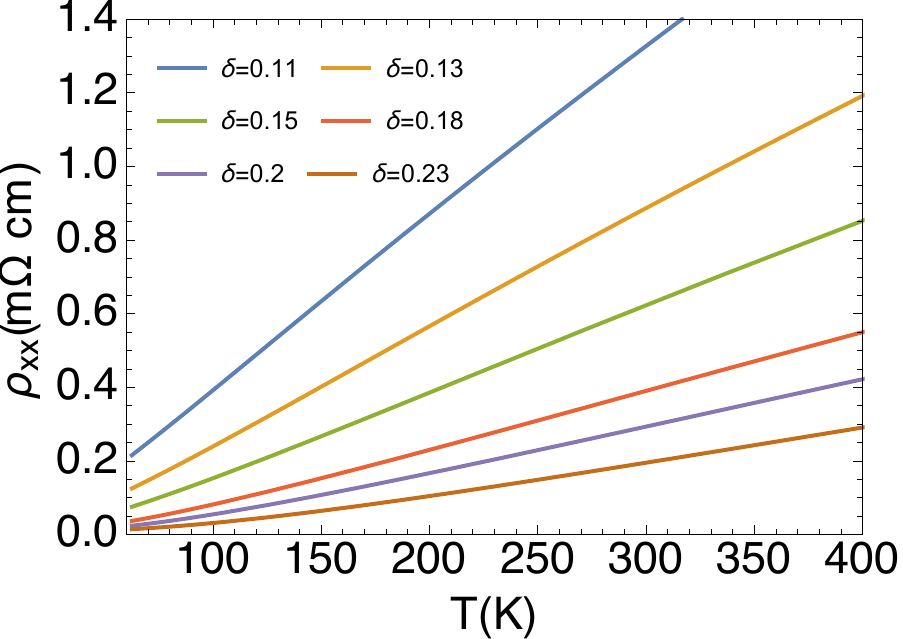}}
\subfigure[\;\; $t'=0.2$]{\includegraphics[width=.49\columnwidth]{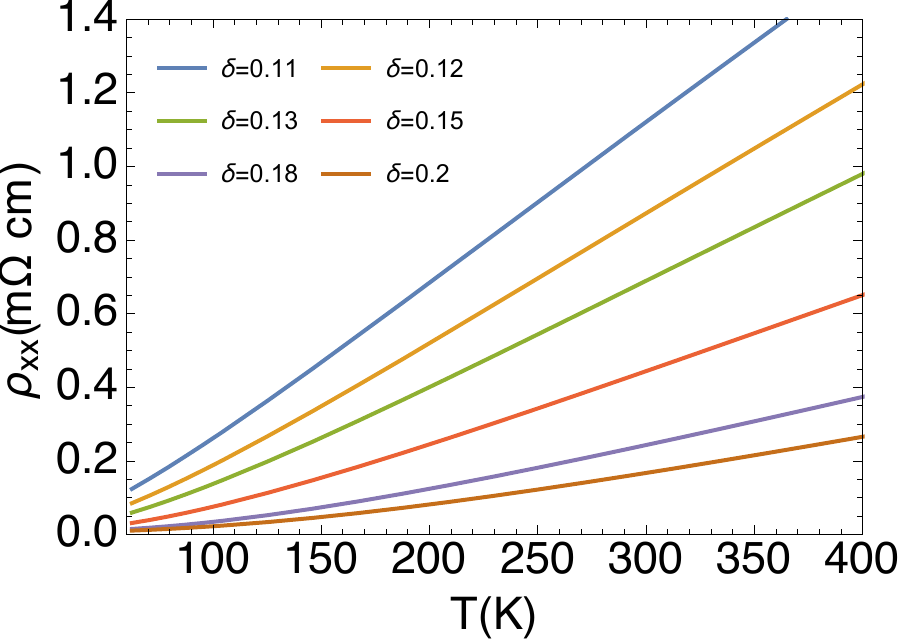}}
\caption{(Color online) Resistivity versus T for varying hole doping $\delta$ and $t'=-0.2, -0.1, 0, 0.2$ (some data in (a) (b) and (d) can be found in \refdisp{SP}).  The curvature tends to change from negative (convex)  to positive (concave) with increasing doping. 
}
\label{resistivity}
\end{figure}

\begin{figure*}[!]
\subfigure[\;\; $t'=-0.2$]{\includegraphics[width=.49\columnwidth]{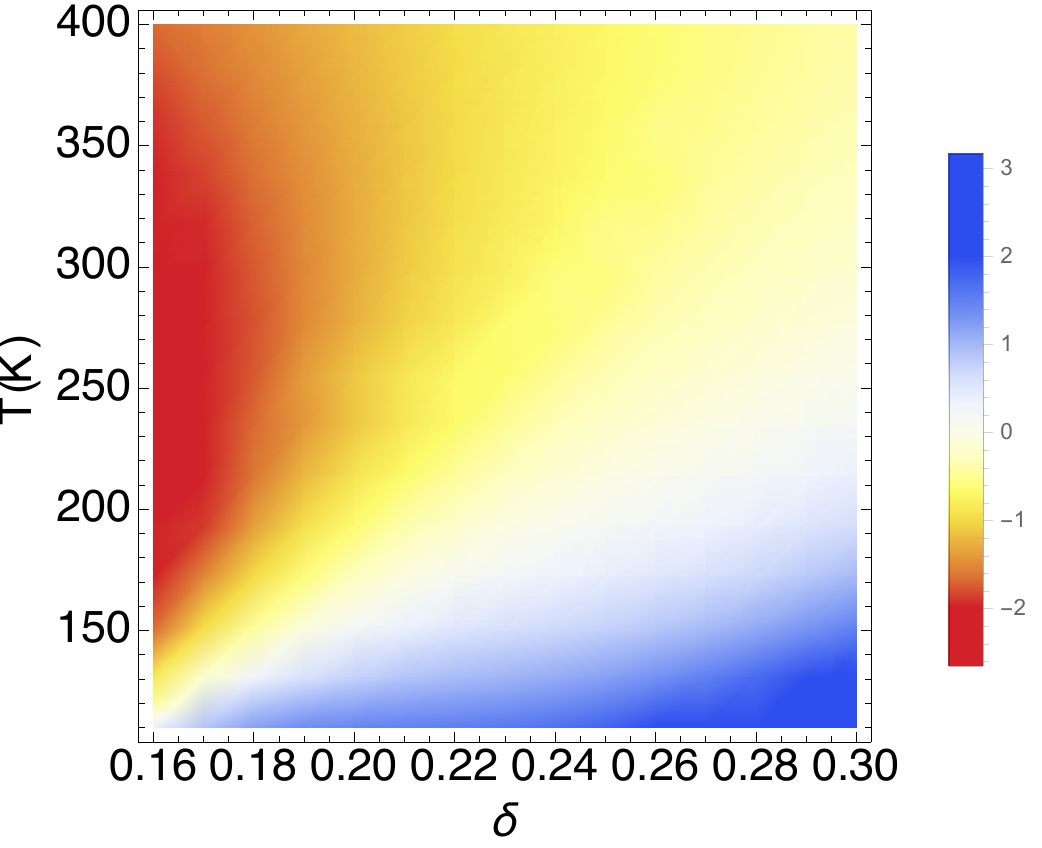}}
\subfigure[\;\; $t'=-0.1$]{\includegraphics[width=.49\columnwidth]{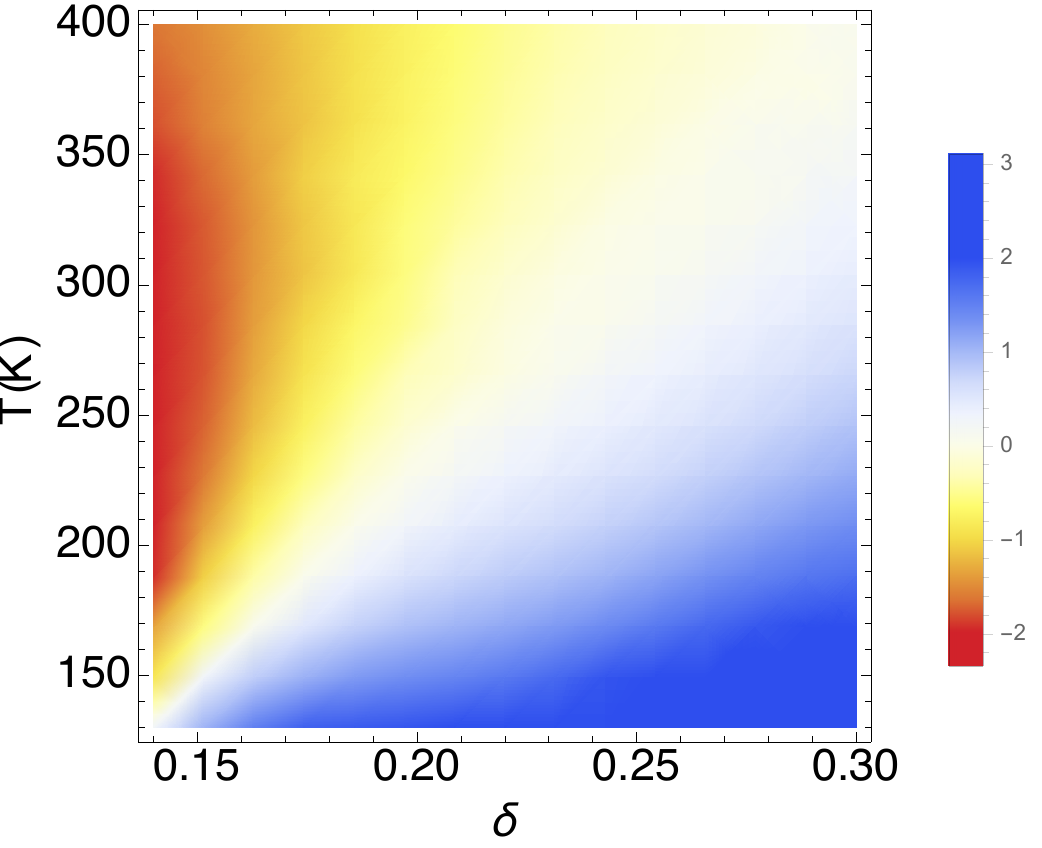}}
\subfigure[\;\; $t'=0$]{\includegraphics[width=.49\columnwidth]{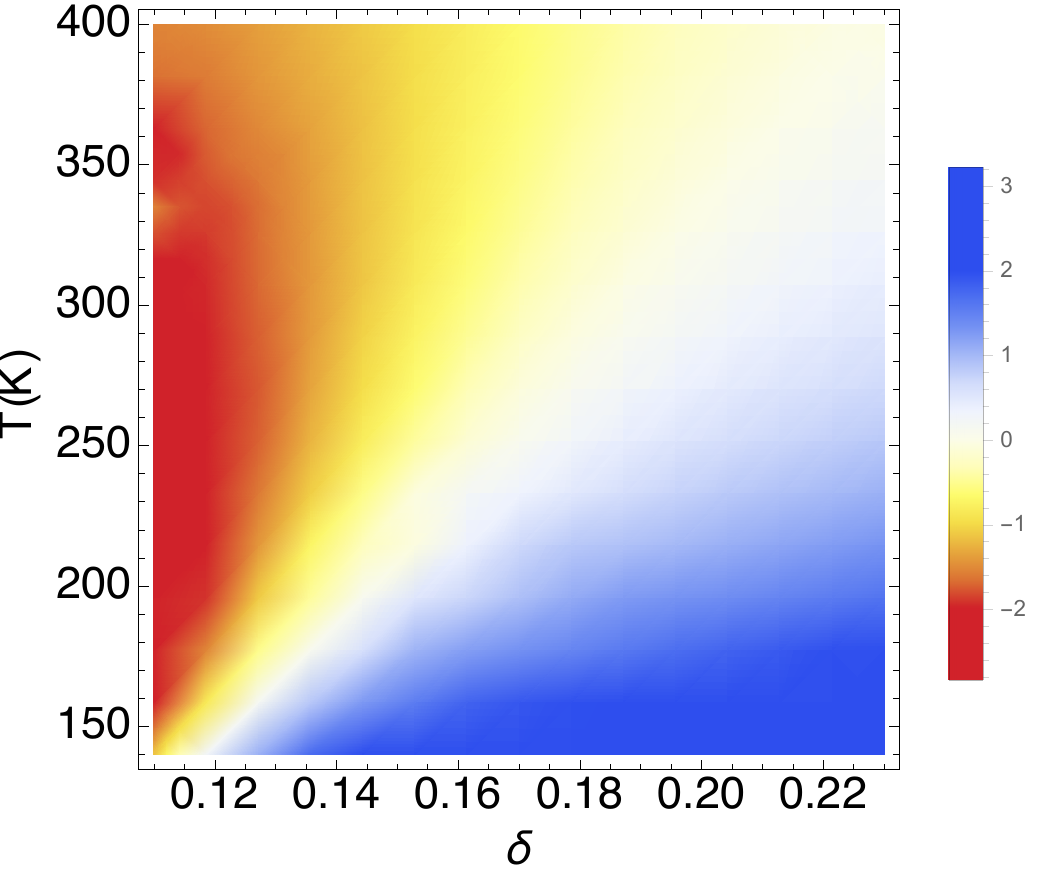}}
\subfigure[\;\; $t'=0.2$]{\includegraphics[width=.49\columnwidth]{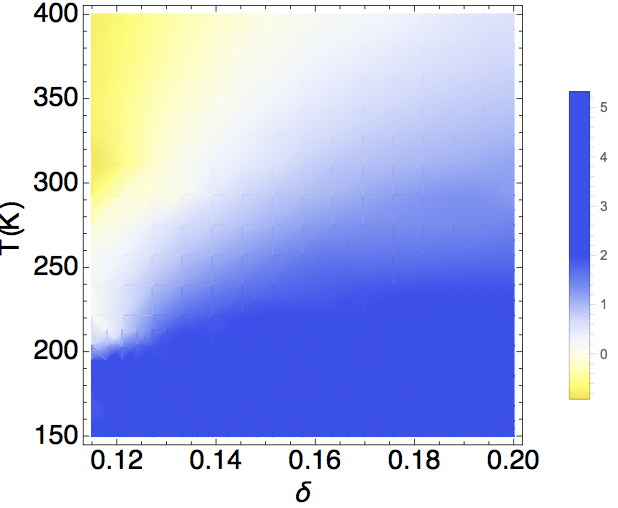}}
\caption{(Color online) Curvature of resistivity versus T for a range of doping $\delta$ and $t'=-0.2, -0.1, 0$ and $0.2$. For most values of  $t'$, there is a blue area towards the right-bottom representing positive (concave) curvature  akin to a Fermi liquid. Towards the left-top we find a  a red area  with negative (convex) curvature resembling a strange (or bad)  metal \cite{WXD}. This trend is consistent with  experimental results\cite{Ando}. }
\label{resistivitycurvature}
\end{figure*}
We next present the resistivity under strong electron-electron interaction. The popular bubble approximation is used and the current correlator is writen as $\langle J(t) J(0) \rangle \sim  \sum_k v_k^2 \G^2(k)$. Here the velocity $\hbar v_k^\alpha=  \frac{\partial \varepsilon_k}{ \partial k_\alpha}$ represents the bare current vertex. In tight binding theory the sign oscillation  in $v_k^\alpha$  leads to a reduction in the  average over the Brillouin zone and therefore diminishes the magnitude of the vertex corrections. Also the weak $k$-dependence of self-energy in \figdisp{rhosigmaEDC} reduces the importance of vertex corrections. 

In our picture of quasi-two dimensional  metal,  there are   2-d  well separated sheets,  by a distance $c_0$ in the c direction. Thus each sheet can be effectively characterized by the 2-d \tJ model. Its DC resistivity $\rho_{xx}$ can be written as follows: 
\beq
 \rho_{xx} = \rho_0 \times \bar{\rho}_{xx} = \frac{\rho_0}{\bar{\sigma}_{xx}}, \label{parallel}
 \eeq
 \beq
  \bar{\sigma}_{xx} = (2 \pi)^2 \int_{-\infty}^\infty d\omega \, (-\frac{\partial f}{\partial \omega}) \; \langle \rho^2_G(\vec{k},\omega) \frac{(\hbar v_k^x)^2}{a^2_0} \rangle_k, \label{parallel}
 \eeq
where $ \bar{\rho}_{xx}$ and $\bar{\sigma}_{xx}$ represents dimensionless resistivity and conductivity respectively; $\rho_0\equiv c_0 h/e^2 $ ($\sim1.718$m$\Omega$ cm) serves as the scale of resistivity; $\langle A \rangle_k \equiv  \frac{1}{N_s}\sum_{\vec{k}} A(\vec{k}) $;  $f
$ is the Fermi distribution function. We present our results in absolute units in \figdisp{resistivity} by putting the measured values of lattice constant into the formula and converting the energy unit using $t=0.45$ ev $\approx 5220K$. The scale of ECFL resistivity is consistent with the experimental findings in cuprates\cite{Ando}.

In our previous study \cite{SP}, a significant finding was that  the curvature of resistivity changes when $t'$ varies. Here we focus more on the $\delta$-dependent behavior of resistivity as shown in \figdisp{resistivity}.  For a given $t'$, decreasing the hole doping changes the curves from concave to linear then to convex and varying $t'$ shifts the crossover doping region. This phenomenon signals a change of the effective Fermi temperature scale. In higher hole doping (lower electron density), there is   less influence of the  Gutzwiller projection. Hence the system has less correlation effectively and displays more Fermi-liquid-like behavior, namely, $T^2$-dependence, and hence positive curvature. In the case with low hole doping, i.e. closer to the Mott insulating limit, the correlation is relatively stronger and suppresses the Fermi liquid state into a much lower temperature region, which is usually masked by superconductivity. In the displayed temperature range of \figdisp{resistivity}, the system shows strange metal or even bad metal behaviors\cite{WXD} instead, and hence negative curvature. The curvature can be explicitly calculated as the second derivative of $\rho_{xx}$ with respect to $T$ shown in \figdisp{resistivitycurvature}, which displays features qualitatively similar to the experiments\cite{Ando,Sam-Martin,Takagi,NCCO-2,Greven}.

To explore the crossover from the Fermi liquid ($\rho_{xx}\propto T^2$) at low $T$ to the strange metal ($\rho_{xx}\propto T$) at higher $T$, we define a simple fitting model:
\beq
 \rho_{approx} = const\times \frac{T^2}{T_{FL}+T}. \label{Rmodel}
\eeq
This fit gives  Fermi liquid behavior for $T\lesssim T_{FL}$ and then crosses over to strange metal linear behavior at $T\gtrsim T_{FL}$. Thus, $T_{FL}$ serves as a crossover scale describing the boundary of Fermi liquid region as well as estimating the strength of correlation. We find our data fits into this model well up to intermediate temperature with fitted coefficient and $T_{FL}$. 

Table.~\ref{TFL} shows the value of $T_{FL}$ in various sets of $\delta$ and $t'$. In all cases, the $T_{FL}$ is considerably smaller than the Fermi temperature in non-interacting case at the order of bandwidth, as a result of strong correlation. In experiment, a small enough $T_{FL}$ prevents the observation of Fermi liquid because at low enough temperature the superconducting state shows up instead\cite{Ando}. Relatively, $T_{FL}$ is further suppressed by smaller second neighbor hopping $t'$ or smaller doping $\delta$, either of which strengthens the effective correlation. Negative $t'$  increases the resistivity and shrinks the temperature region for Fermi liquid. 
\begin{table}[h!]
\begin{centering}
\begin{tabular}{ |c |c|c|c|c|c|}\hline 
\multicolumn{6}{|c|}{Fermi liquid temperature $T_{FL}~(K)$}\\  \hline
 $\downarrow\delta$ \; $\rightarrow t'$&$-0.2$& $-0.1$& $0$& $0.1$& $0.2$\\  \hline 
 0.12 &10.0&18.4&33.1&68.2&117.6\\ 
 0.15 &15.8&31.1&66.3&135.4&218.0\\
 0.18 &24.4&53.7&117.4&245.2&420.9\\ 
 0.21 &37.3&78.8&189.5&360.3&618.4\\
 0.24 &56.8&145.2&274.4&569.5&820.5\\\hline 
\end{tabular}
\caption{The Fermi liquid temperature $T_{FL}$ obtained from fitting the data with Eq. \ref{Rmodel}. Increasing either $t'$ (horizontally) or doping $\delta$ (vertically) increases $T_{FL}$, signaling weaker correlations.}
\label{TFL}
\end{centering}
\end{table}
In this sense, decreasing $t'$ turns up the effective correlation by depressing the hopping process. On the other hand, decreasing doping leaves less space for electron movement, which also effectively increases the correlation and suppresses $T_{FL}$. $\delta$ and $t'$ both control the effective correlation strength and hence $T_{FL}$, as shown in Table.~\ref{TFL}. Their similar role can also be understood in the fact that they both change the geometry of the Fermi surface which determines the conductivity at $T\ll W$, where $W=8t$ is the bare bandwidth. In general, either increasing $\delta$ with fixed $t'$ or increasing $t'$ with fixed $\delta$ changes the Fermi surface from hole-like to electron-like.

\subsection{Hall number}

\begin{figure}[h!]
\subfigure[\;\; $t'=-0.4$]{\includegraphics[width=.49\columnwidth]{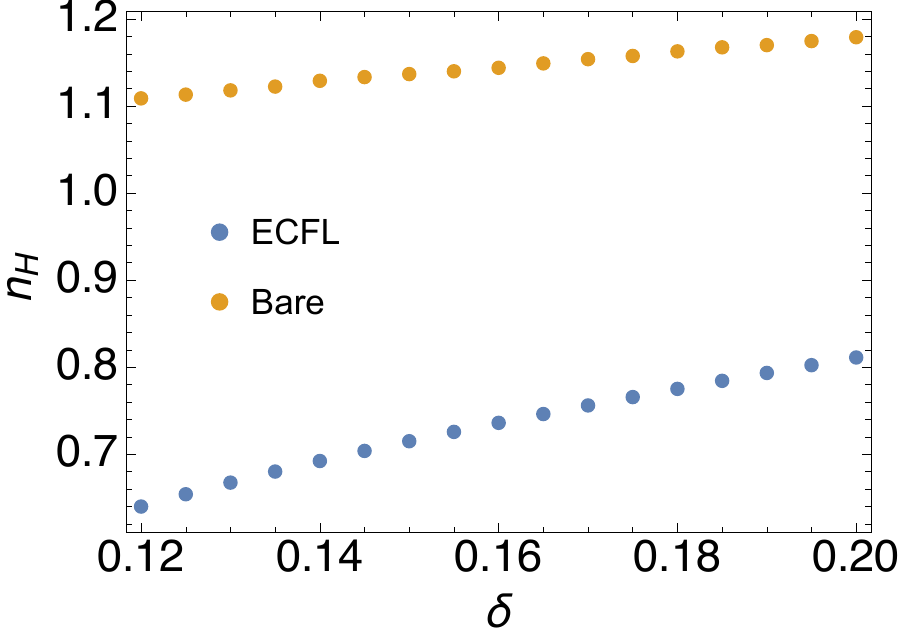}}
\subfigure[\;\; $t'=-0.3$]{\includegraphics[width=.49\columnwidth]{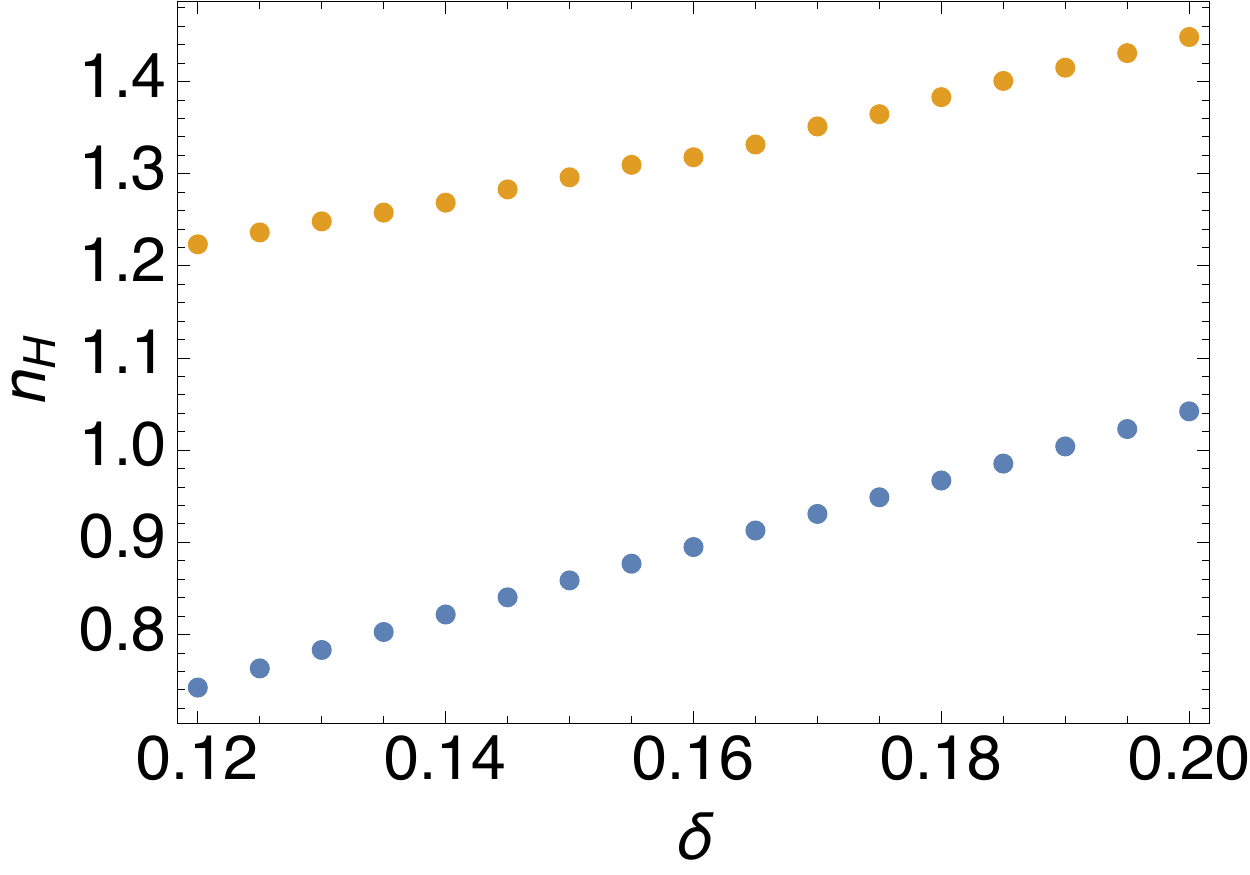}}
\subfigure[\;\; $t'=-0.25$]{\includegraphics[width=.49\columnwidth]{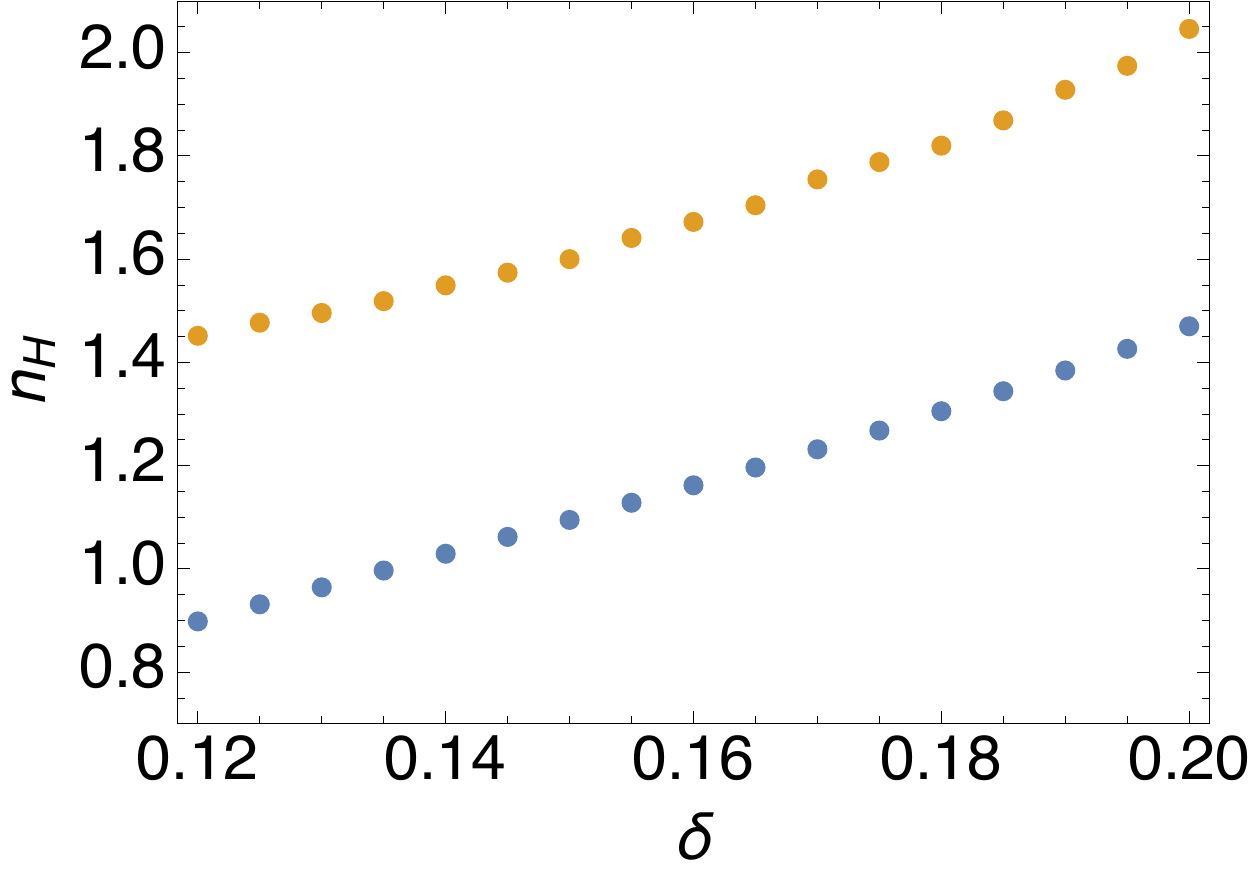}}
\subfigure[\;\; $t'=-0.2$]{\includegraphics[width=.49\columnwidth]{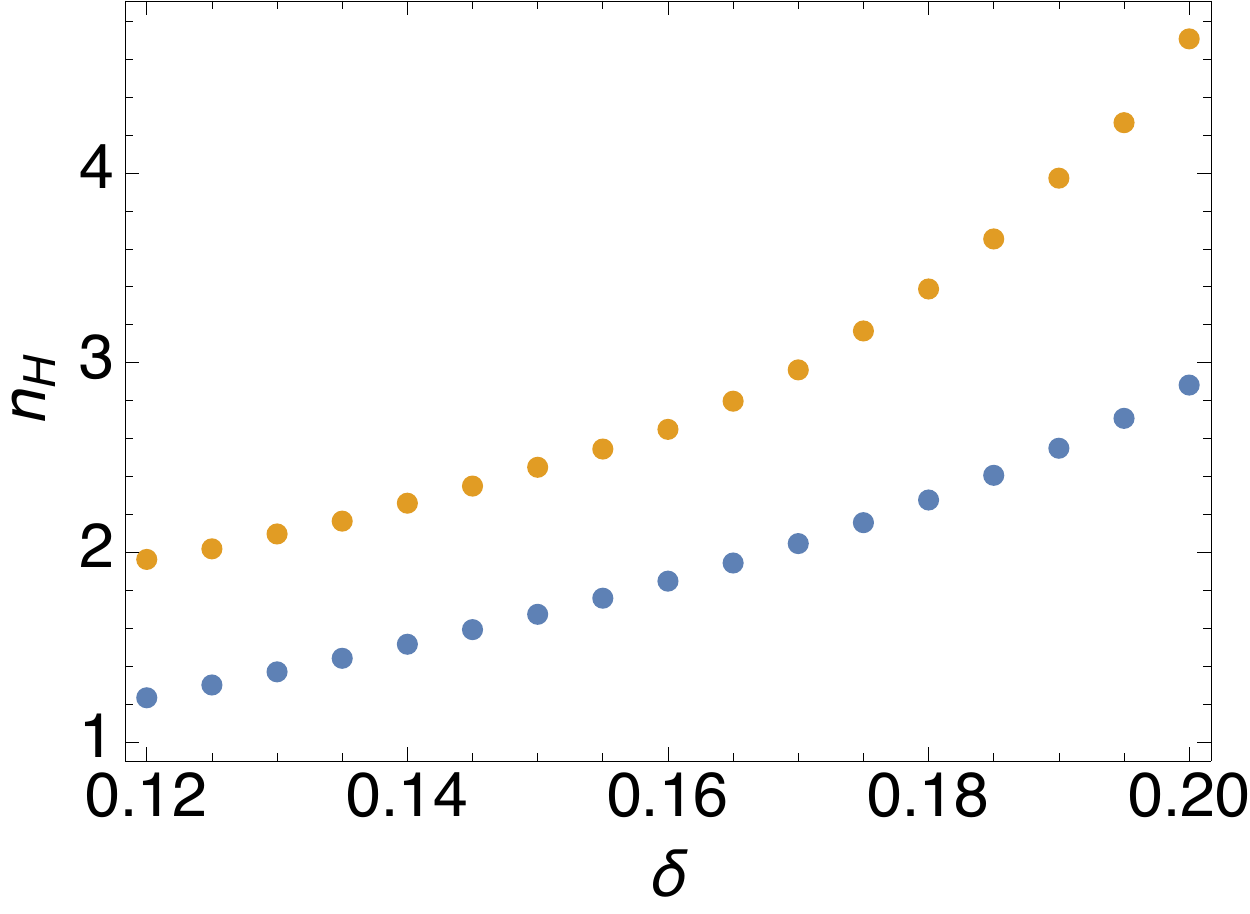}}
\caption{(Color online) Hall number vs doping at different $t'$ and $T=105K$, where $t'$ controls the scale of $n_H$. }
\label{Hallnumber}
\end{figure}

Within the  bubble scheme,  we also calculate the Hall conductivity\cite{voruganti,hall-extra,Tremblay,HFL} as $ \sigma_{xy}= - 2 \pi^2/{\rho_0} \times (\frac{\Phi}{\Phi_0}) \times \; \bar{\sigma}_{xy} $.   The dimensionless conductivity can be written as:
  \beq 
 \bar{\sigma}_{xy}&=& \frac{4 \pi^2 }{3 }   \int_{-\infty}^\infty d\omega \, (- {\partial f}/{\partial \omega}) \langle \rho^3_G(k,\omega) \eta(k) \rangle_k, \;\;\;\; \label{hall1}
 \eeq
 where $\eta(k)= \frac{ \hbar^2}{a_0^4} \{ (v_k^x)^2  \frac{\partial^2 \varepsilon_k}{\partial k_y^2}- (v_k^x v_k^y) \frac{\partial^2 \varepsilon_k}{\partial k_x \partial k_y} \}$; $\Phi=B a_0^2$ is  the  flux\cite{lattice}, and $\Phi_0= hc/(2 |e|)$ is the flux  quantum. In these terms, we can compute the Hall number as
 \beq
 n_H=-\frac{1}{4\pi^2} \; \frac{\bar{\sigma}_{xx}^2}{\bar{\sigma}_{xy}}. \label{number-Hall}
 \eeq
Note that in this definition, the sign of the Hall number is opposite to that in \refdisp{SP}. In this definition, $n_H$ shares the same sign with Hall coefficient $R_H$, consistent to the experimental convention\cite{Takagi,Hwang,Ando,NCCO-2,Sam-Martin,Ando-Hall,Greven,Boebinger1,Boebinger2,NCCO-Hall}. We present the ECFL Hall number $n_H$ in \figdisp{Hallnumber} together with the non-interacting one $n_{H0}$ for comparison. In all cases of different $t'$, $n_H$ is around 60$\%$ of $n_{H0}$ and decreasing $t'$ suppresses the scale of $n_H$. It indicates the reduction of effective charge carrier due to strong correlation. Therefore, the Hall number increases when the effective correlation turns down either by increasing $t'$ or increasing $\delta$, as shown in \figdisp{Hallnumber}. In Panel d, $n_{H}$ remains smooth when crossing the Lifshitz transition $\delta\approx0.17$, where the Fermi surface changes from opened to closed as presented in Section. \ref{spectralproperties}, while $n_{H0}$ shows a crossover to a steeper region. 

\subsection{Spin susceptibility and the NMR relaxation rate}
\begin{figure}[ht]
\subfigure[\;\; t'=-0.2]{\includegraphics[width=.49\columnwidth]{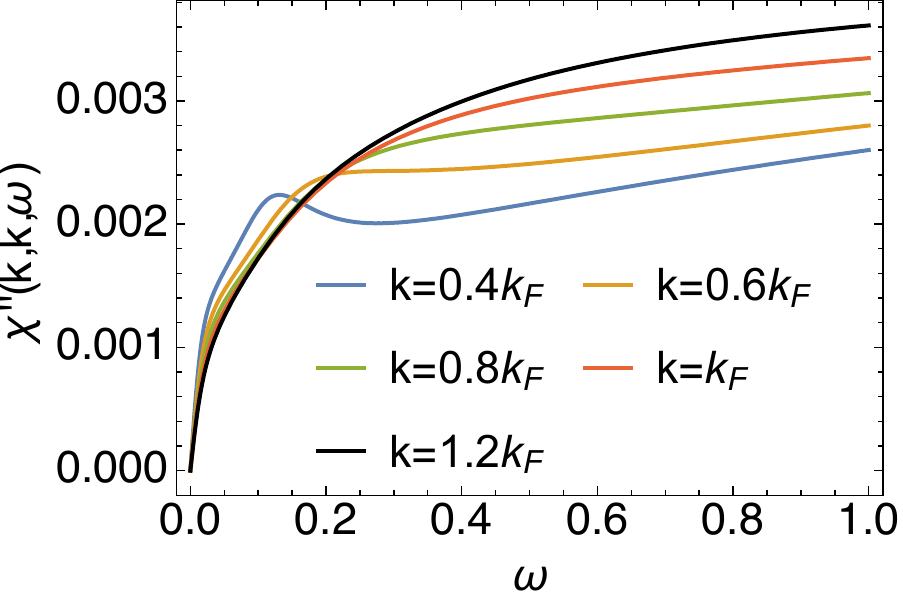}}
\subfigure[\;\; t'=0.2]{\includegraphics[width=.49\columnwidth]{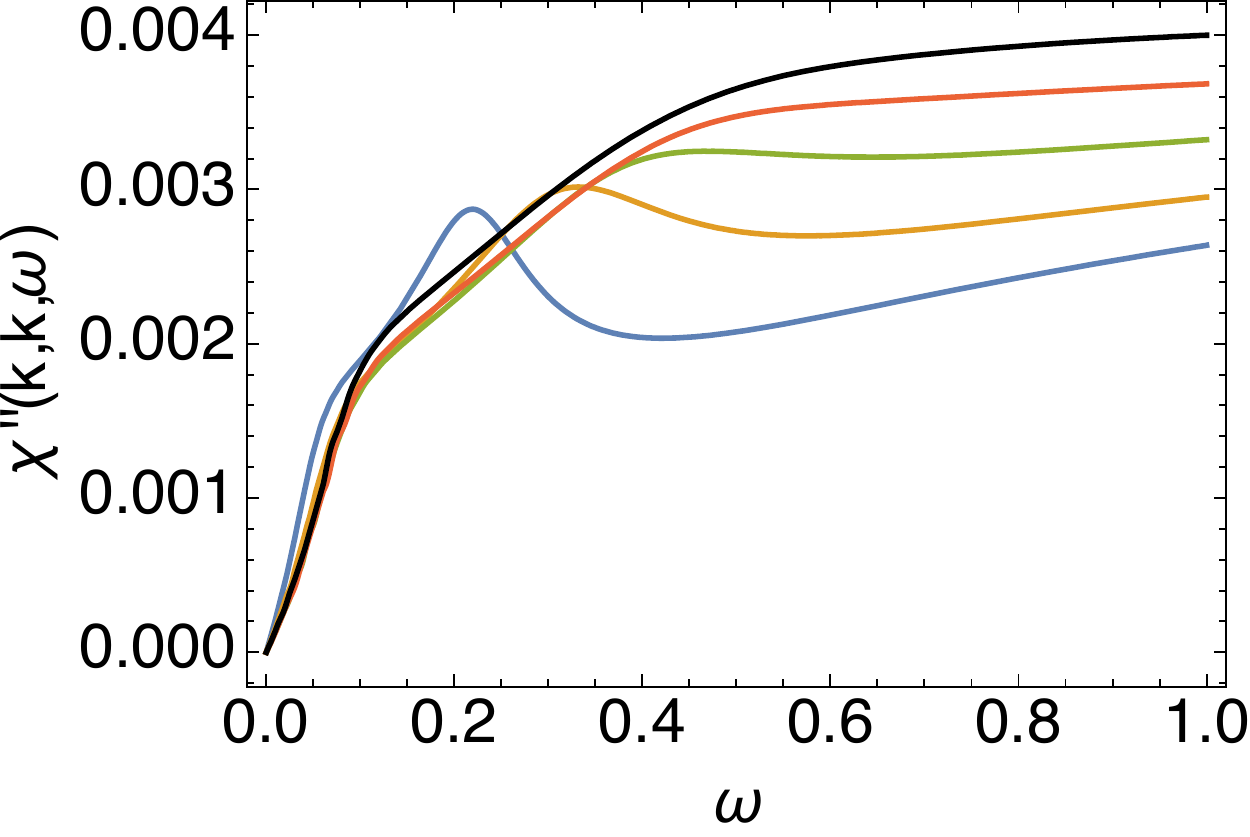}}
\caption{$\chi''$ at different $k$ for $\delta=0.15$, $T=63K$ and $t'=\pm 0.2$.}\label{chipp}
\end{figure}

The imaginary part of spin susceptibility can also be calculated in the Bubble approximation:
\beq
\begin{split}
\chi''(k,\omega)=&\int_{-\infty}^\infty dy \langle \rho_G(p,y) \rho_G(p+k,y+\omega) \rangle_p \\&(f(y)-f(y+\omega))
\end{split}
\eeq
while the real part $\chi'$ can be obtained from calculating the Hilbert transform of $\chi''$. $\chi''$ is shown in \figdisp{chipp} for hole-doped ($t'=-0.2$) and electron-doped ($t'=0.2$) cases at various fixed $k$. In both cases, we see the quasi-elastic peaks in the occupied region for small $k$ which disappears gradually as $k$ increases.

\begin{figure}[ht]
\subfigure[\;\; $t'=-0.2$]{\includegraphics[width=.49\columnwidth]{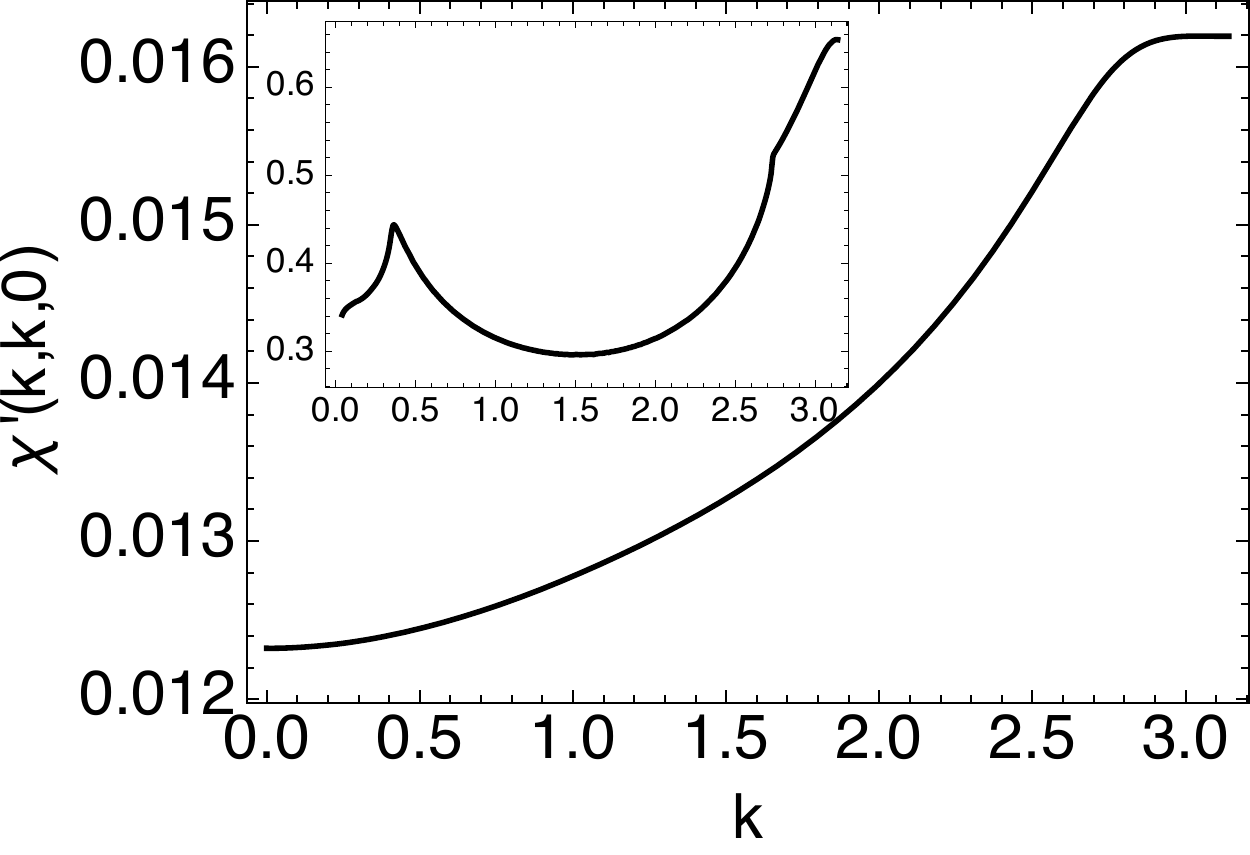}}
\subfigure[\;\; $t'=0.2$]{\includegraphics[width=.49\columnwidth]{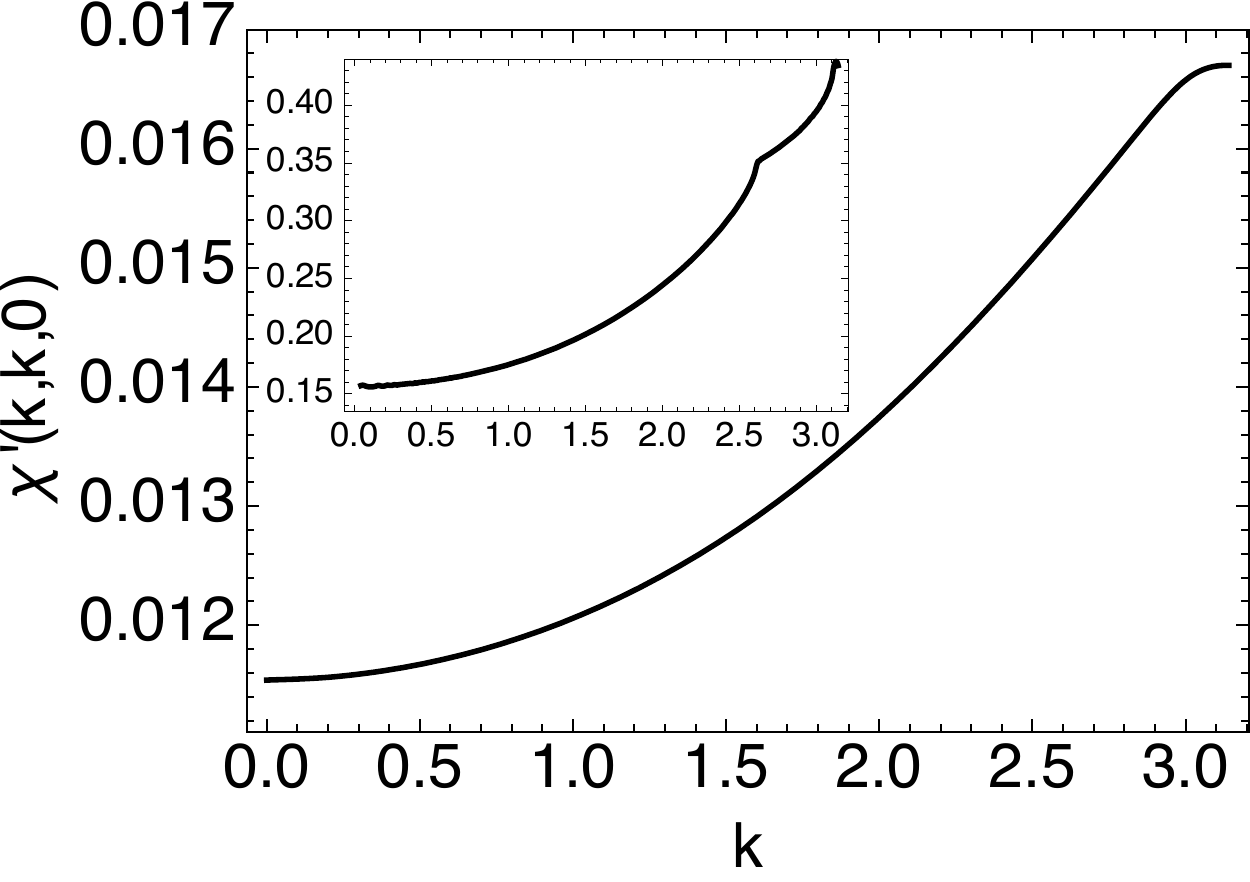}}
\caption{$\chi'$ at $\omega=0$ for $\delta=0.15$, $T=63K$ and $t'=\pm 0.2$. Inset shows the corresponding non-interacting $\chi_0'$. $\chi'$ is largely suppressed from the bare case due to strong interaction.}\label{chip}
\end{figure}

\figdisp{chip} presents the k-dependent $\chi'$ at zero frequency, in comparison with the non-interacting $\chi'_0$ in the inset. We observe that $\chi'$ is much smaller than $\chi'_0$ due to the broadening in the spectral function as a result of strong interaction. Despite the scale difference, the $k$-dependent $\chi$ seems closer to $\chi_0$ in the electron-doped case ($t'=0.2$) than the hold-doped case ($t'=-0.2$), consistent to the previous discussion that the system is more Fermi-liquid-like for positive $t'$. 
The knight shift $\chi'(k=0,\omega=0)$ of the system is almost independent of temperature and therefore not shown specifically in figure.

\begin{figure}[ht]
\subfigure[\;\; $\delta=0.15$, varying $t'$]{\includegraphics[width=.49\columnwidth]{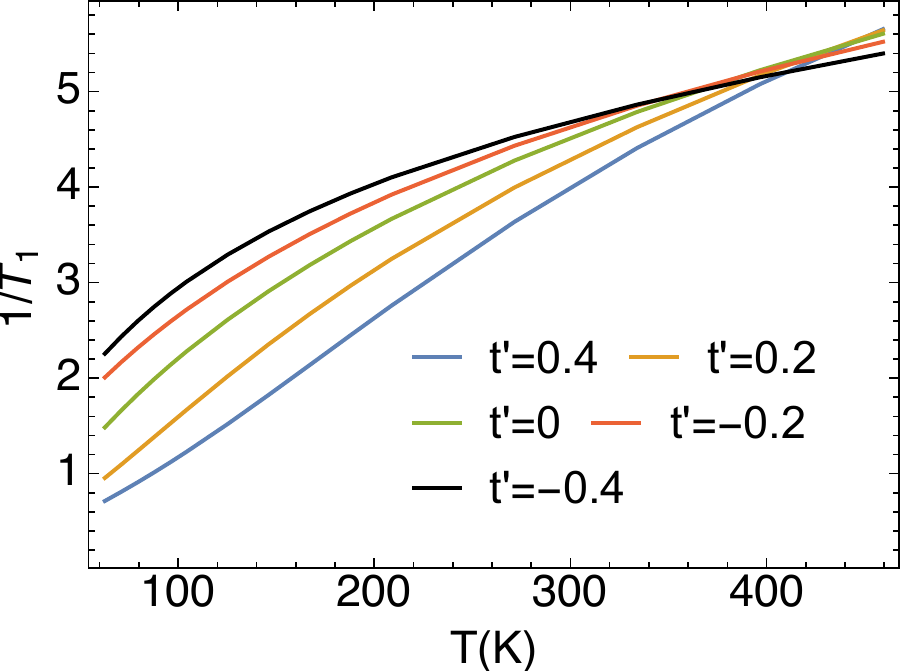}}
\caption{Relaxation rate from \disp{T1} (arb units) at $\delta=0.15$ and different $t'$. The curve becomes more sublinear as $t'$ decreases from positive to negative. The sub-linear curve at $t'=-0.2$ looks similar to the Copper relaxation rate in \refdisp{Walstedt}. }\label{relaxationrate}
\end{figure}

 The relaxation rates for cuprates are given by\cite{Walstedt-Book,Walstedt,spins}
\beq
\frac{1}{T_1}= \frac{\gamma^2 k_BT}{\mu_B^2} \sum_q A_q^2 \frac{\chi''(q,\omega_0)}{\omega_0} \label{T1}
\eeq
where $A_q$ is a  form factor  that is determined by the local geometry of the nucleus\cite{Walstedt-Book,Walstedt,spins}, and  $\omega_0$ is nuclear frequency which is assumed to be very small. Our scheme of calculation  is not yet refined enough to capture the detailed difference between the Copper and Oxygen relaxation rates in cuprates. Hence, we will content ourselves by presenting the case with $A_q=1$, which should correspond to the inelastic neutron scattering (INS) derived relaxation rate in \refdisp{Walstedt} from Walstedt {\em et. al.}. We plot $1/T_1$ vs $T$ at $\delta=0.15$ and various $t'$ in \figdisp{relaxationrate}. For $t'=-0.2$, $1/T_1$ increases sub-linearly with temperature.   It  shows roughly the same trend as the Copper rates shown in \refdisp{Walstedt}, but  is somewhat steeper than the derived INS rate therein. 


\subsection{$J$ variation}

\begin{figure}[!]
\subfigure[\;\; $J=0$, EDC]{\includegraphics[width=.44\columnwidth]{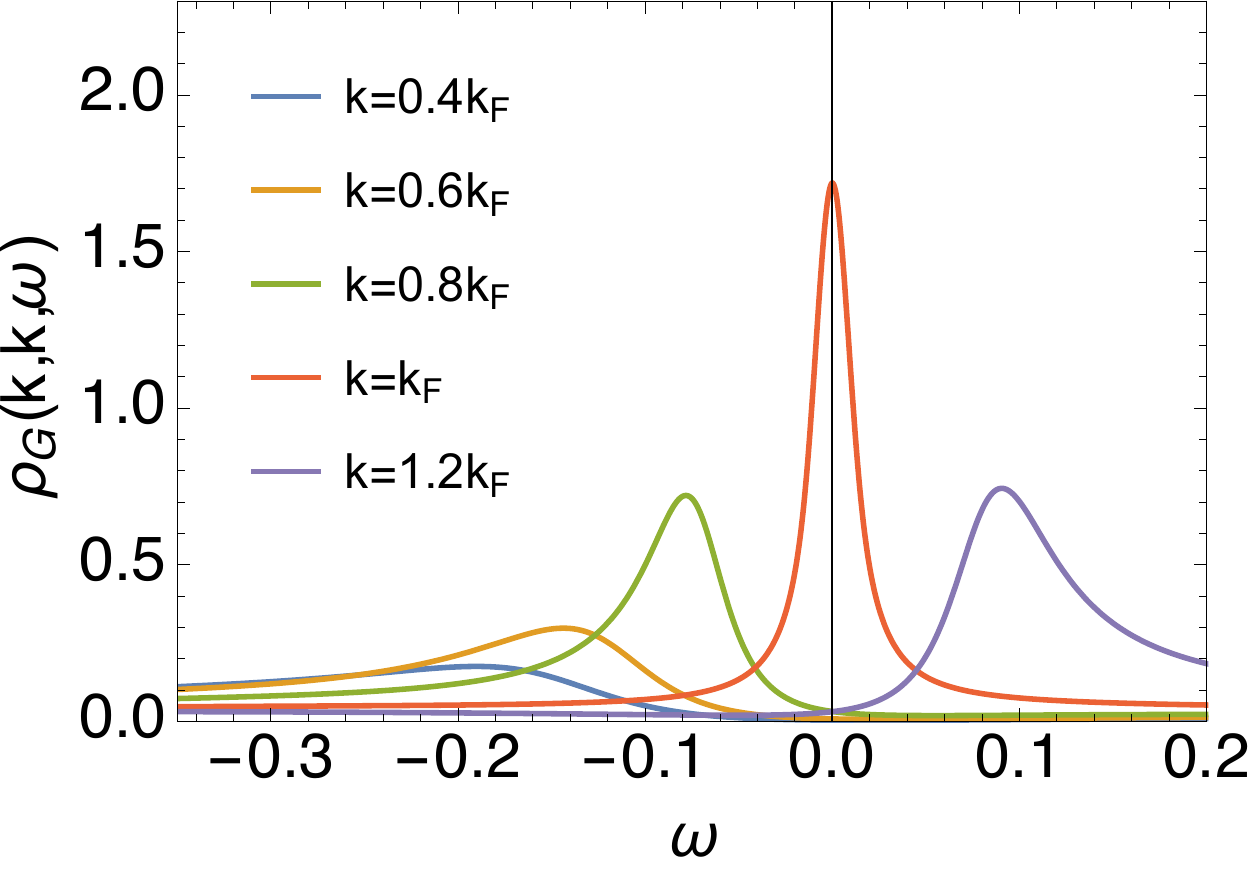}}
\subfigure[\;\; $J=0$, MDC]{\includegraphics[width=.44\columnwidth]{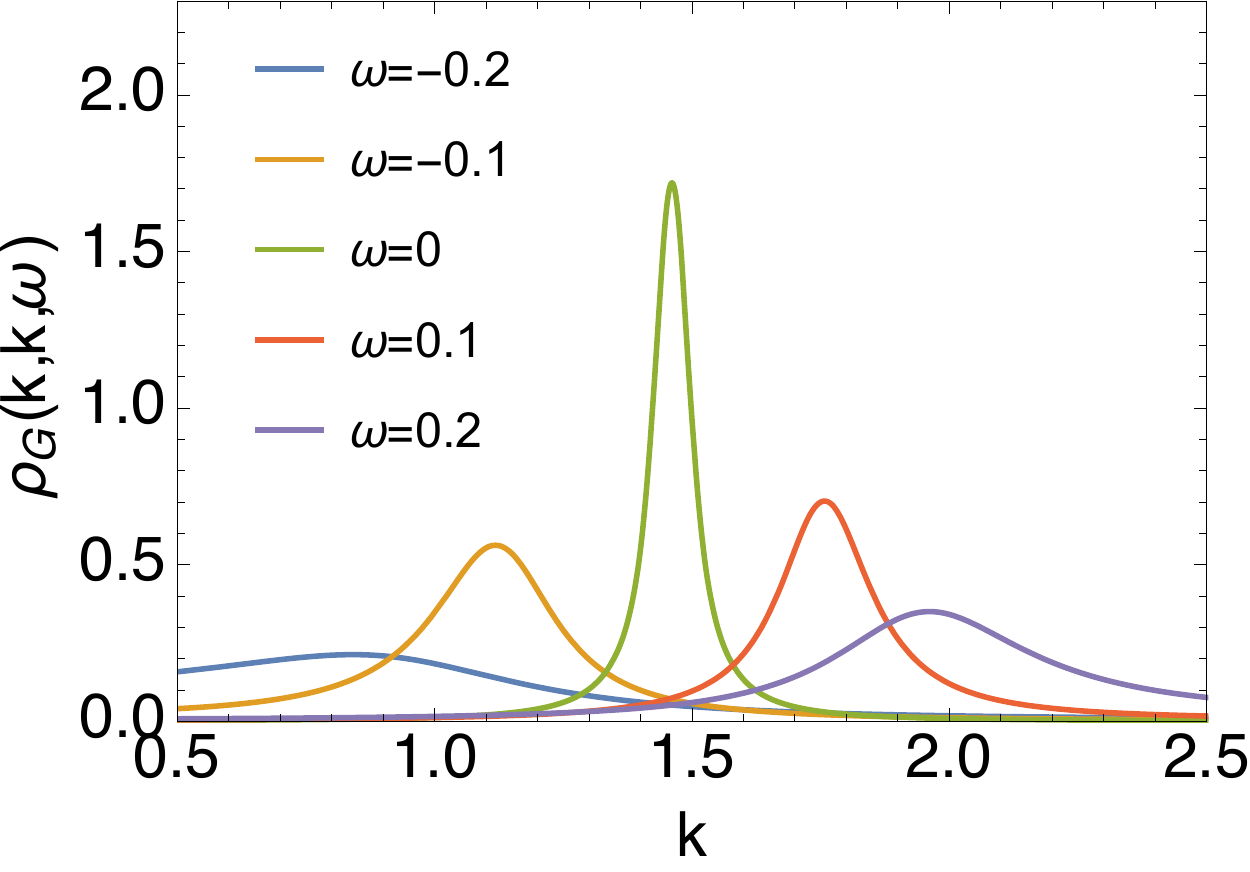}}
\subfigure[\;\; $J=0.17$, EDC]{\includegraphics[width=.44\columnwidth]{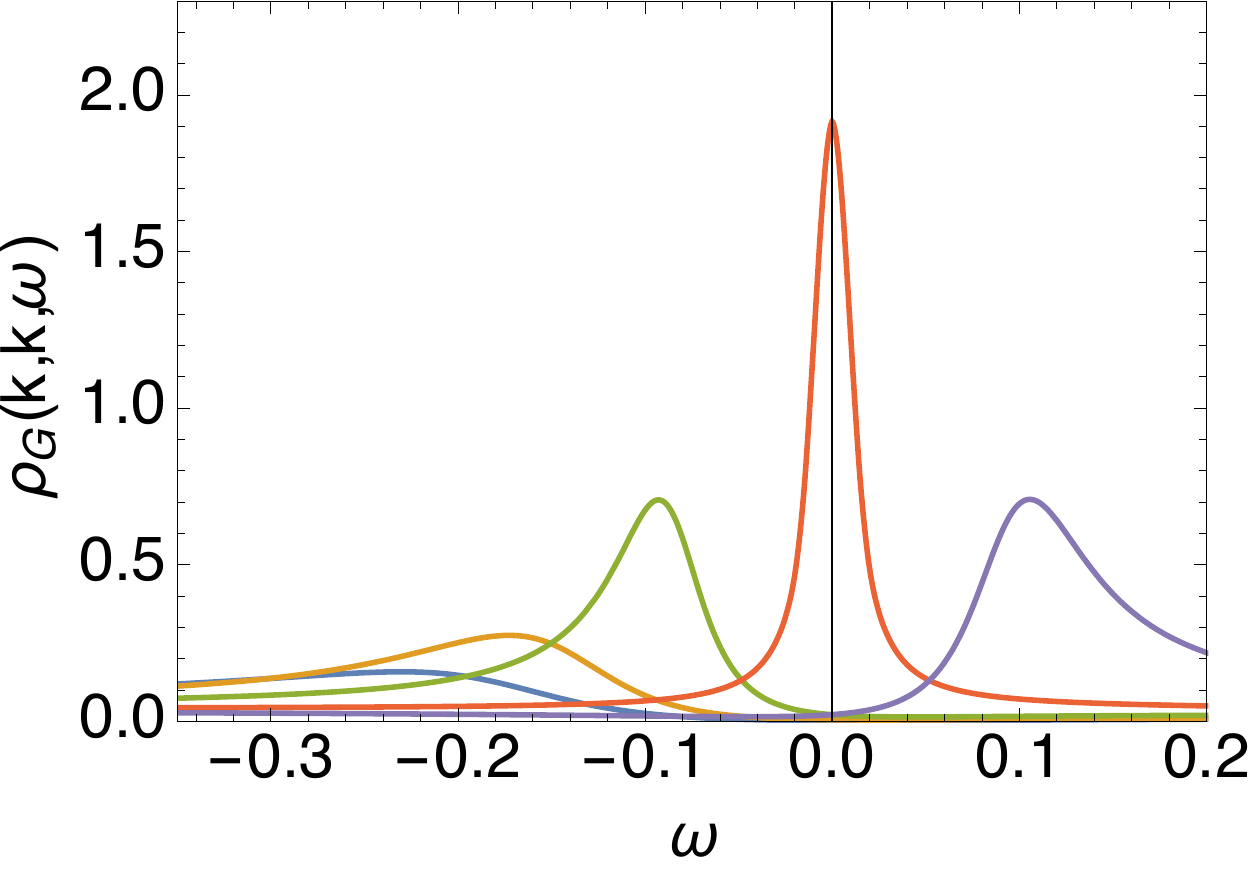}}
\subfigure[\;\; $J=0.17$, MDC]{\includegraphics[width=.44\columnwidth]{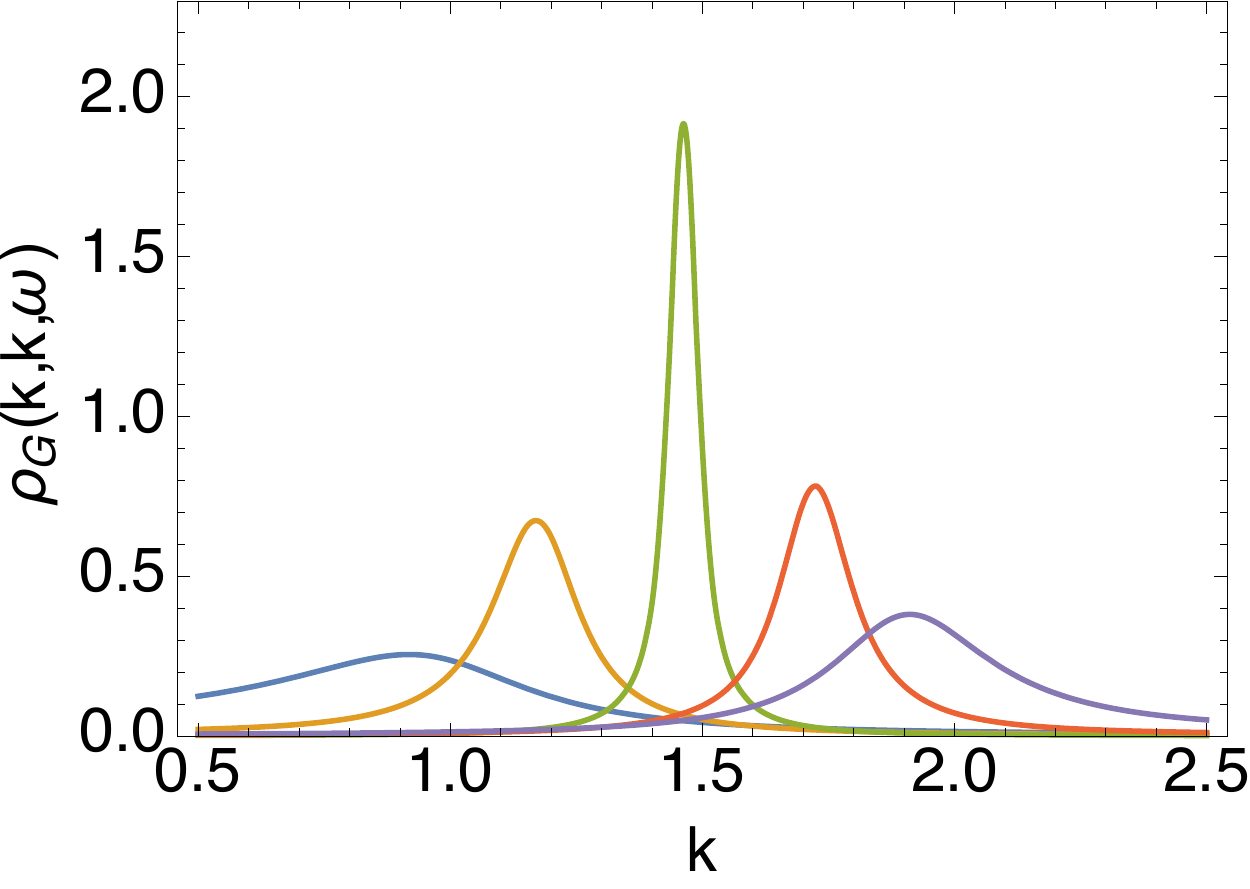}}
\subfigure[\;\; $J=0.4$, EDC]{\includegraphics[width=.44\columnwidth]{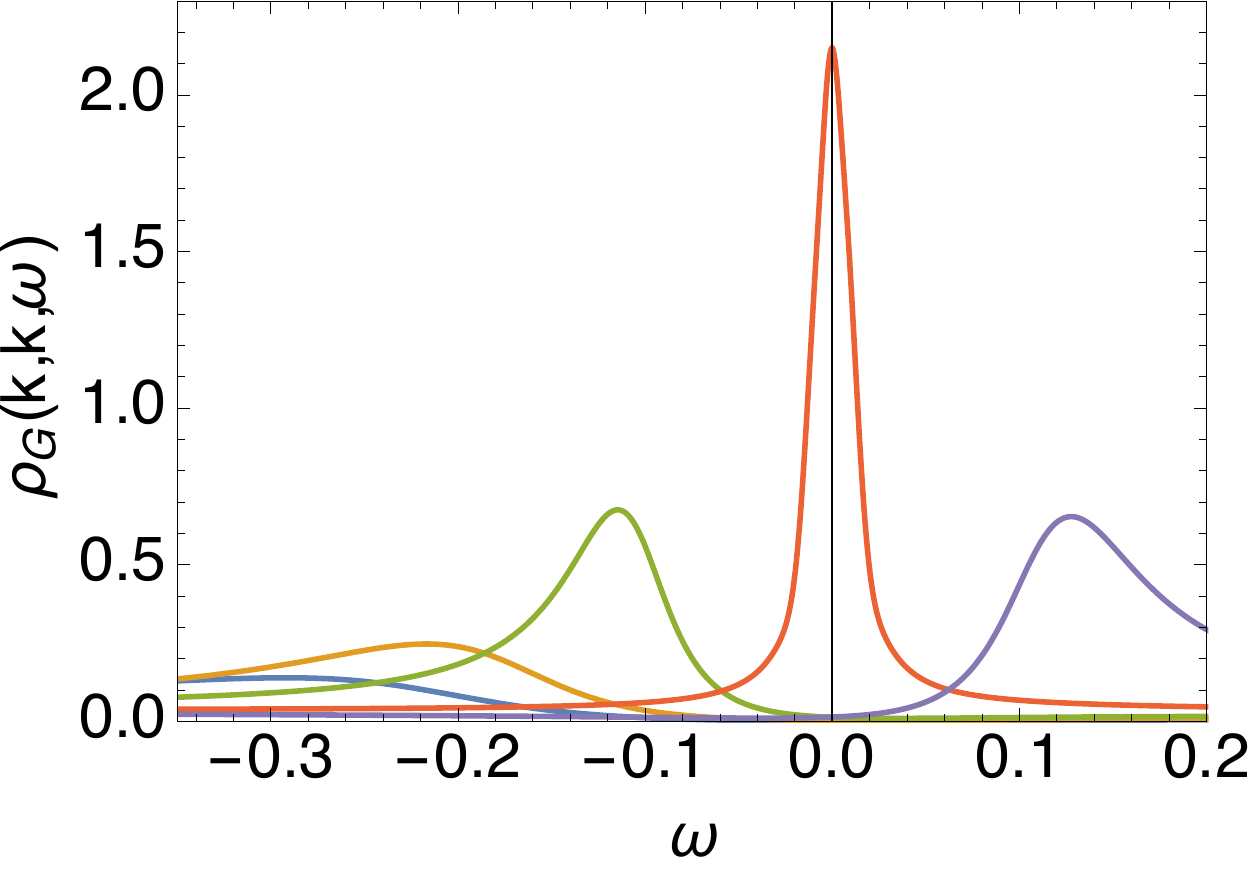}}
\subfigure[\;\; $J=0.4$, MDC]{\includegraphics[width=.44\columnwidth]{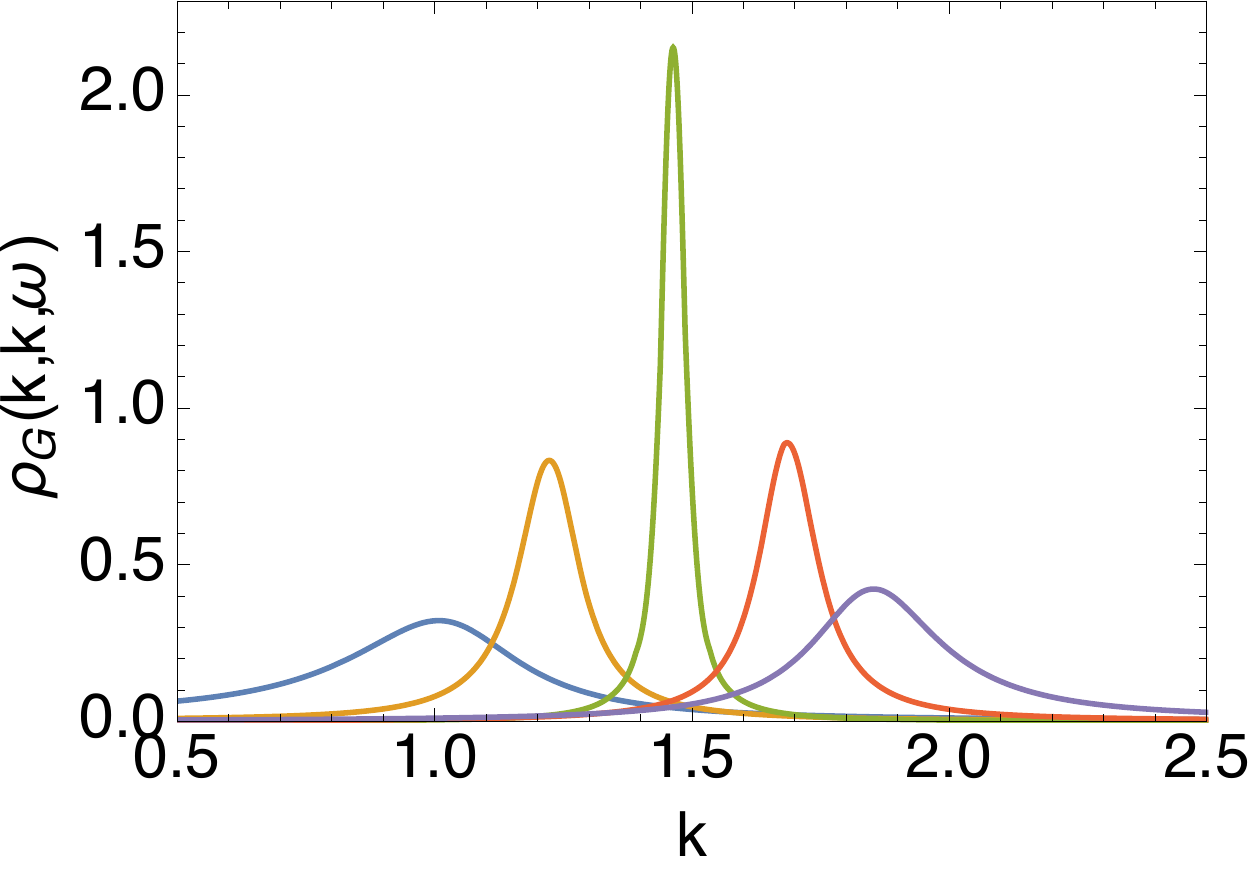}}
\caption{(Color online) EDC and MDC line shapes at different values of superexchange $J$. All EDC figures (a, c, e) or MDC (b, d, f) figures share the same legend respectively. Here the parameters are set as $\delta=0.15$, $t'=0$, $T=105K$ and $J=0, 0.17, 0.4$, in the nodal ($\Gamma\rightarrow X$) direction. Increasing $J$ the peak at the chemical potential  becomes somewhat higher, but it remains qualitatively  similar at all $J$. Besides, increasing $J$ separates the EDC lines further away from $k=k_F$ and brings the MDC lines closer to $\omega=0$.}
\label{EDCMDCJvar}
\end{figure}

\begin{figure}[ht]
\subfigure[\;\; EDC dispersion]{\includegraphics[width=.49\columnwidth]{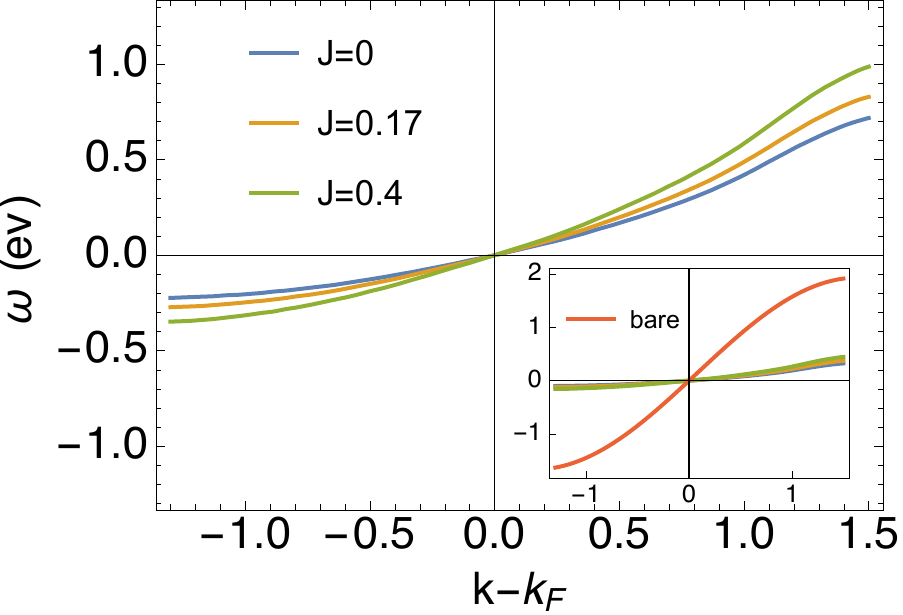}}
\subfigure[\;\; MDC dispersion]{\includegraphics[width=.49\columnwidth]{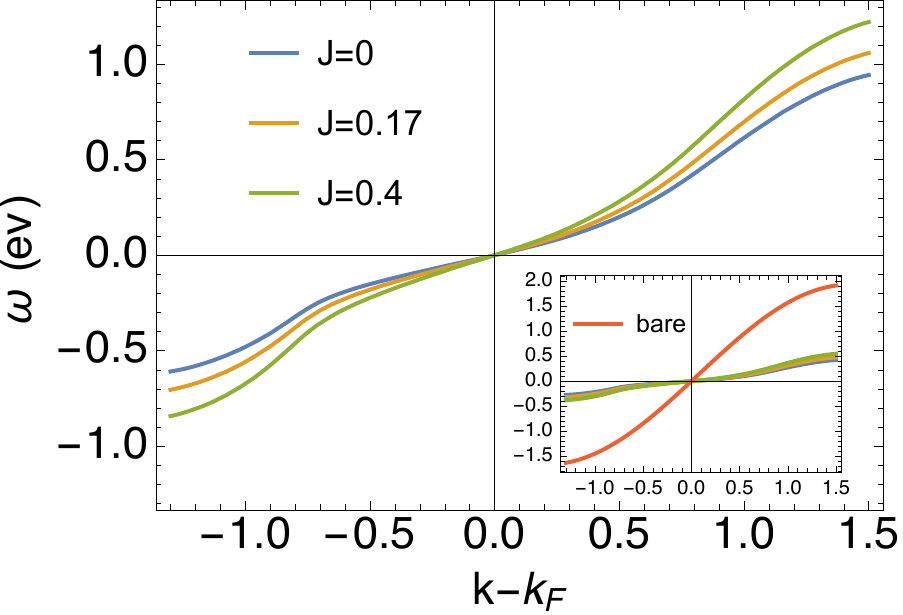}}
\caption{(Color online) EDC and MDC dispersion relation at different values of superexchange $J$. In both cases, increasing $J$ expands the renormalized bandwidth, consistent to \figdisp{EDCMDCJvar} of EDC and MDC lines. Both insets show that the renormalized band is strongly suppressed by correlation compared with the bare one. The energy and $k$ resolution in the present study is not fine enough to deduce the detailed properties of the low energy kinks
(for $\omega \sim .07$ eV)  discussed phenomenologically within ECFL in \refdisp{Matsuyama}.}
\label{dispersionJvar}
\end{figure}

\begin{figure}[!]
\subfigure[\;\; $J=0$, $T=105K$]{\includegraphics[width=.49\columnwidth]{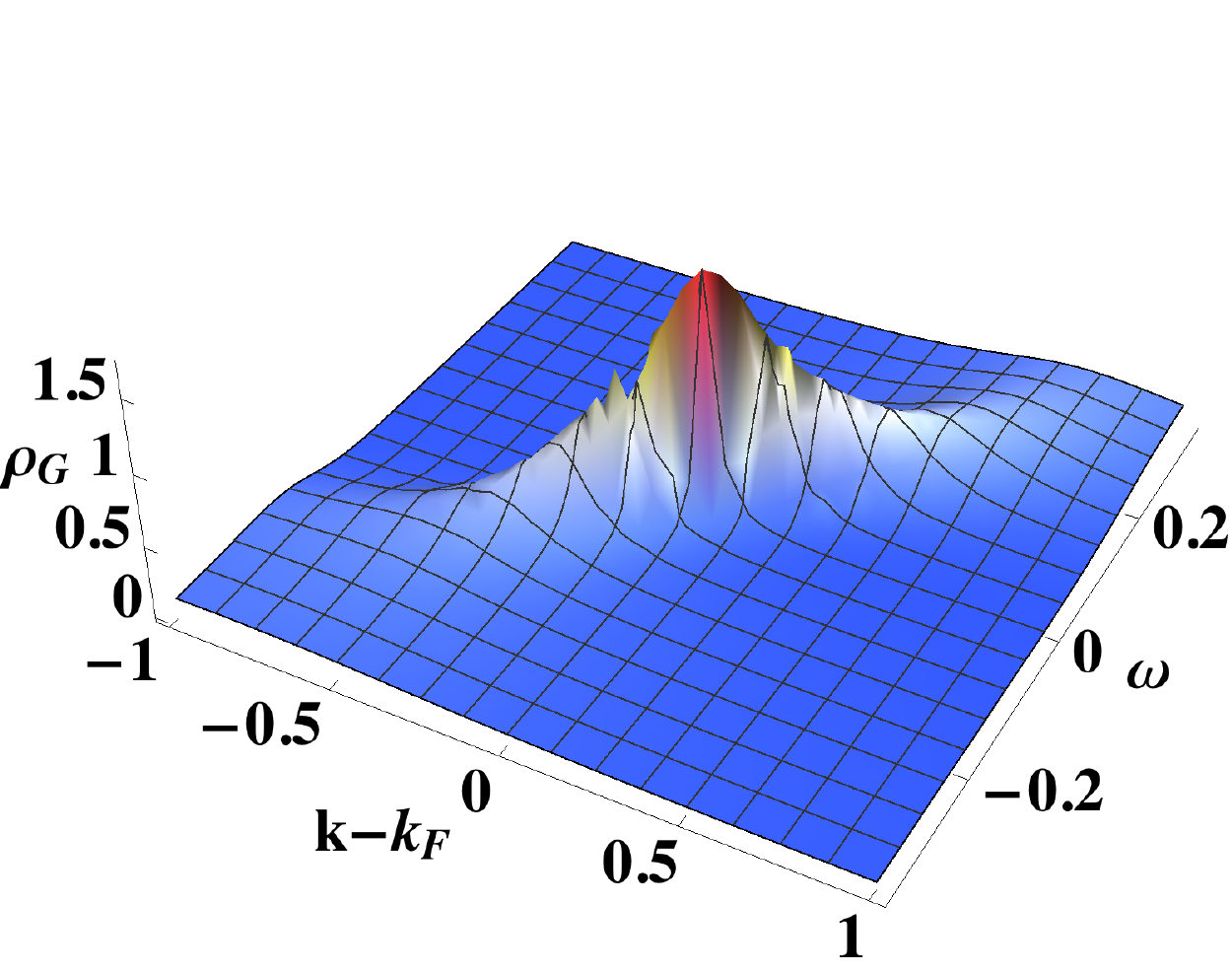}}
\subfigure[\;\; $J=0$, $T=400K$]{\includegraphics[width=.49\columnwidth]{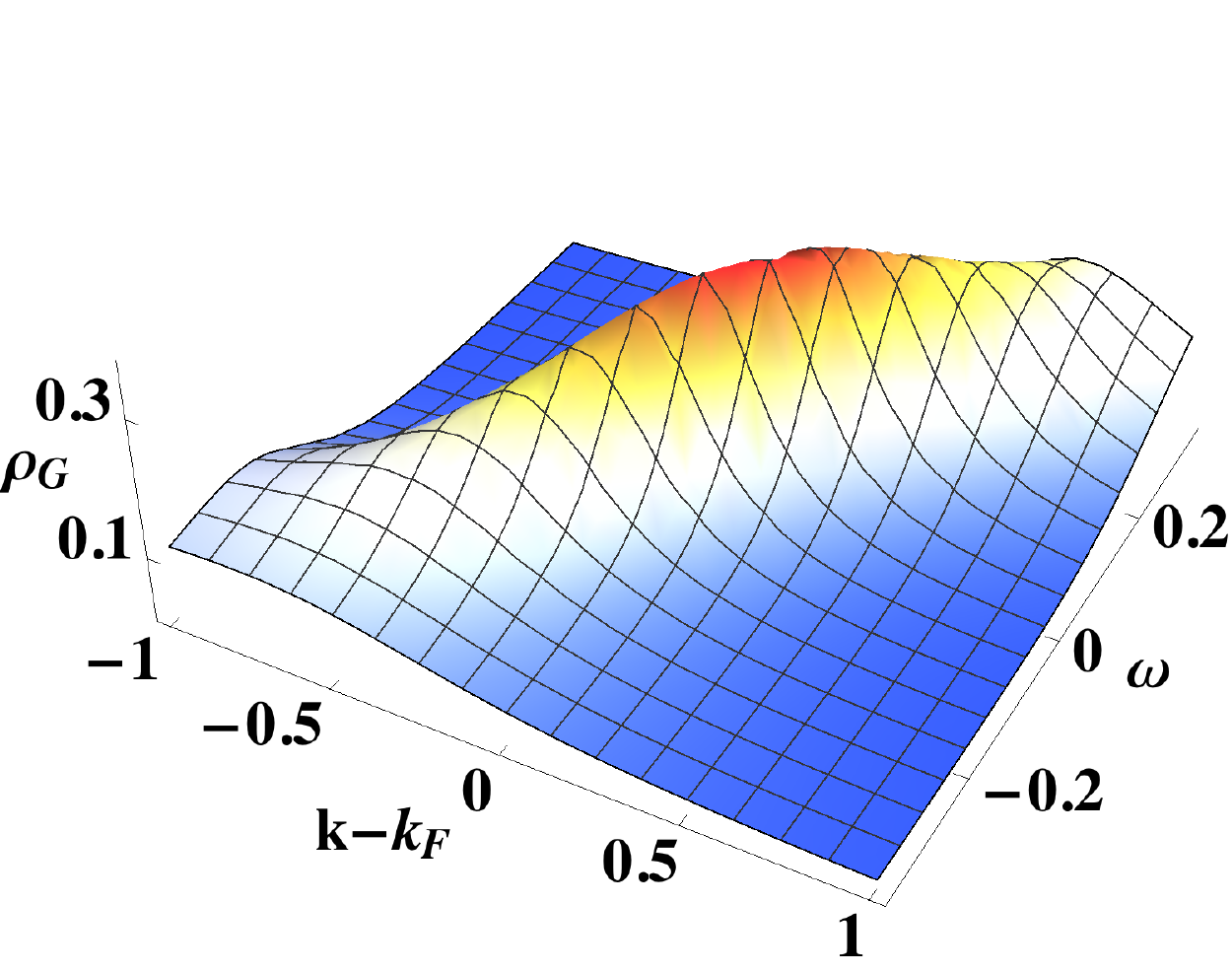}}
\subfigure[\;\; $J=0.4$, $T=105K$]{\includegraphics[width=.49\columnwidth]{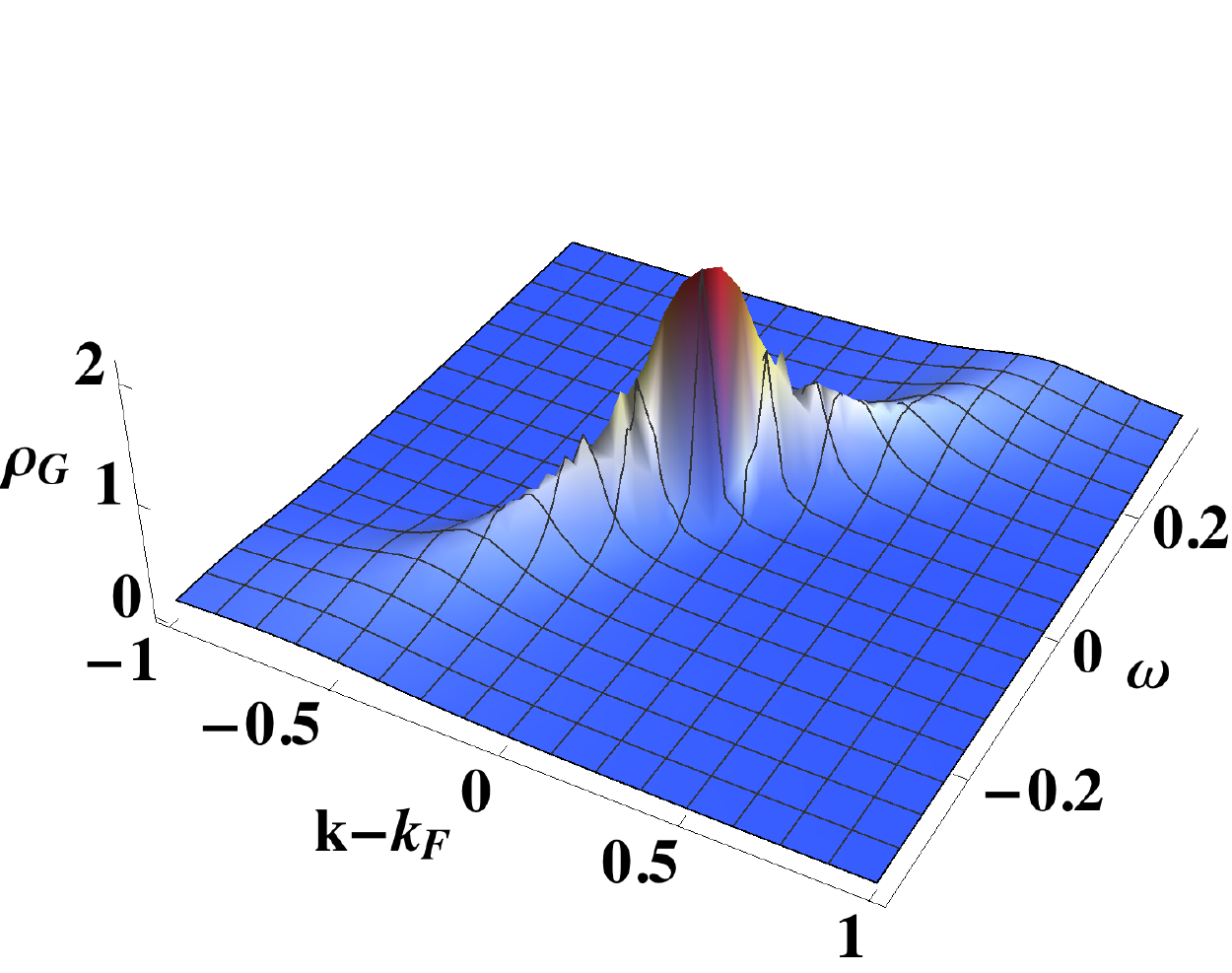}}
\subfigure[\;\; $J=0.4$, $T=400K$]{\includegraphics[width=.49\columnwidth]{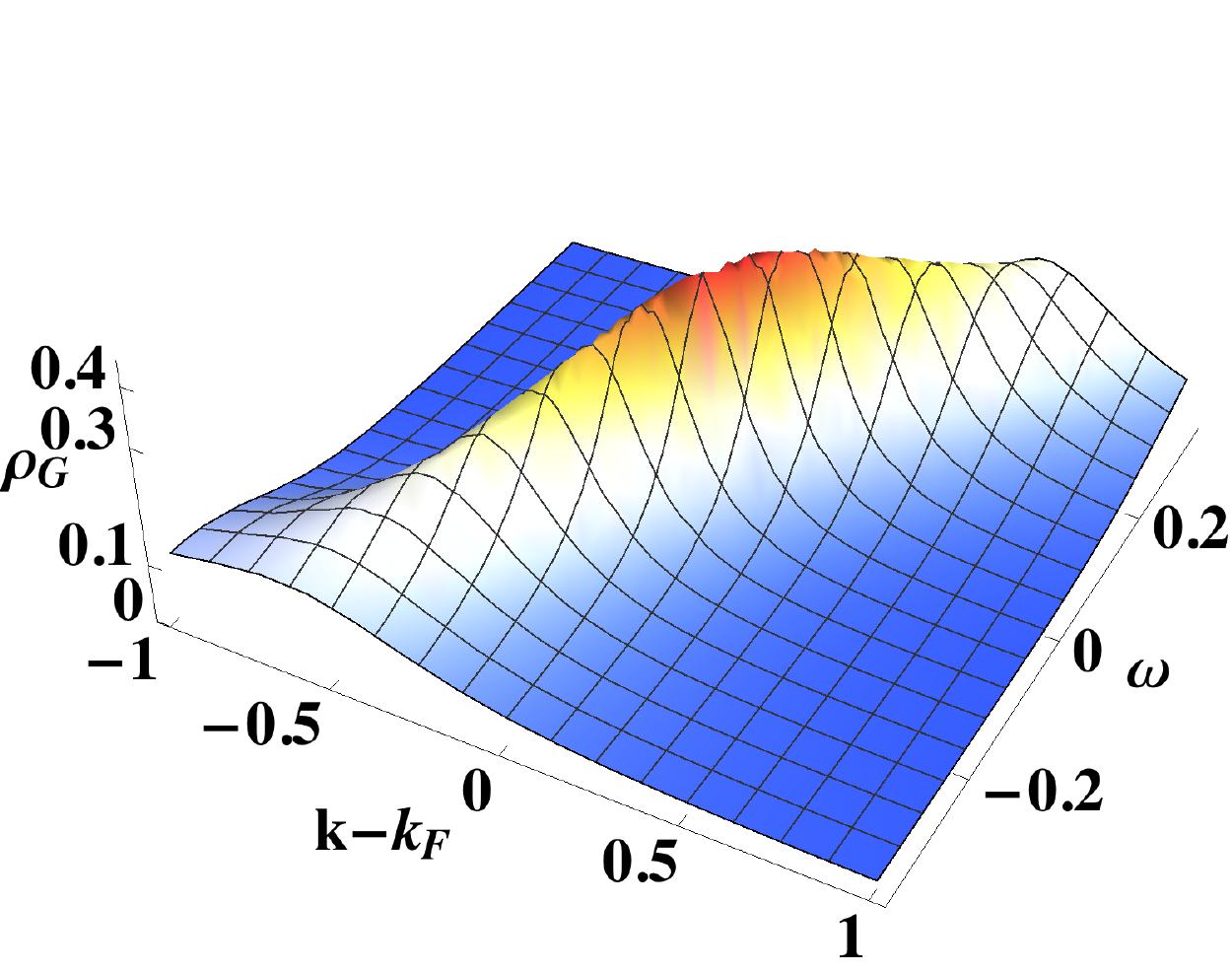}}
\caption{(Color online) 3-d plot of the nodal direction spectral function $\rho_G(k,k,\omega)$. Consistent with \figdisp{EDCMDCJvar}, turning on $J$ increases the peak height and rotates $\rho_G$ counterclockwise with respect to the $z$ axis with $k=k_F$ and $\omega=0$ if viewed from above. This is another facet of the steeper dispersion with $J$ noted in \figdisp{dispersionJvar}.}
\label{3dEDCMDCJvar}
\end{figure}

Above we have discussed the ECFL results at $J=0.17$. We next address the  question of   variation with $J$.
\figdisp{EDCMDCJvar} shows the EDCs and MDCs at different $J$ fixing $t'=0$. Turning on $J$ raises the peak in EDC (a$\rightarrow$c$\rightarrow$e) and MDC (b$\rightarrow$d$\rightarrow$f) slightly. Also, increasing $J$ separates the other  EDC lines further away from $k=k_F$ while brings the other MDC lines closer to $\omega=0$.

We find that  $J$ has an effect on the effective bandwidth. This  can be seen   in the EDC and MDC dispersion relation in \figdisp{dispersionJvar}. As $J$ increases, the EDC and MDC  band separate out more widely, though they are still very  narrow (due to strong correlations) compared to the bare bandwidth. The MDC dispersion shows a high energy feature, namely the  kink (or waterfall). Due to the finite lattice size and to  second order approximation made in the present work, the low energy kink discussed in \refdisp{Matsuyama} cannot be resolved clearly.  Another angle to view the  effect of $J$ is through the 3D-plot of the nodal direction spectral function $\rho_G(k,k,\omega)$ in \figdisp{3dEDCMDCJvar}. It appears that turning on $J$ rotates the spectral function counterclockwise with respect to the $z$ axis with $k=k_F$ and $\omega=0$ if viewed from above. In other words, increasing $J$ extended the renormalized bandwidth with no effect on the Fermi surface location since all curves cross at the same $k_F$. That said, small variation of J does not change the system behavior qualitatively,  and only slightly in quantitative detail. Therefore it is reasonable to set $J=0.17$ from experiment as a representative number and to explore the $k$, $\omega$, $t'$ and $\delta$-dependence of the system.


\begin{figure}[!]
\subfigure[\;\; $t'=-0.4$]{\includegraphics[width=.49\columnwidth]{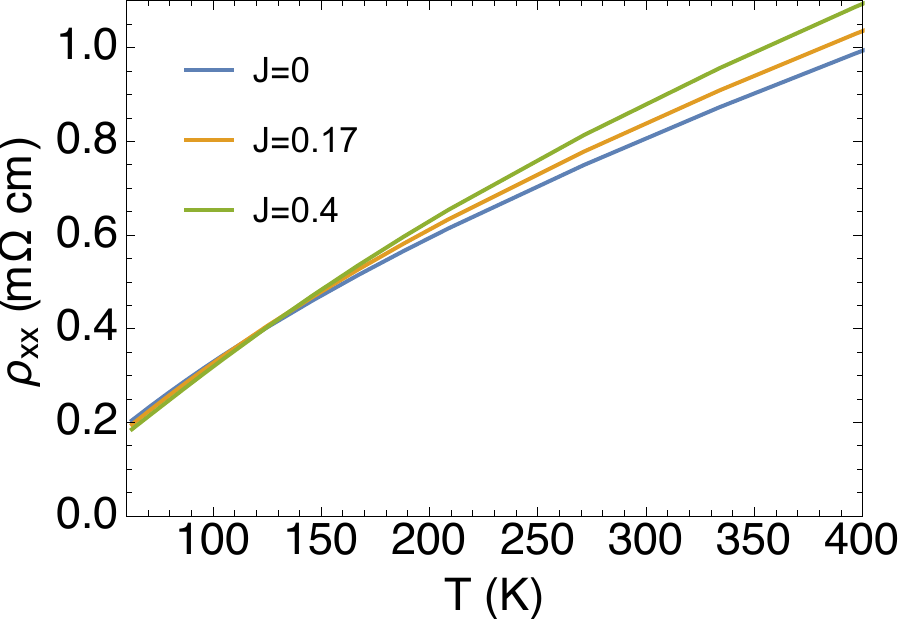}}
\subfigure[\;\; $t'=-0.2$]{\includegraphics[width=.49\columnwidth]{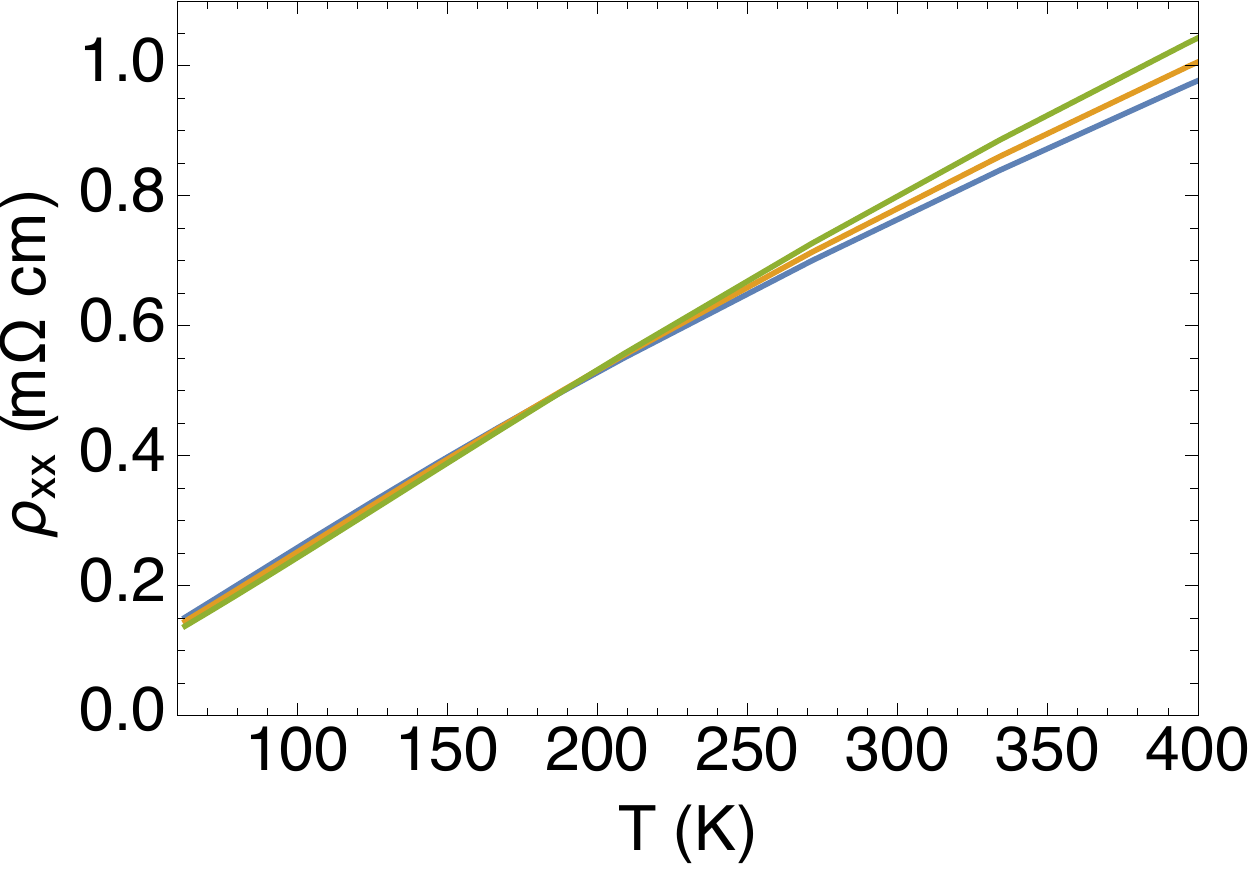}}
\subfigure[\;\; $t'=0$]{\includegraphics[width=.49\columnwidth]{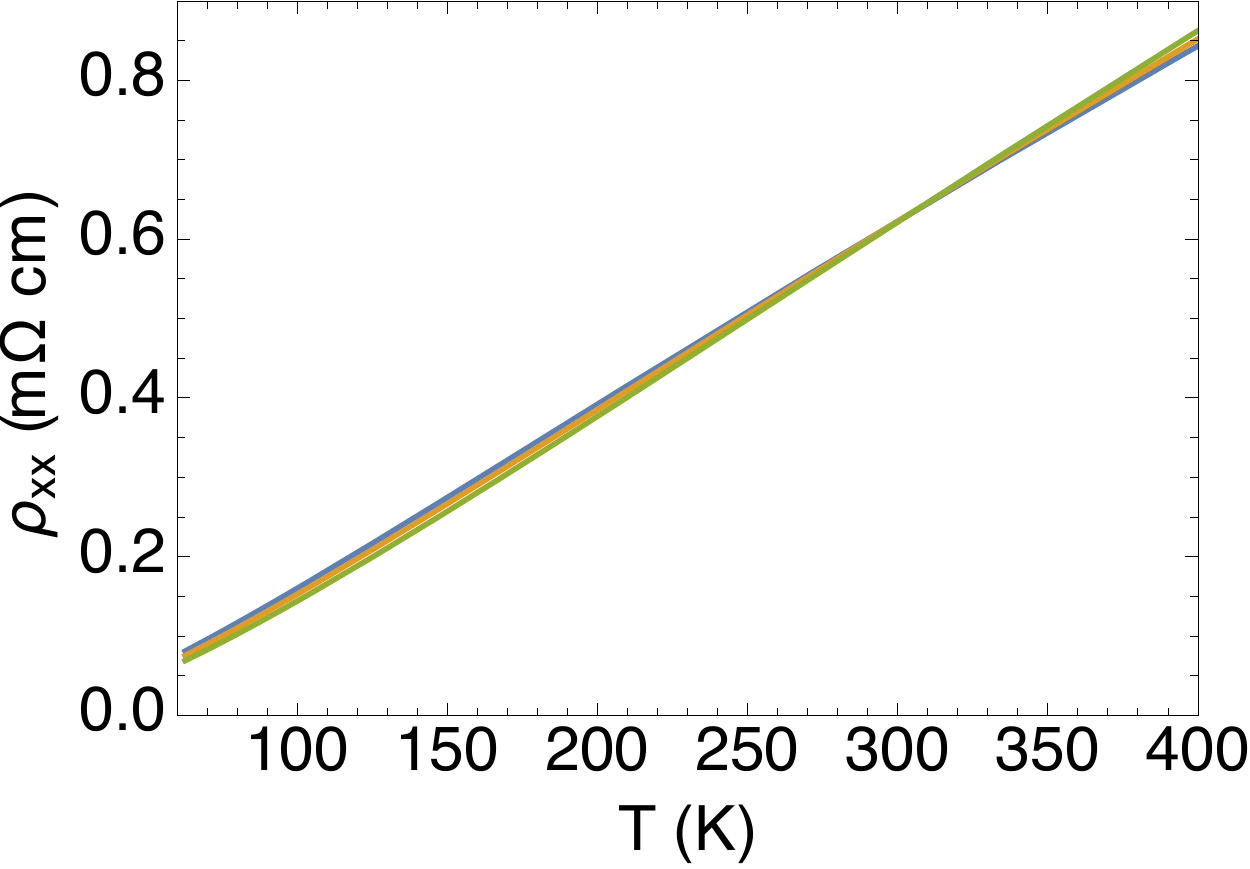}}
\subfigure[\;\; $t'=0.2$]{\includegraphics[width=.49\columnwidth]{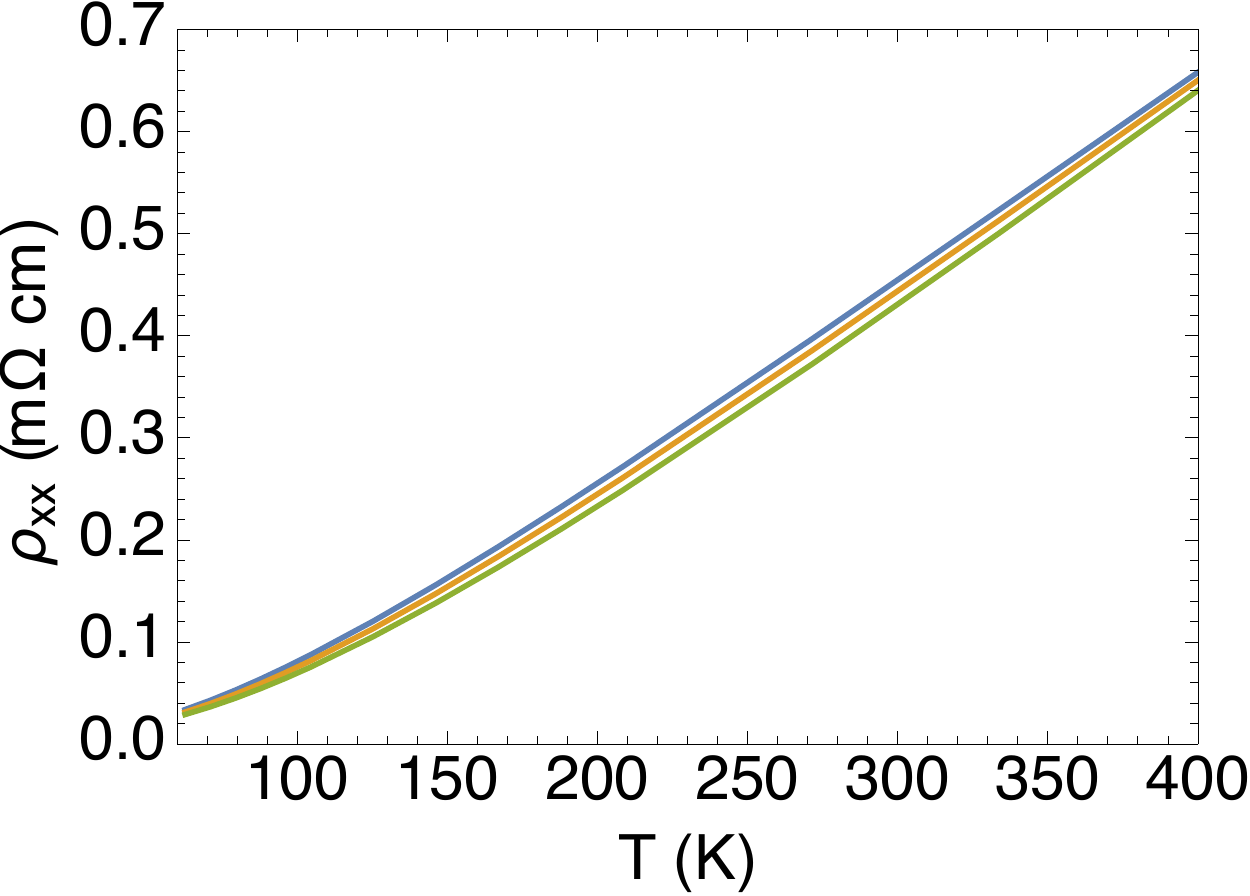}}
\subfigure[\;\; $t'=0.4$]{\includegraphics[width=.49\columnwidth]{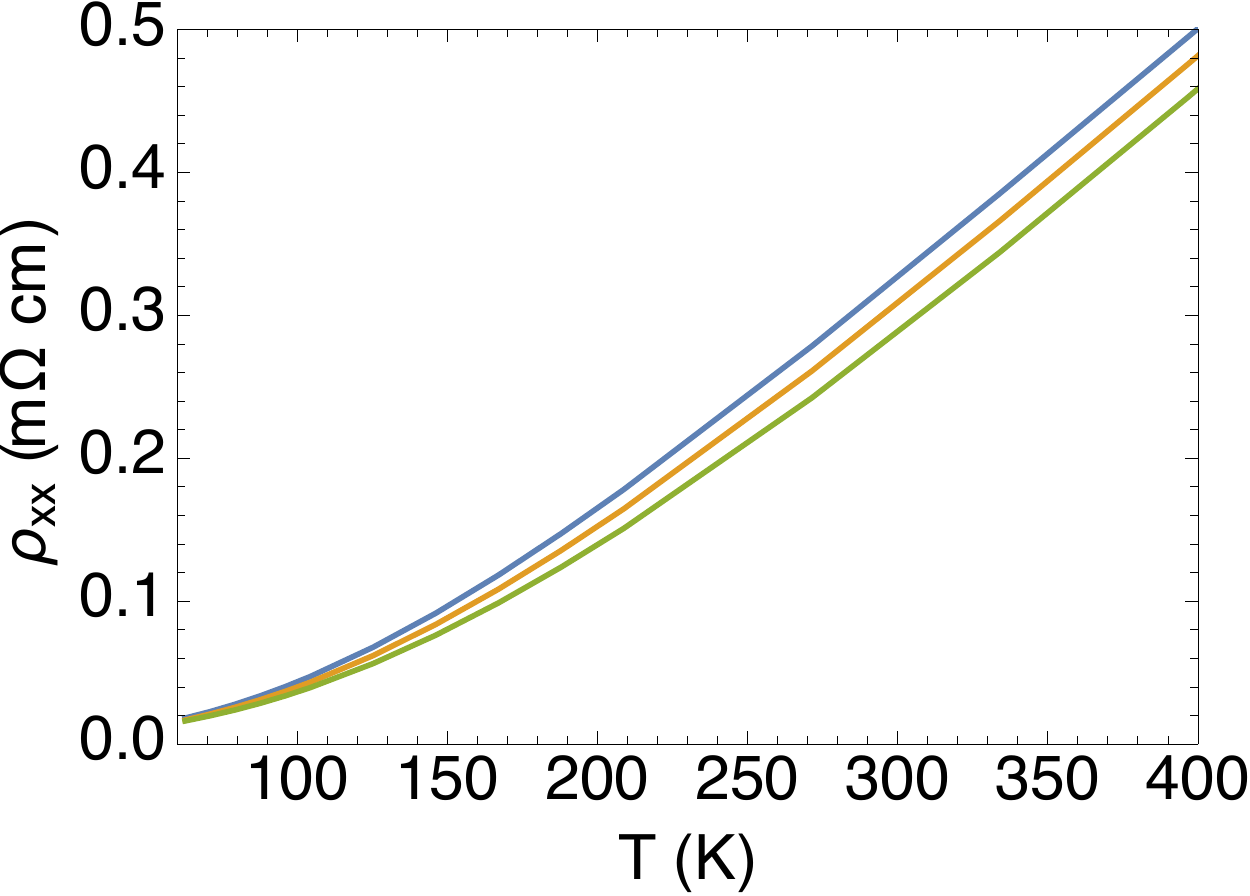}}
\caption{(Color online) Resistivity at $\delta=0.15$ versus T for various $J$ and $t'$ (same legend for all panels). In all $t'$, we observe $J$ variation of the resistivity is small.  As $|t'|$ becomes large  $J$ has a somewhat larger influence on the resistivity. }
\label{resistivityvaryJ}
\end{figure}

From the discussion above, we expect the $k$-average physical quantity like resistivity with significant contribution from the area around the Fermi surface to be insensitive to $J$ variation. \figdisp{resistivityvaryJ} shows the resistivity at different $J$ for fixed $t'$. As expected,  varying $J$ from $0$ to $0.4$ does not make a qualitative difference in the resistivity of the normal state, although it has a relatively stronger effect on the case with larger $|t'|$. 

\section{Conclusion} \label{conclusion}
We apply the recently developed second order ECFL scheme \cite{Sriram-Edward,SP} into studying the 2-d \tJ model with second nearest neighbor hopping $t'$. We have presented the spectral function, self-energy, LDOS, resistivity, Hall number and dynamical susceptibility at low and intermediate temperatures, with $t'$ varying from $-0.4$ to $0.4$ and within a large density region around optimal doping. 

The spectral properties are shown to be consistent with ARPES experiment\cite{ARPES1,ARPES2,ARPES3,ARPES4,ARPES5} on correlated material. The asymmetric EDCs and more symmetric MDCs are observed as expected from the previous study on the phenomenological model of simplified ECFL theory\cite{Gweon}. The weak $k$-dependence of self-energy indicates the relative unimportance of vertex corrections at the densities considered, and gives credence to the use of the bubble approximation for transport.

The curvature change on the resistivity $\rho-T$ curve arises from varying $t'$ and $\delta$, signaling different strength of effective correlation. Both $t'$ and $\delta$ affect the effective electron-electron correlation because $t'$ controls second neighbor hopping process and $\delta$ leaves more or less space for electron movement. As a feature in 2-d, the combination of them determines the geometry of the Fermi surface and therefore the low energy behaviors.

\section{Acknowledgement}
We thank Edward Perepelitsky and Michael Arciniaga for helpful comments on the manuscript.
We thank Shawdong Dong for  helpful suggestions with the computations. The work at UCSC was supported by the U.S. Department of Energy (DOE), Office of Science, Basic Energy Sciences  under Award \# DE-FG02-06ER46319. Computations reported here used the XSEDE Environment\cite{xsede} (TG-DMR170044) supported by National Science Foundation grant number ACI-1053575.



\begin{thebibliography}{99}
\bibitem{hall-extra} For this we additionally assume that the magnetic field vertex also assumes  its  bare value. This assumption requires further  validation in 2-dimensions within the \tJ model,  hence the results for the Hall conductivity are less reliable than the longitudinal conductivity.
\end{thebibliography}
\end{document}